\documentclass[a4paper,fleqn,usenatbib]{mnras}
\usepackage{newtxtext,newtxmath}
\usepackage[T1]{fontenc}
\usepackage{ae,aecompl}

\usepackage{graphicx}	
\usepackage{amsmath}	    
\usepackage{amssymb}	    
\usepackage{tikz}

\title[Singularities in Gravitational Lensing]{Finding Singularities in Gravitational Lensing}

\author[Meena and Bagla]{
Ashish Kumar Meena,$^{1}$\thanks{E-mail: ashishmeena@iisermohali.ac.in}
J. S. Bagla,$^{1}$\thanks{E-mail: jasjeet@iisermohali.ac.in}
\\
$^{1}$Indian Institute of Science Education and Research Mohali,
Knowledge City, Sector 81, Sahibzada Ajit Singh Nagar, Punjab 140306,
India\\}

\date{Accepted XXX. Received YYY; in original form ZZZ}

\pubyear{2019}

\begin{document}
\label{firstpage}
\pagerange{\pageref{firstpage}--\pageref{lastpage}}
\maketitle

\begin{abstract}
  The number of strong lens systems is expected to increase
  significantly in ongoing and upcoming surveys. 
  With an increase in the total number of such systems we expect to
  discover many configurations that correspond to unstable caustics. 
  In such cases, the instability can be used to our advantage for 
  constraining the lens model.  
  We have implemented algorithms for detection of different types of
  singularities in gravitational lensing. 
  We apply our approach on a variety of lens models and then go on to
  test it with the inferred mass distribution for Abell 697 as an
  example application. 
  We propose to represent lenses using $A_{3}$-lines and singular
  points ($A_4$ and $D_4$) in the image plane.
  We propose this as a compact representation of complex lens
  systems that can capture all the details in a single snapshot. 
\end{abstract}


\section{Introduction}
\label{sec:Introduction}

Strong gravitational lenses are unique probes of the Universe.
By producing multiple images, they provide constraints on the lens
mass distribution \citep{2011A&ARv..19...47K}. 
The high magnification due to lensing gives us the opportunity to look
further into the history of the Universe by observing magnified
sources which otherwise would have remained unobserved, see e.g.,
\citet{2018MNRAS.479.5184A}. 
In a given lensing system, the observed configuration and
magnification of multiple images depends on properties of the lens and
the location of the source with respect to the lens.
The set of all points in the plane of the lens is called the image
plane: here we are working in the small angle approximation. 
Each point on the image plane can be mapped to a plane at the source
redshift, the so called source plane.
For a given lens, and the distance to the source, there is a set of
directions where the magnification is formally infinite.
The set of points on the image plane representing these directions
form the critical curves.
As all sources have a finite size, magnification is always
finite. 
The critical curves, mapped to the source plane form the caustics. 
High magnification images are formed if the source lies on or close to
a caustic \citep{1989Sci...245..824B,Schneider1992}.

We have mentioned above that the critical curves and caustics
correspond to infinite magnification.
This happens because the lens mapping at these points is singular: a
finite solid angle element in the image plane gets mapped to a line or
a point in the source plane.
The structure of the caustic depends on the form of the singularity:
singularities of the lensing map can be classified using catastrophe
theory \citep{1980PrOpt..18..257B,PostonStewart 1978,Gilmore 1981}. 
The use of catastrophe theory in gravitational lensing was first
discussed by \citet{BlandfordNarayan1986} in case of elliptical
lenses.
Later this was discussed by \citet{rajaram_1990}, \citet{Kassiola 1992},
\citet{Schneider1992} (hereafter SEF) and \citet{Petters 2001}.
Independently, classification of singularities in the same map in the
context of Zel'dovich approximation was done by \citet{Arnold 1982}.

One can divide singularities of the lensing map into two types: stable
(fold and cusp) and unstable singularities (swallowtail, umbilics).
Stable caustics are called so because a small perturbation in the
lensing potential leads to a correspondingly small shift in the
location of fold and cusp.
On the other hand, the so called unstable caustics may disappear
entirely on introduction of a small perturbation.
In view of this, the focus of most of the studies has been on stable
caustics with only a few efforts to improve our understanding of image
formation and characteristics of unstable singularities in realistic
lens maps \citep{Bagla 2001,Orban 2009} though these caustics have
been known and studied theoretically \citep{Schneider1992}. 

In this work we propose that unstable caustics can be potentially useful
to constrain lens models much more strongly than the stable
caustics. 
The unstable caustics have a stronger variation of magnification
around the singular points as compared to stable caustics.
Further, if we can predict the location of unstable singularities in the
image plane then these regions may be targeted for deep surveys to
look for highly magnified sources \citep{Yuan 2012,Zheng 2012,Coe
  2013,Mcleod 2015,Ebeling 2018}.  
The high magnification comes with a characteristic image formation,
and due to the unstable nature of the singularity the characteristic
image formation is visible only for a small range of source redshift. 
With upcoming facilities like EUCLID \citep{Laureijs 2009}, LSST
\citep{Ivezic 2008}, Dark Energy Survey (DES)
\citep{2016MNRAS.460.1270D}, JWST \citep{gardner_2006}, WFIRST \citep{2019arXiv190205569A},
the number of strong lenses will increase by more than an order of
magnitude in the next decade.
Thus the possibility of observing lensing near unstable singularities
is higher and therefore it is timely that we carry out a detailed
study.
Preliminary results of this study were reported in \citet{Bagla 2001}.
We use algorithms described briefly in that work.
We have developed and refined these algorithms further, and used 
them in case of simple lens models. 
The algorithms make use of the definitions of singularities, e.g., see
\citep{Arnold 1982} and are similar to those reported in
\citet{Hidding 2014} for the case of Zel'dovich approximation in two
dimensions. 
These algorithms allow us to locate {\sl all} singularities of the
lensing map in the image plane starting from the lensing potential. 
We then proceed to analyse lens models with one or two major
components and study the singularities.
We also study variation in singularities in presence of perturbing
shear.
We illustrate characteristic image formations for each type of
unstable singularity. 
This effort is complementary to an atlas of observed images in 
exotic lenses \citep{Orban 2009} and makes the task of predicting
possibility of such image formations much easier.

This paper is organized as follows.
In \S{2} we review the basics of the gravitational lensing and
introduce the quantities that are useful for the following discussion.
In \S{3} we review the classification of singularities and their
properties.
\S{4} contains a description of the algorithm used.
Results are given in \S{5} for a variety of lenses.
Summary and conclusions are presented in \S{6}.
We discuss possibilities for future work in this section.

\section{Theory}
\label{sec:Theory}

In this section we review the basics of gravitational lensing that are
relevant for the following discussion. 
We use the formalism given in SEF.
This is followed by an introduction to singularities in gravitational
lensing.

The lens equation is a map between the image plane and the source
plane.  
It can be written in a dimensionless form as:
\begin{equation}
\mathbf{y} = \mathbf{x}-\mathbf{\zeta}(\mathbf{x}),
\label{eq:(lens equation)}
\end{equation}
where $\mathbf{x}$ (in the image plane) and $\mathbf{y}$ (in the
source plane) are two-dimensional vectors with respect to the optic
axis.
The choice of optic axis is arbitrary.
And $\mathbf{\zeta}(\mathbf{x})$ is the scaled deflection angle for
a light ray in lens plane at $\mathbf{x}$. 
The scaled deflection angle $\mathbf{\alpha}(\mathbf{x})$ is related
to the projected lensing potential $\psi(\mathbf{x})$: 
$\mathbf{\zeta}(\mathbf{x})$ = $\left( D_{ds}/D_{s}
\right)\nabla\psi(\mathbf{x})$.
The projected lensing potential is given by,
\begin{equation}
\psi(\mathbf{x}) = \frac{1}{\pi} \int d^{2}x' \kappa(\mathbf{x}')
\ln|\mathbf{x}-\mathbf{x}'|, 
\label{eq:(delection potential)}
\end{equation}
where 
\begin{equation}
\kappa = \frac{\Sigma\left(\mathbf{x}\right)}{\Sigma_{cr}},\quad
\Sigma_{cr} =\frac{c^{2}}{4\pi G D_{d} }. 
\label{eq:(critical density)}
\end{equation}
The convergence, $\kappa$ represents the dimensionless surface mass
density of the lens and $\Sigma_{cr}$ denotes the critical density for
a source at infinity.
$D_{s}, D_{d}, D_{ds}$ represent the angular diameter distances to the
source, to the lens (sometimes referred to as the deflector) and from
lens to source.  

The properties of the lens mapping~(\ref{eq:(lens equation)}) can be
described by the Jacobian matrix: 
\begin{equation}
A(\mathbf{x})=\frac{\partial\mathbf{y}}{\partial\mathbf{x}}=
\delta_{ij}-\left(\frac{D_{ds}}{D_{s}} \right)\psi_{ij} 
\label{eq:(jacobian)}
\end{equation}
where subscripts of deflection potential denote partial derivatives,
i.e.,
\begin{displaymath}
  \psi_{ij}=\frac{\partial^2\psi}{\partial x_{i}\partial x_{j}} .
\end{displaymath}
This is also known as the deformation tensor and it describes the
distortion of the observed images:
\begin{equation}
\psi_{ij} = \left(\begin{array}{cc} \kappa+\gamma_{1} & \gamma_{2}\\
                    \gamma_{2} & \kappa-\gamma_{1} \end{array}\right), 
\label{eq:(deformation tensor)}
\end{equation}
where we have introduced convergence $\kappa$ and the components of
the shear tensor $\gamma\equiv\gamma_{1}+\iota\gamma_{2}$, which can
be written in terms of derivatives of projected lensing potential as, 
\begin{equation}
\kappa =\frac{1}{2}\left(\psi_{11}+\psi_{22}\right),
\label{eq:(convergence)}
\end{equation}
\begin{equation}
\gamma_{1} =\frac{1}{2}\left(\psi_{11}-\psi_{22}\right),\quad
\gamma_{2} = \psi_{12}. 
\label{eq:(shear)}
\end{equation}
The convergence $\kappa$ introduces isotropic distortion in the
image, i.e., the image will be rescaled by a constant factor in all
directions.
Shear, as it derives from the traceless part of the
deformation tensor, distorts the image by
stretching it in one direction while compressing it in the other
direction. 
As a result, for $\gamma\neq0$, a circular source will have an
elliptical image.  
The magnification factor for an image formed at $\mathbf{x}$ is given
by: 
\begin{equation}
\mu\left(\mathbf{x}\right) \equiv \frac{1}{det
  A\left(\mathbf{x}\right)} =
\frac{1}{\left(1-a\alpha\right)\left(1-a\beta\right)}, 
\label{eq:(magnification)}
\end{equation}
where $\alpha$ and $\beta$ are eigenvalues of the deformation
tensor($\alpha \geq \beta$) and $a={D_{ds}}/{D_{s}}$. 
The magnification (formally) goes to infinity at points where either
$\alpha=1/a$ or $\beta=1/a$ or both $\alpha=1/a=\beta$.
As mentioned in the introduction, the finite size of real sources
leads to a finite magnification.   
These points with infinite magnification are singularities of the lens
mapping.
These singularities form smooth closed curves in the image plane, known as
critical curves.
The corresponding curves (not necessarily smooth) in the source plane
are known as caustics.
Following equation~(\ref{eq:(magnification)}), one can see that the
critical curves are eigenvalue contours with a value $1/a$.
This implies that for a given lens system, the position of critical
curves in the image plane can be completely determined by the
deformation tensor.
The following section uses the deformation tensor in terms of its
eigenvalues and eigenvectors to classify the different kind of
singularities that can occur in strong gravitational lensing. 

\begin{figure*}
\centering
\begin{tikzpicture}
\node[anchor=south west,inner sep=0] (image) at (0,0) {\includegraphics[width=\textwidth,height=5.5cm,width=5.5cm]{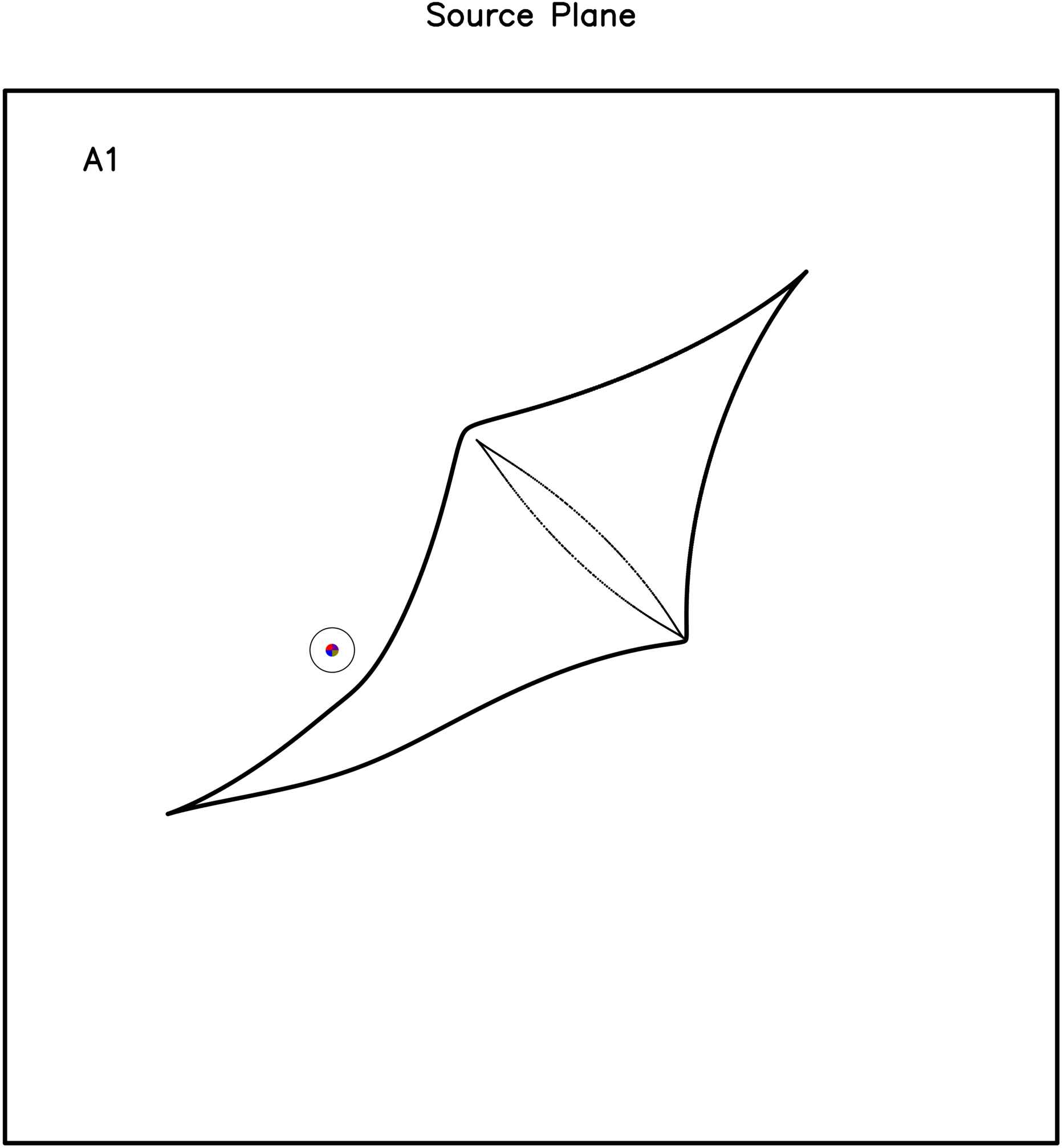}};
\node[anchor=south west,inner sep=0] (image) at (5.5,0) {\includegraphics[width=\textwidth,height=5.5cm,width=5.5cm]{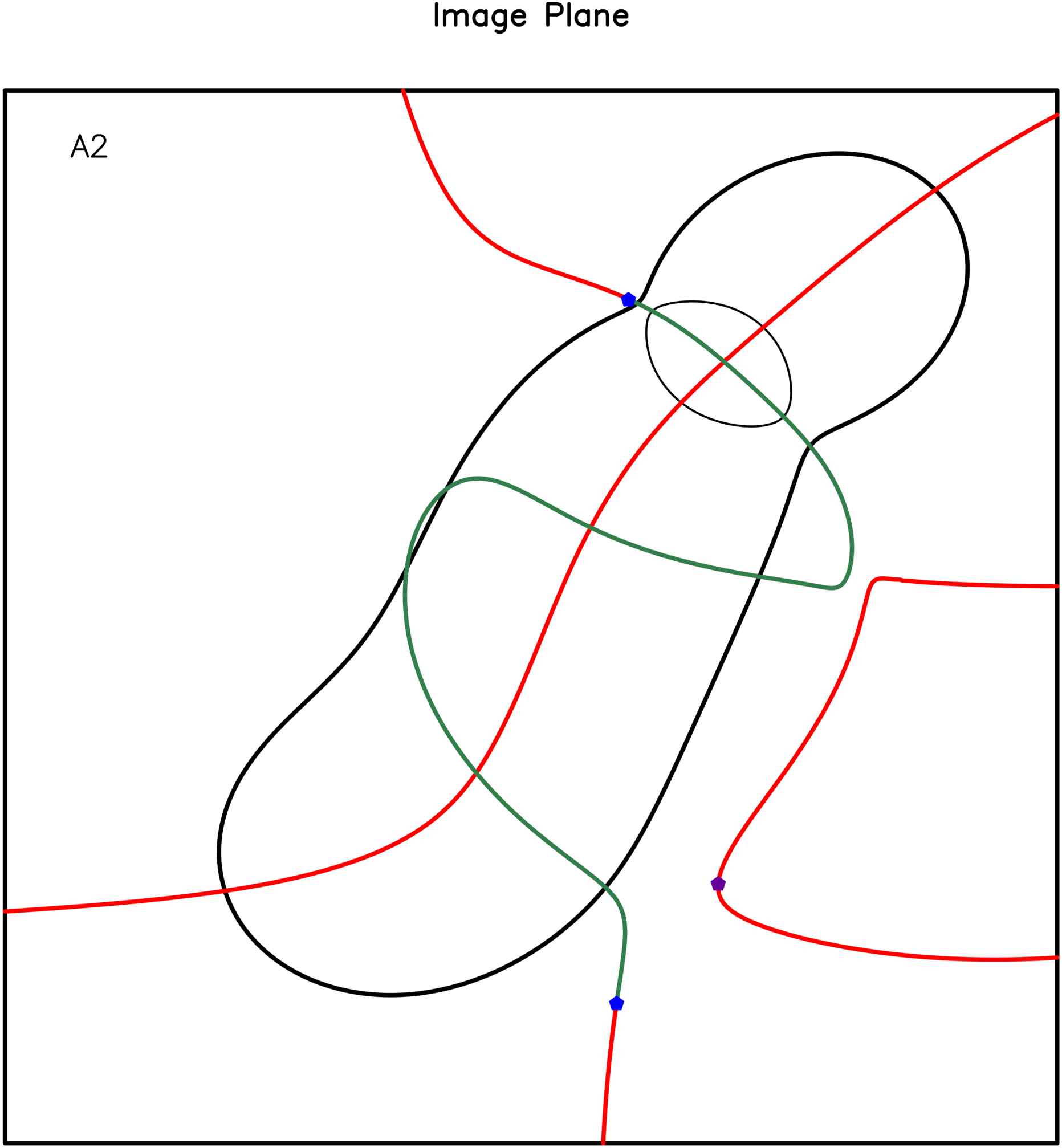}};
\node[anchor=south west,inner sep=0] (image) at (11.0,0) {\includegraphics[width=\textwidth,height=5.5cm,width=5.5cm]{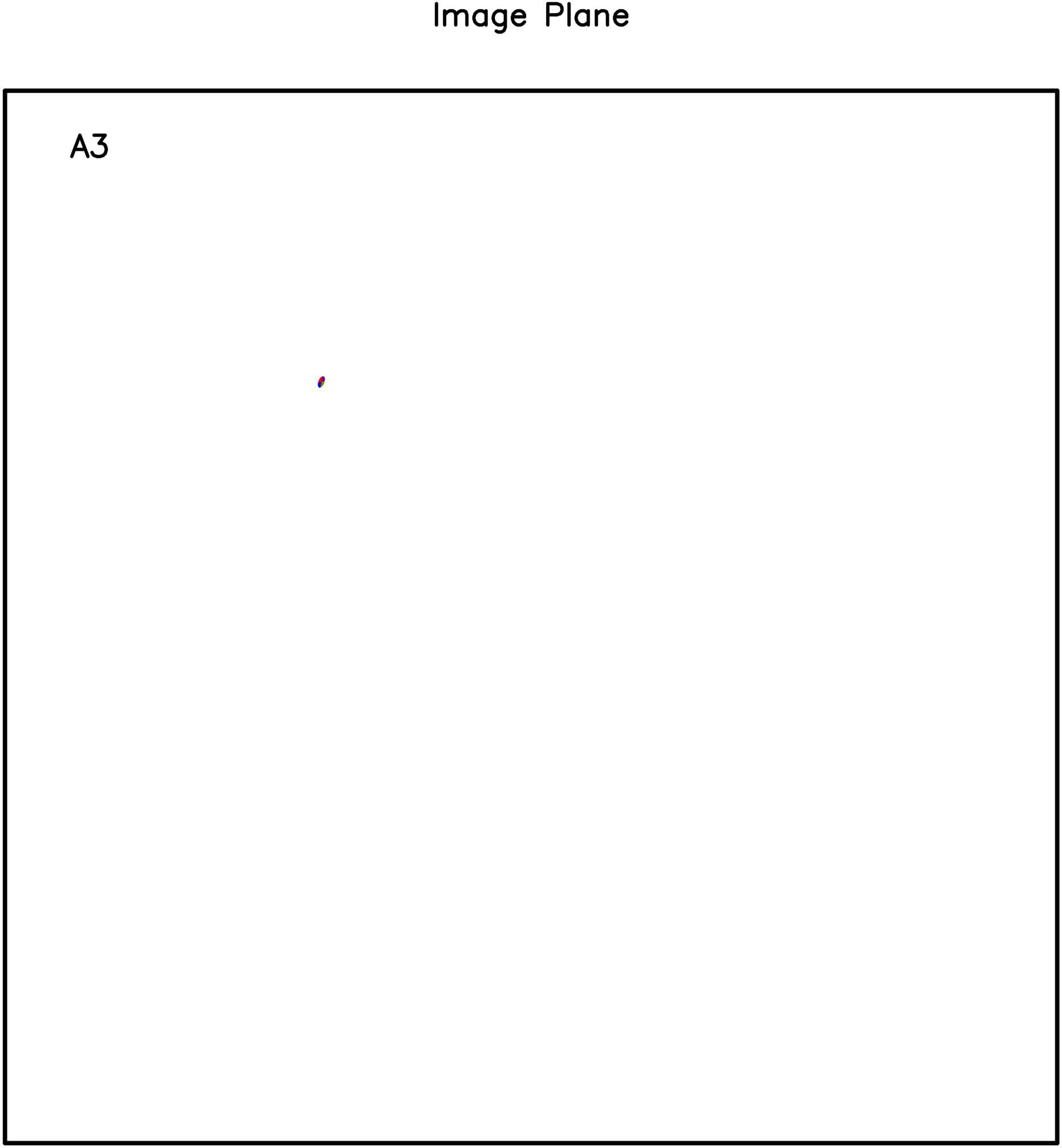}};
\end{tikzpicture}
\end{figure*}
\begin{figure*}
\centering
\begin{tikzpicture}
\node[anchor=south west,inner sep=0] (image) at (0,0) {\includegraphics[width=\textwidth,height=5.5cm,width=5.5cm]{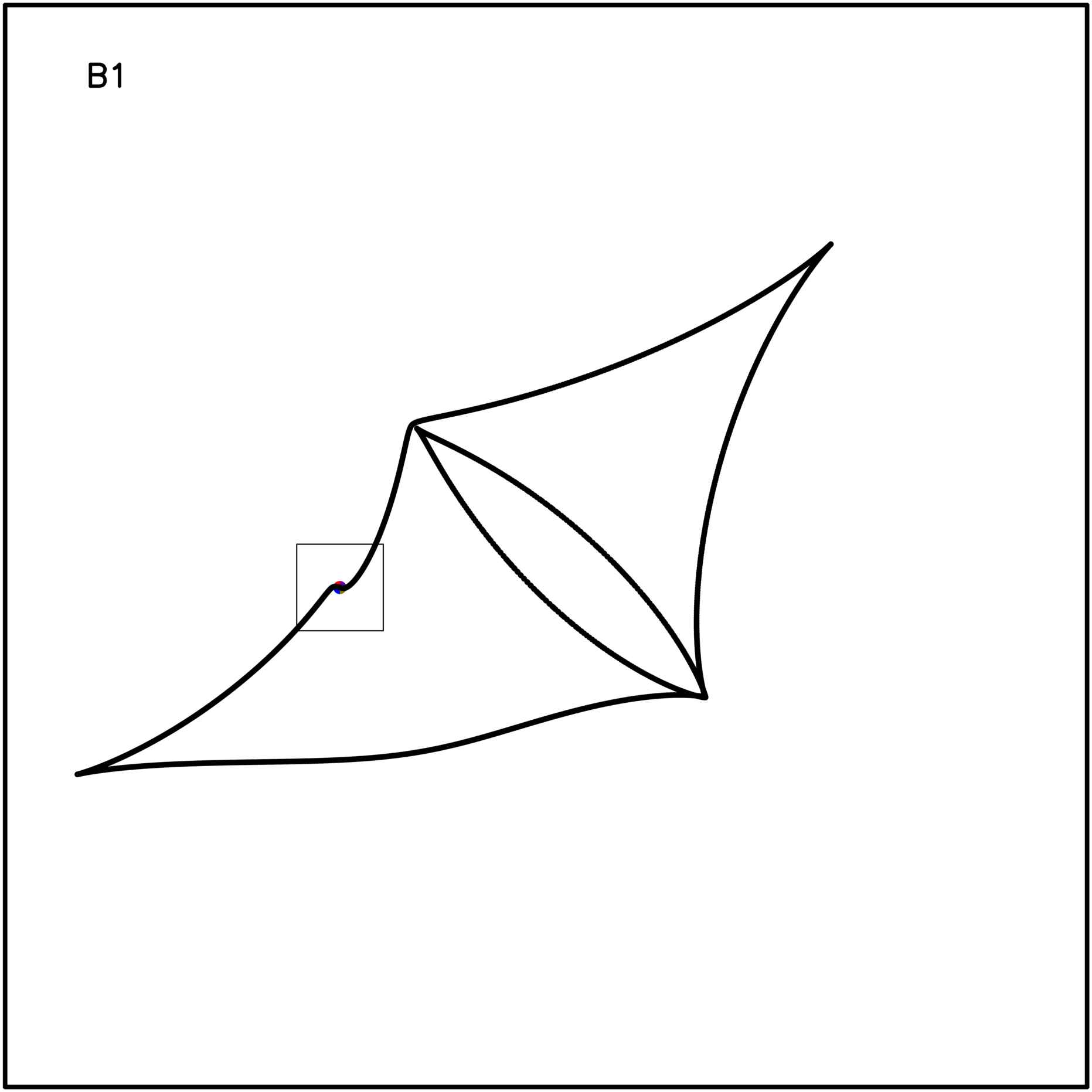}};
\begin{scope}[x={(image.south east)},y={(image.north west)}]
\node[anchor=south west,inner sep=0] (image) at (0.7,0.12) {\includegraphics[width=\textwidth,height=1cm,width=1cm]{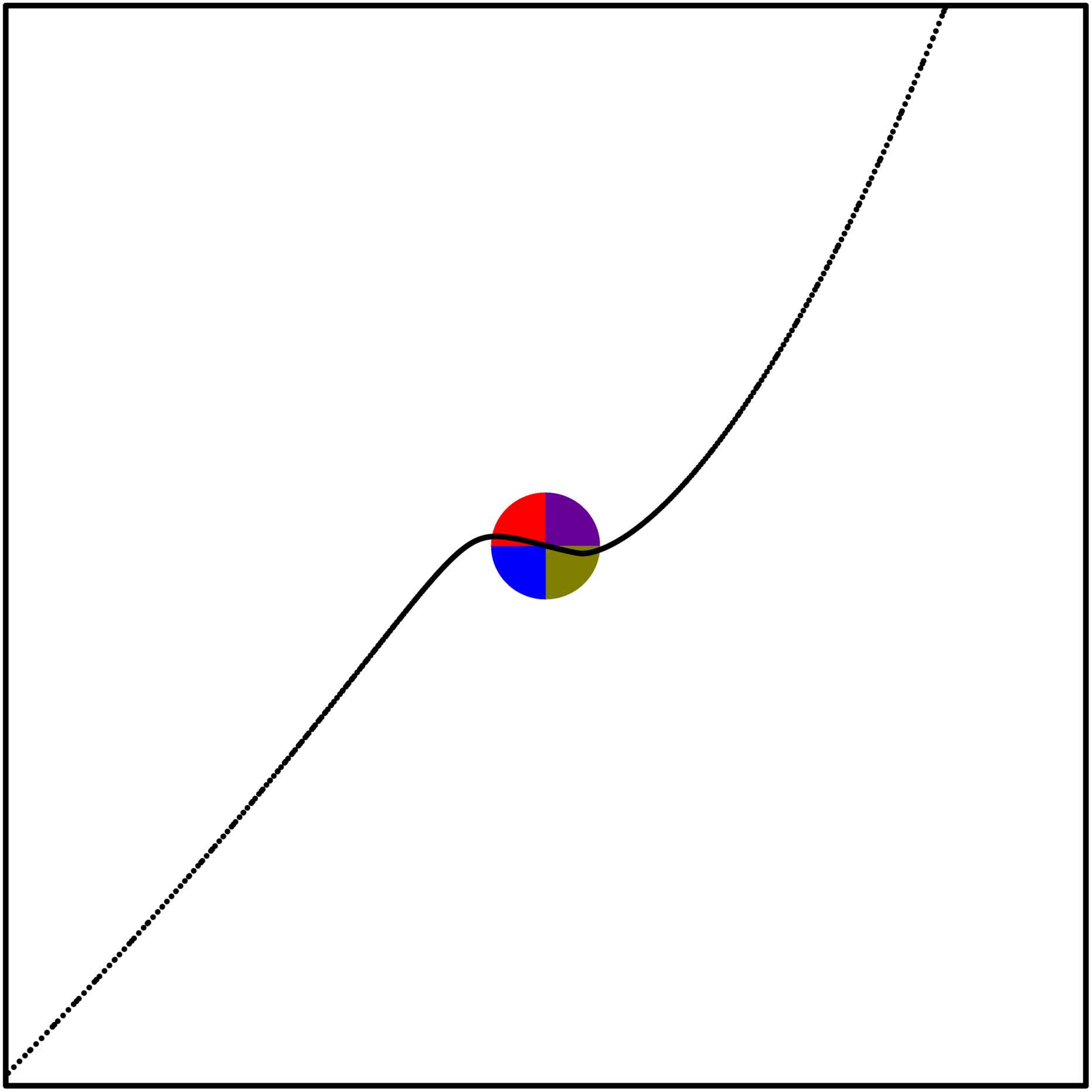}};
\end{scope}
\node[anchor=south west,inner sep=0] (image) at (5.5,0) {\includegraphics[width=\textwidth,height=5.5cm,width=5.5cm]{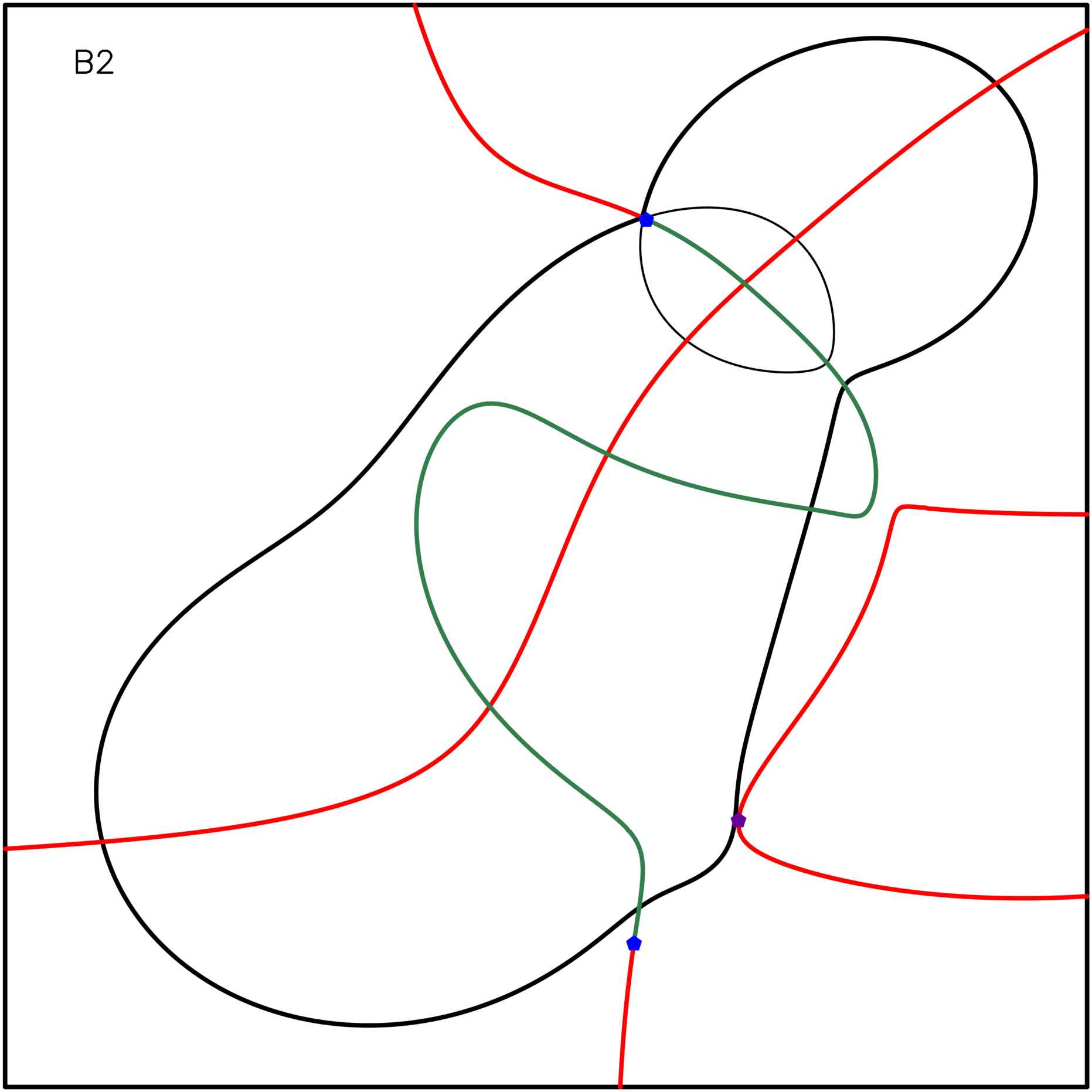}};
\node[anchor=south west,inner sep=0] (image) at (11.0,0) {\includegraphics[width=\textwidth,height=5.5cm,width=5.5cm]{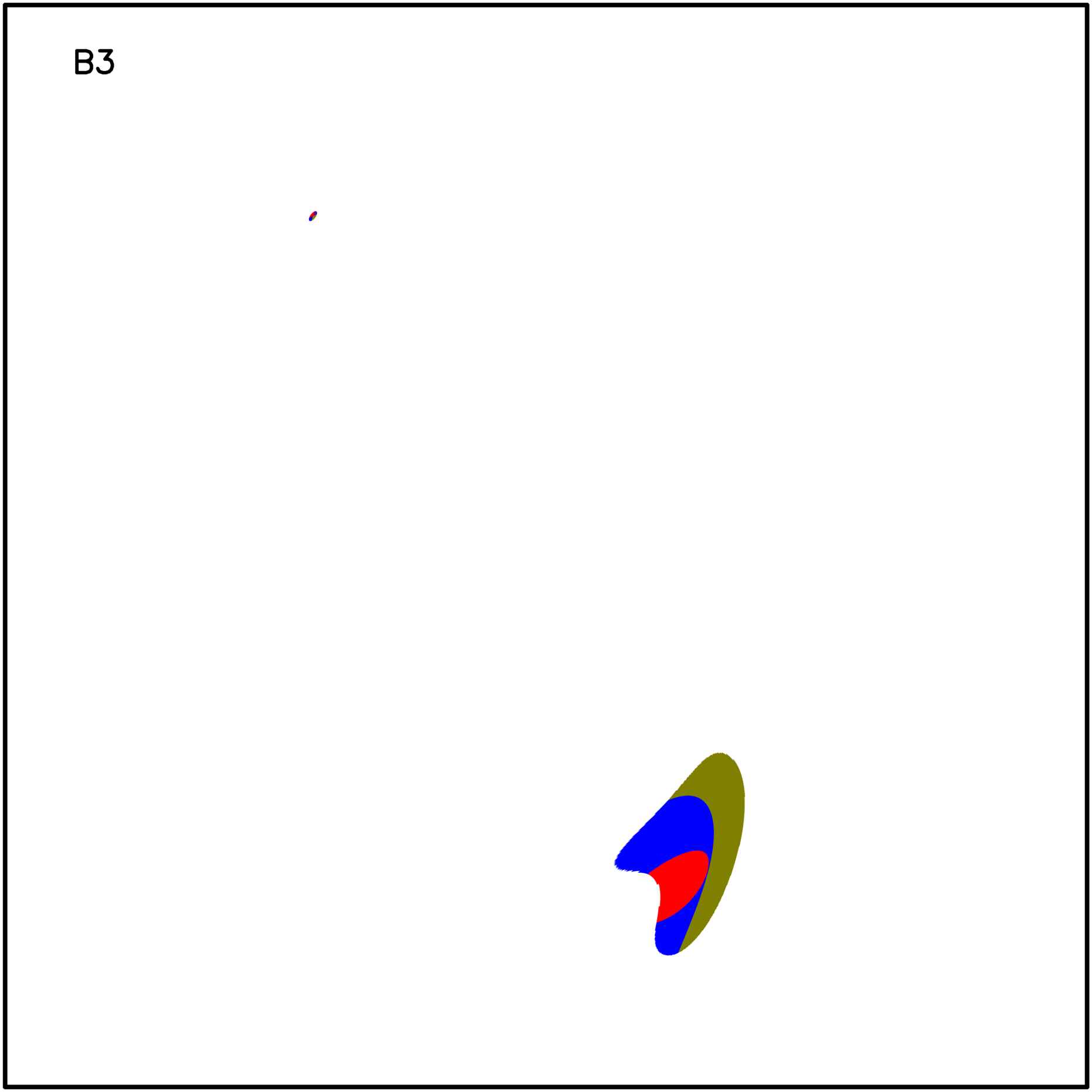}};
\end{tikzpicture}
\end{figure*}
\begin{figure*}
\centering
\begin{tikzpicture}
\node[anchor=south west,inner sep=0] (image) at (0,0) {\includegraphics[width=\textwidth,height=5.5cm,width=5.5cm]{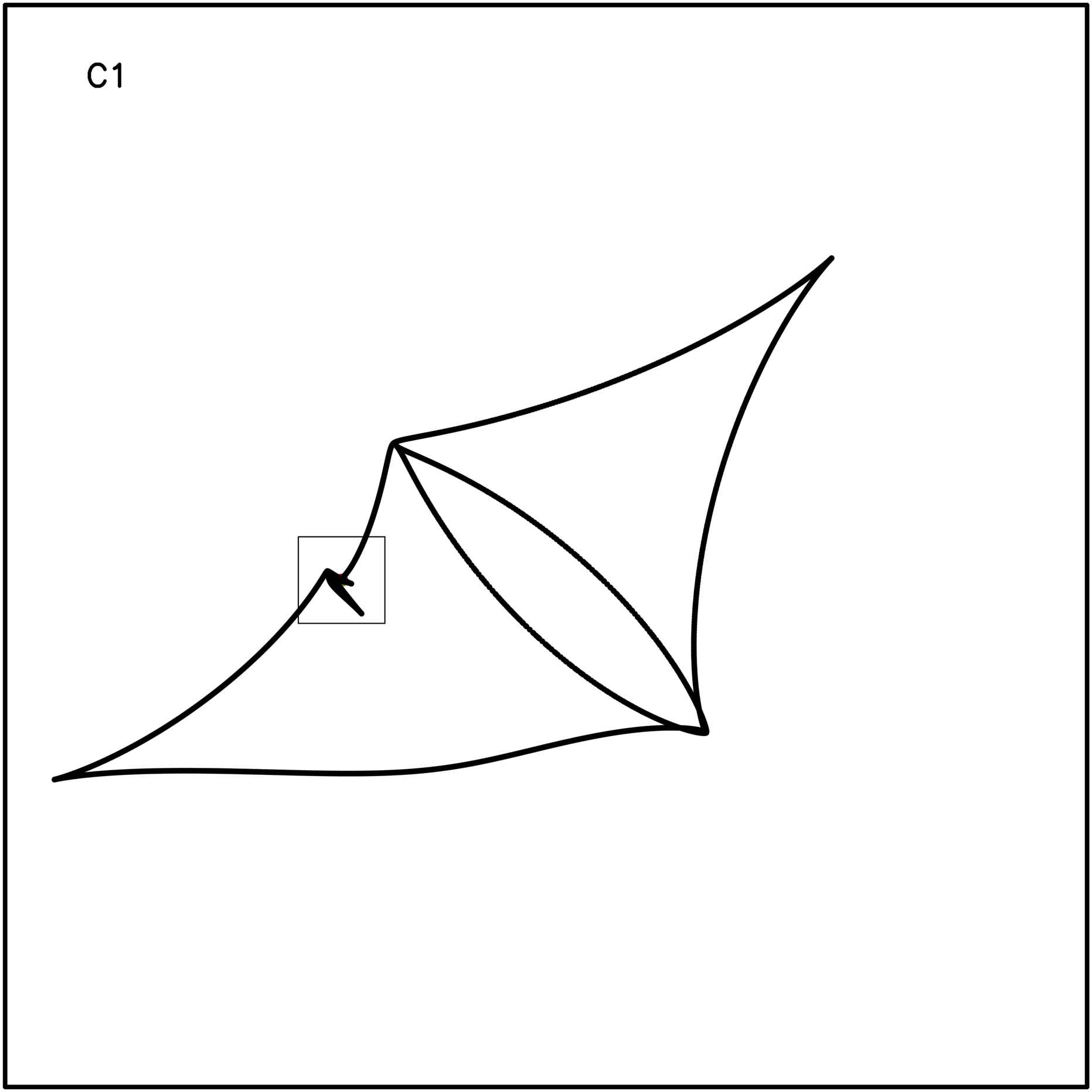}};
\begin{scope}[x={(image.south east)},y={(image.north west)}]
\node[anchor=south west,inner sep=0] (image) at (0.7,0.12) {\includegraphics[width=\textwidth,height=1cm,width=1cm]{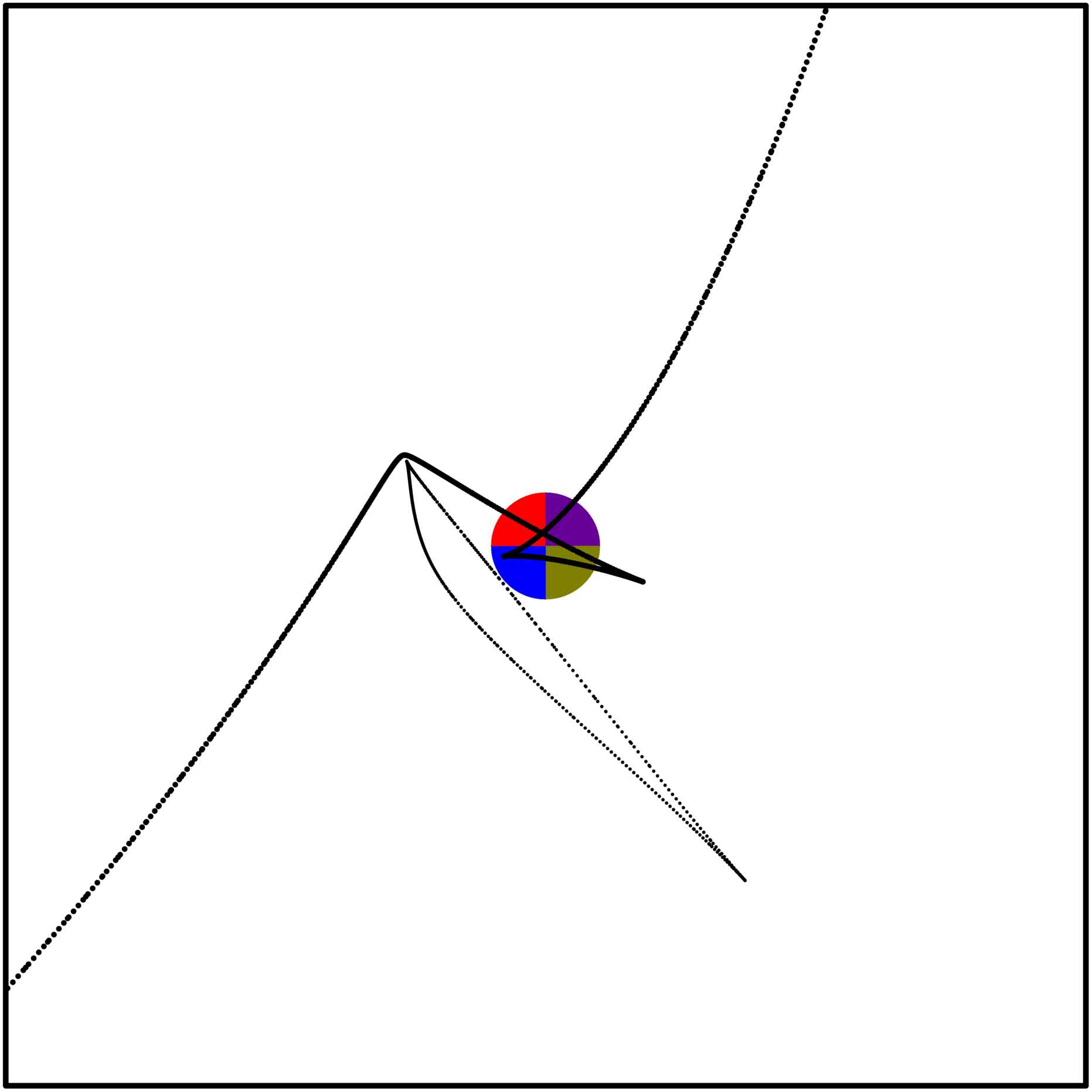}};
\end{scope}
\node[anchor=south west,inner sep=0] (image) at (5.5,0) {\includegraphics[width=\textwidth,height=5.5cm,width=5.5cm]{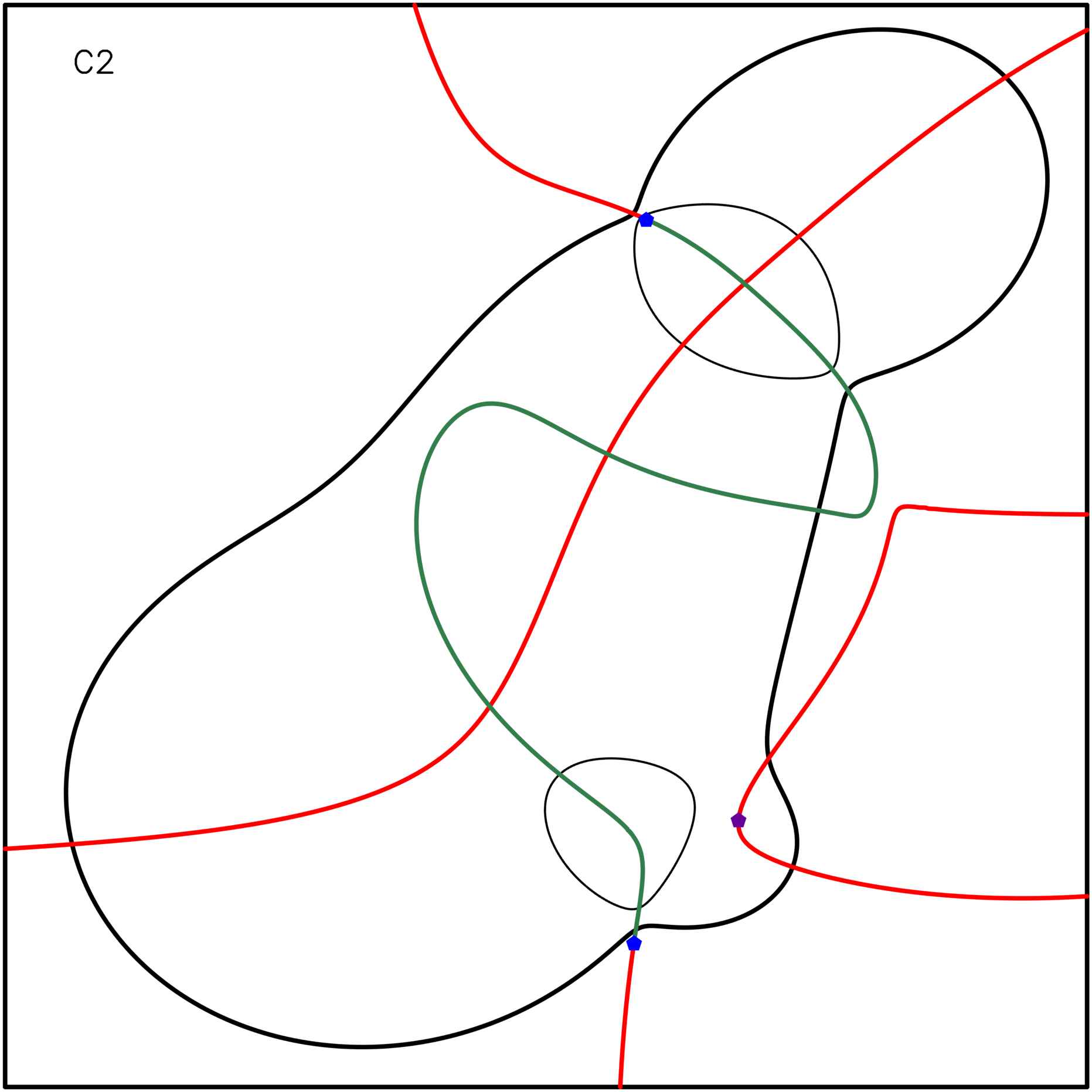}};
\node[anchor=south west,inner sep=0] (image) at (11.0,0) {\includegraphics[width=\textwidth,height=5.5cm,width=5.5cm]{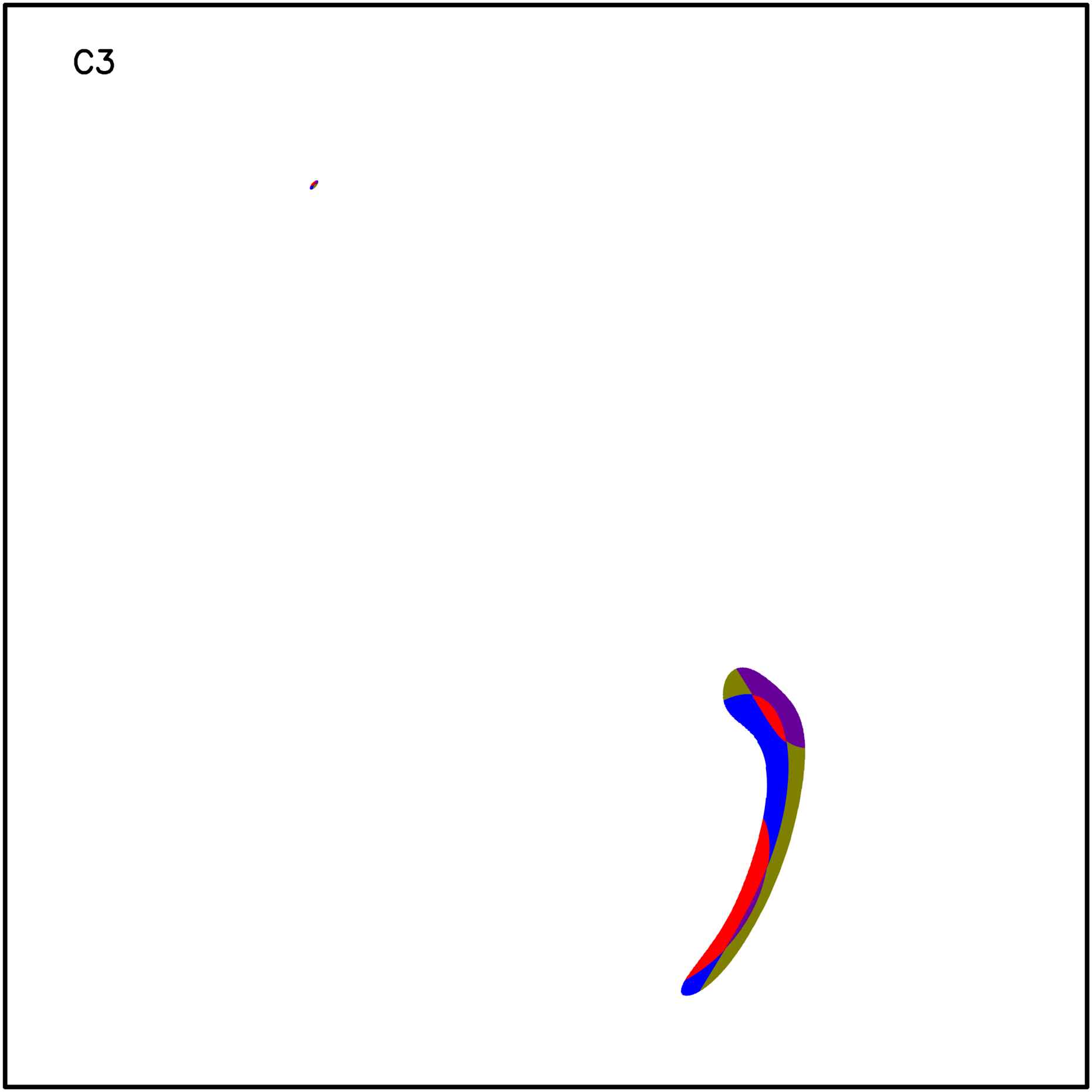}};
\end{tikzpicture}
\caption{Evolution of caustics and critical lines around a
    swallowtail singularity. The left column (A1, B1, C1) shows the
    caustics in source plane for three different redshifts including
    redshift at which swallowtail singularity becomes critical (panel
    B1). The middle column (A2, B2, C2) shows the corresponding
    critical lines and the singularity map including $A_{3}$-lines
    (red and dark green lines), swallowtail (violet point), hyperbolic
    umbilics (blue points). And the right column (A3, B3, C3) shows
    the image formation.  The {\sl source} used here is a circle with
    different colours in four quadrants.  This helps us in seeing the
    parity of images.}
\label{fig:swallowtail_figure}
\end{figure*}
\begin{figure*}
\centering
\begin{tikzpicture}
\node[anchor=south west,inner sep=0] (image) at (0,0) {\includegraphics[width=\textwidth,height=5.5cm,width=5.5cm]{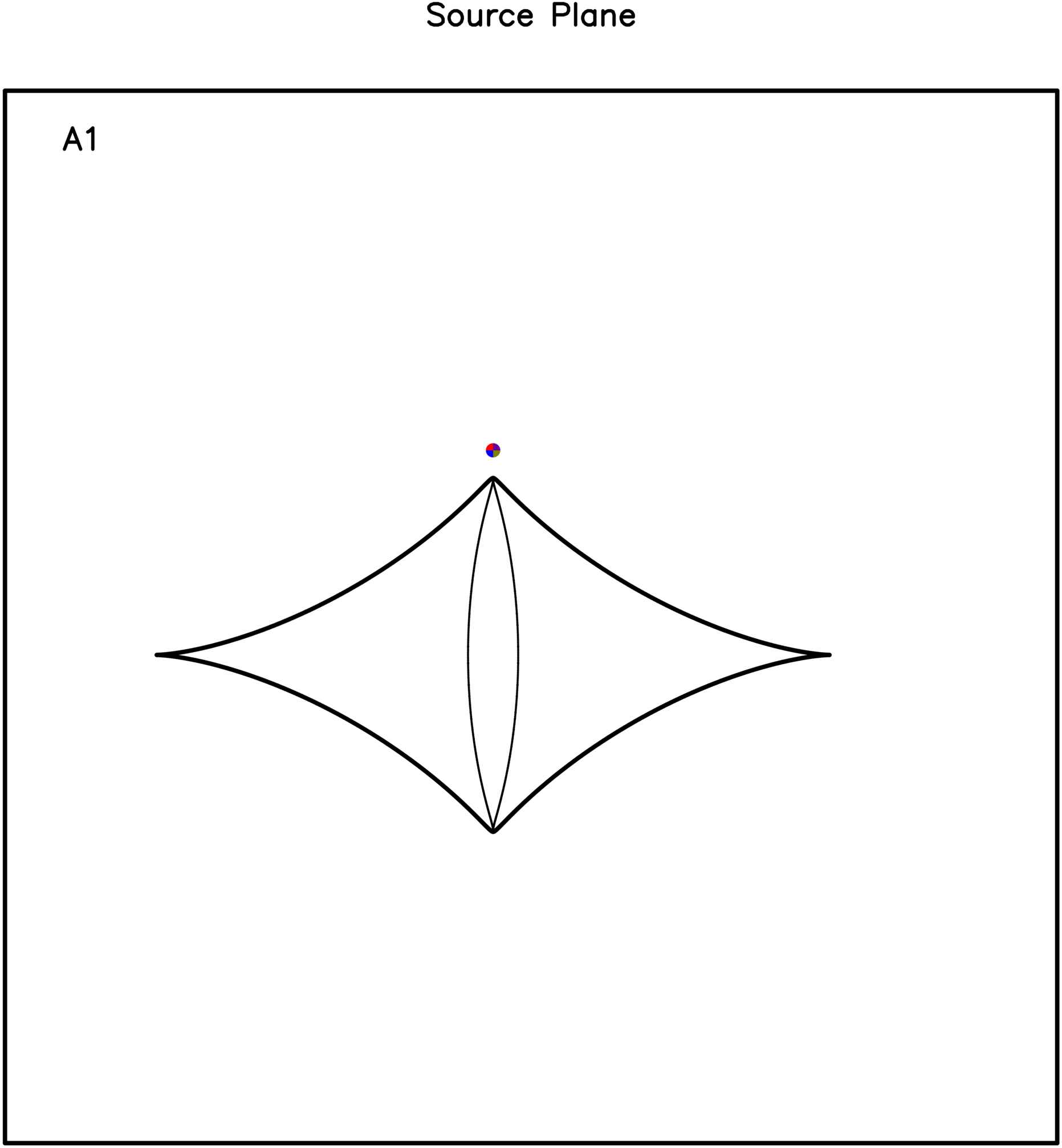}};
\begin{scope}[x={(image.south east)},y={(image.north west)}]
\node[anchor=south west,inner sep=0] (image) at (0.7,0.12) {\includegraphics[width=\textwidth,height=1cm,width=1cm]{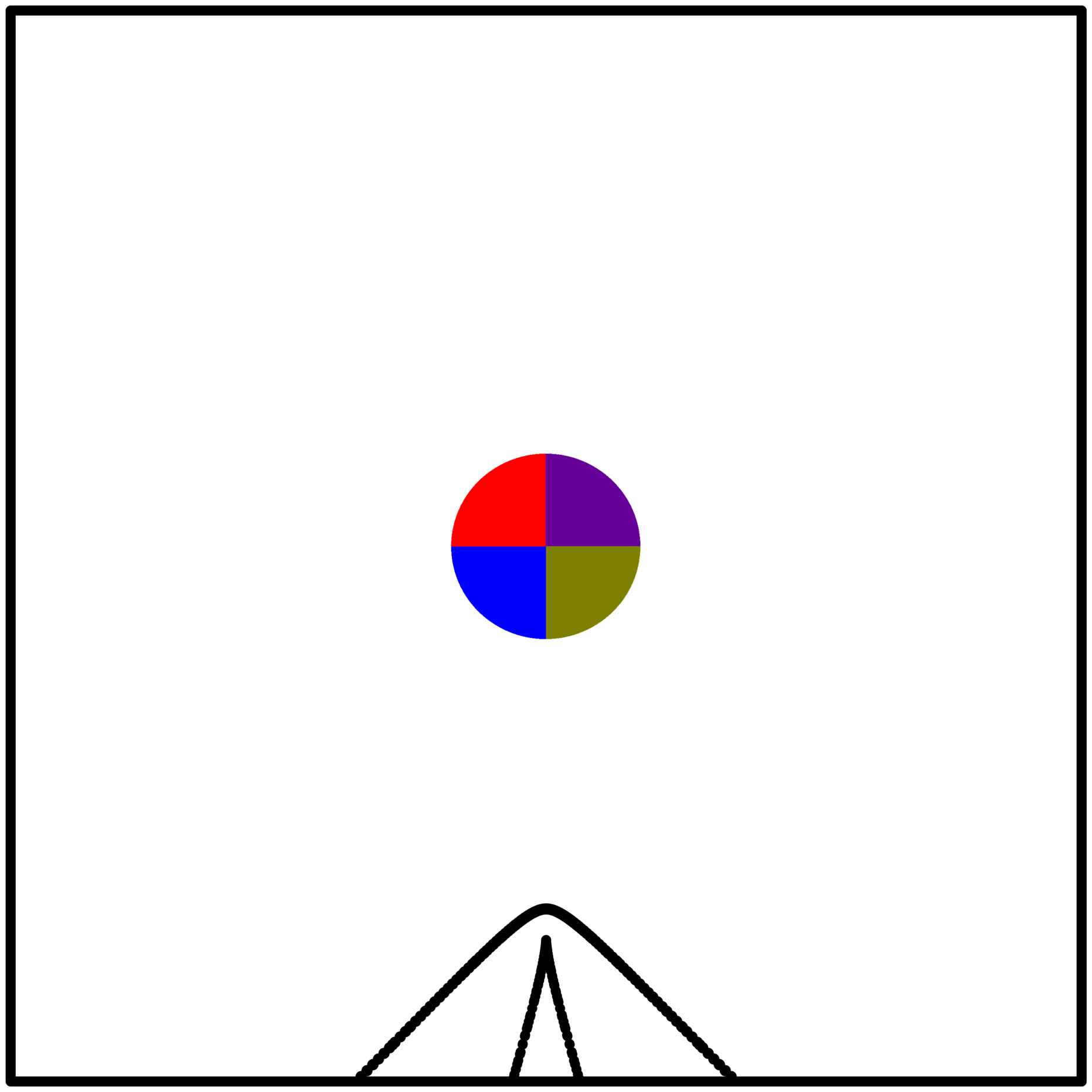}};
\end{scope}
\node[anchor=south west,inner sep=0] (image) at (5.5,0) {\includegraphics[width=\textwidth,height=5.5cm,width=5.5cm]{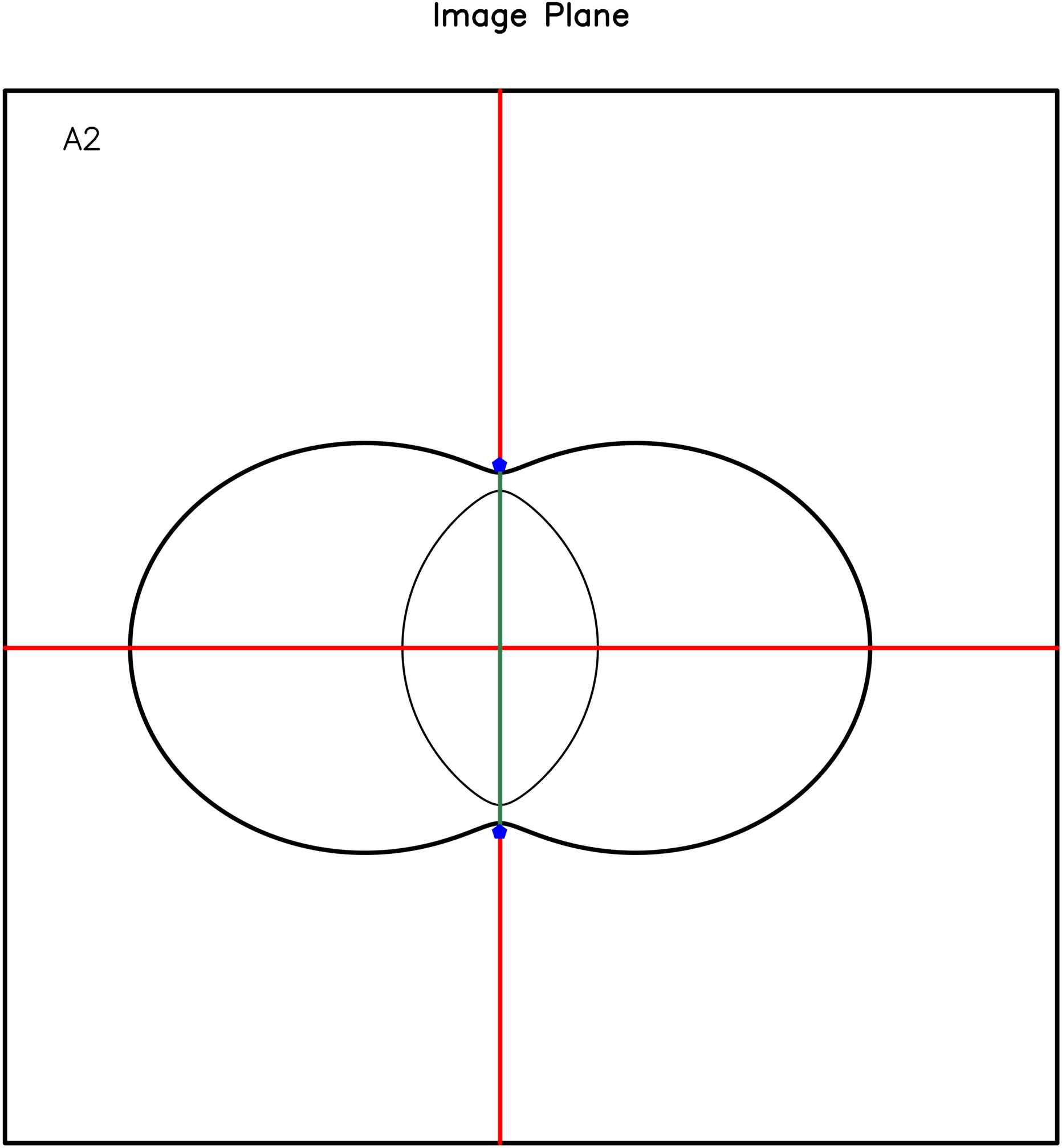}};
\node[anchor=south west,inner sep=0] (image) at (11.0,0) {\includegraphics[width=\textwidth,height=5.5cm,width=5.5cm]{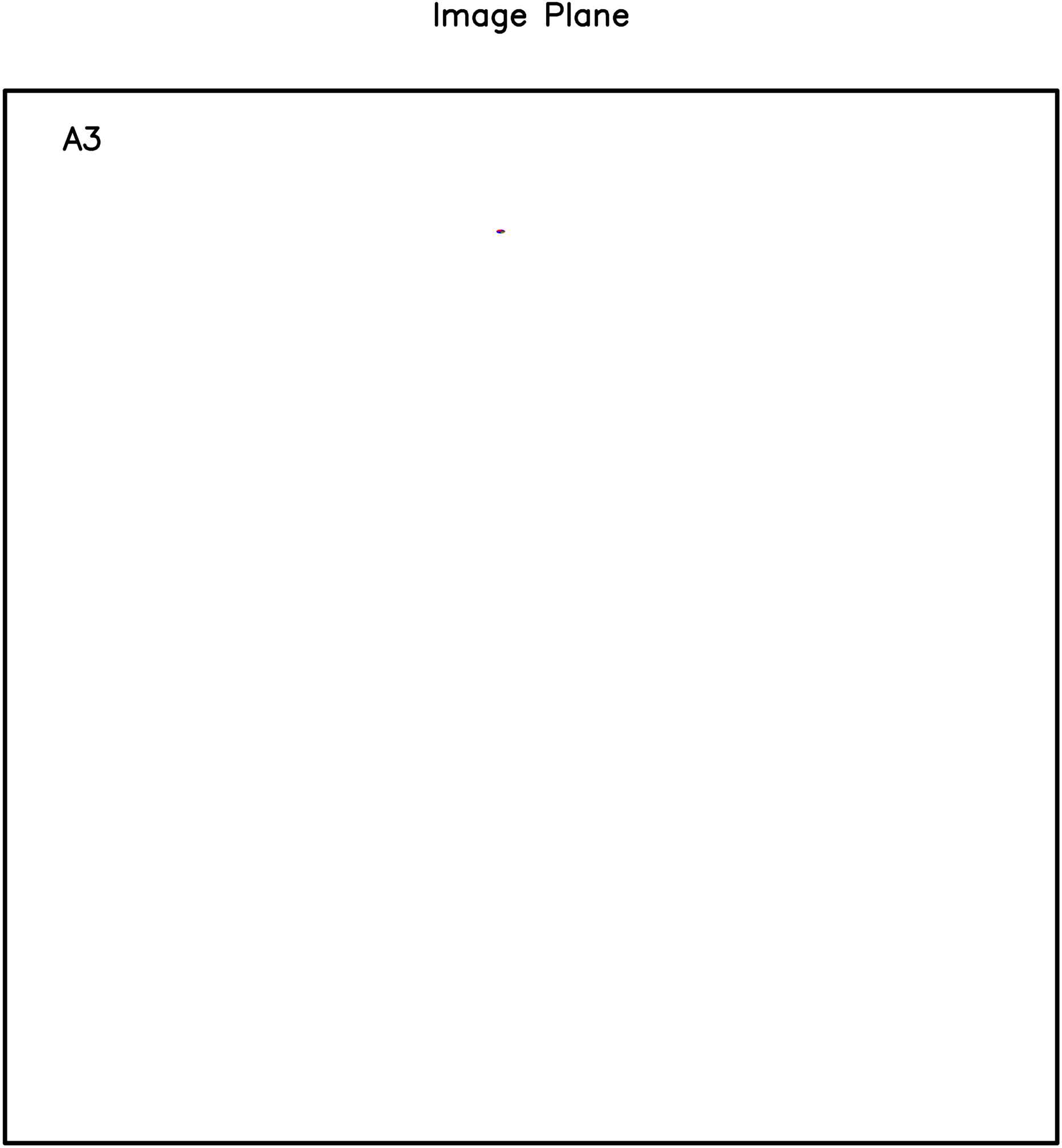}};
\end{tikzpicture}
\end{figure*}
\begin{figure*}
\centering
\begin{tikzpicture}
\node[anchor=south west,inner sep=0] (image) at (0,0) {\includegraphics[width=\textwidth,height=5.5cm,width=5.5cm]{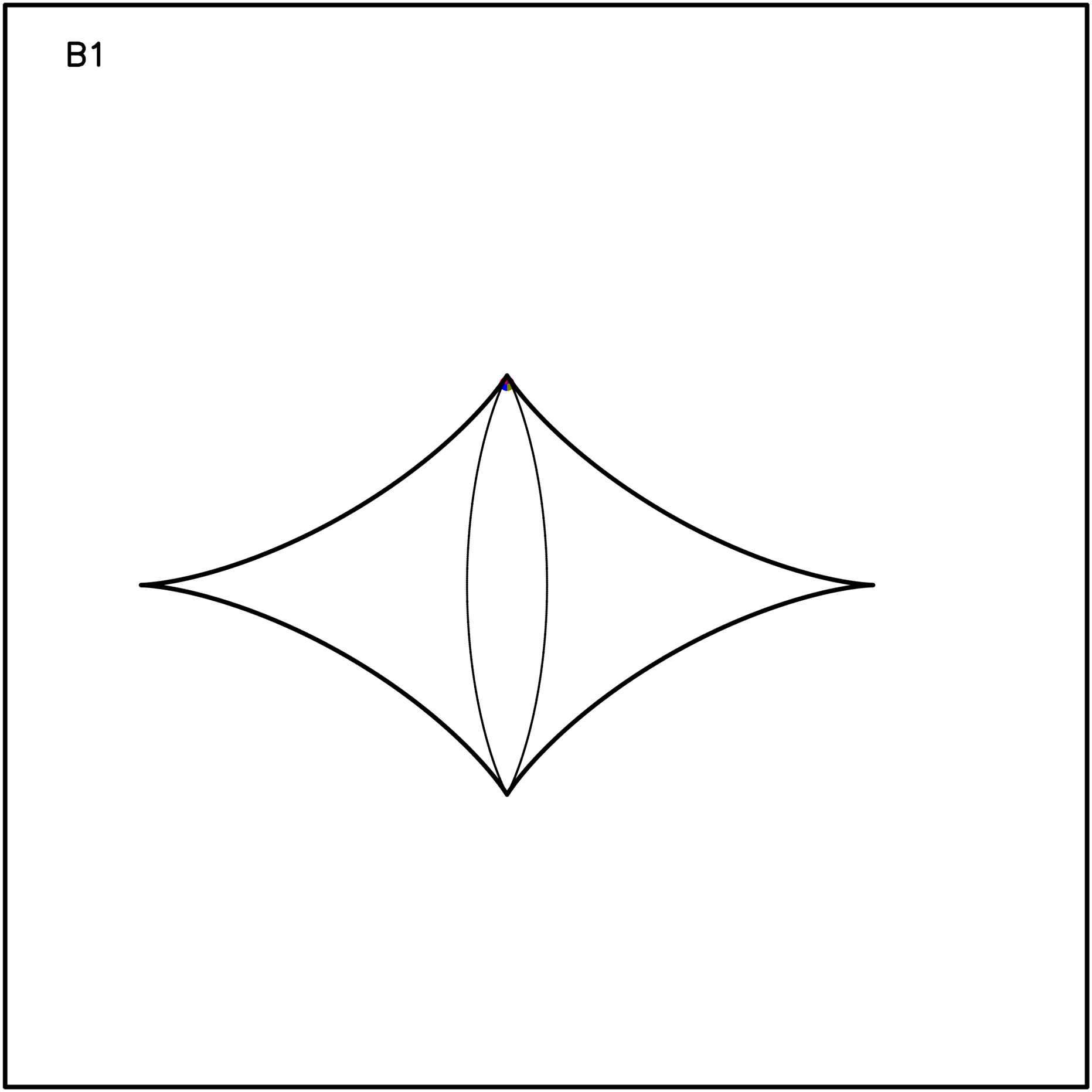}};
\begin{scope}[x={(image.south east)},y={(image.north west)}]
\node[anchor=south west,inner sep=0] (image) at (0.7,0.12) {\includegraphics[width=\textwidth,height=1cm,width=1cm]{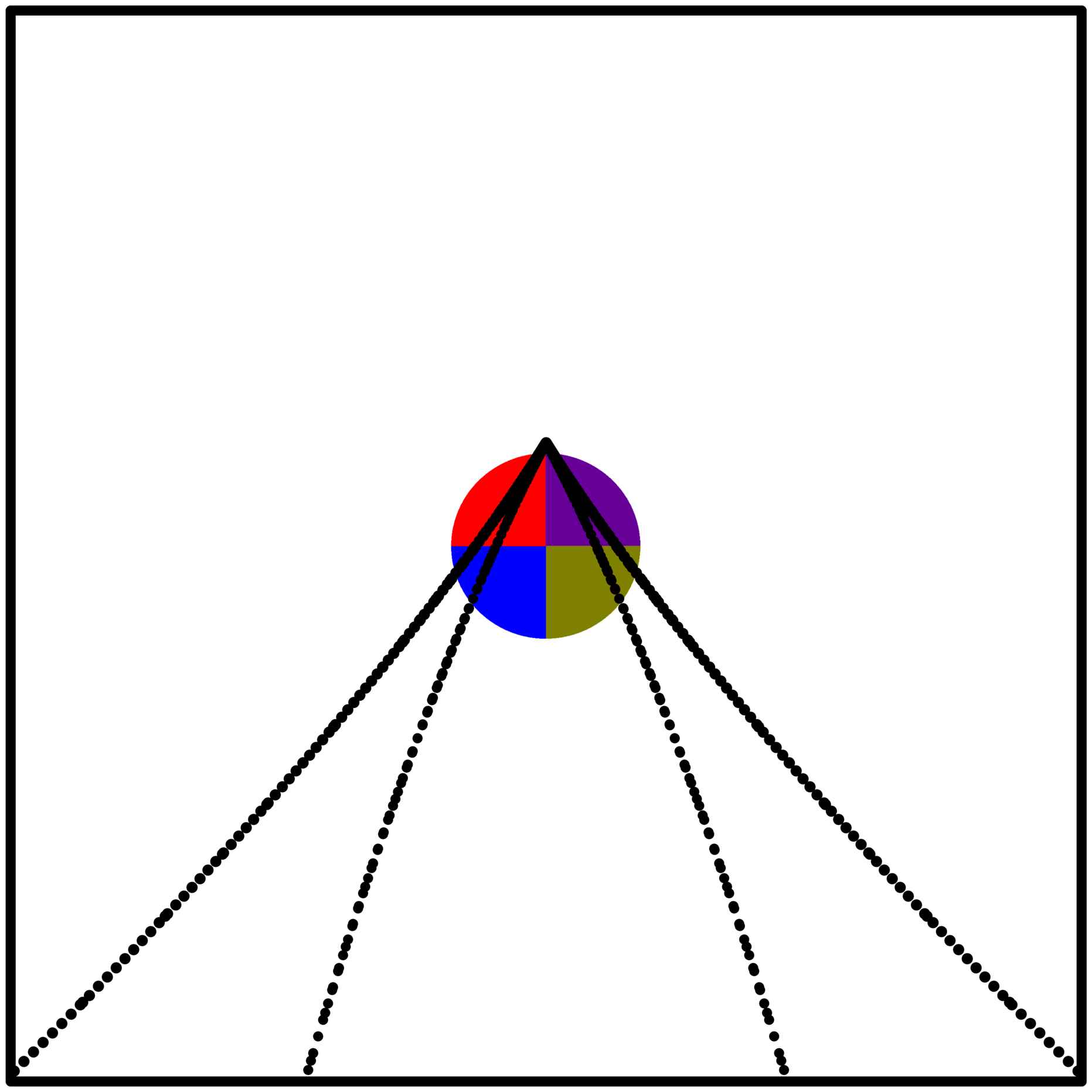}};
\end{scope}
\node[anchor=south west,inner sep=0] (image) at (5.5,0) {\includegraphics[width=\textwidth,height=5.5cm,width=5.5cm]{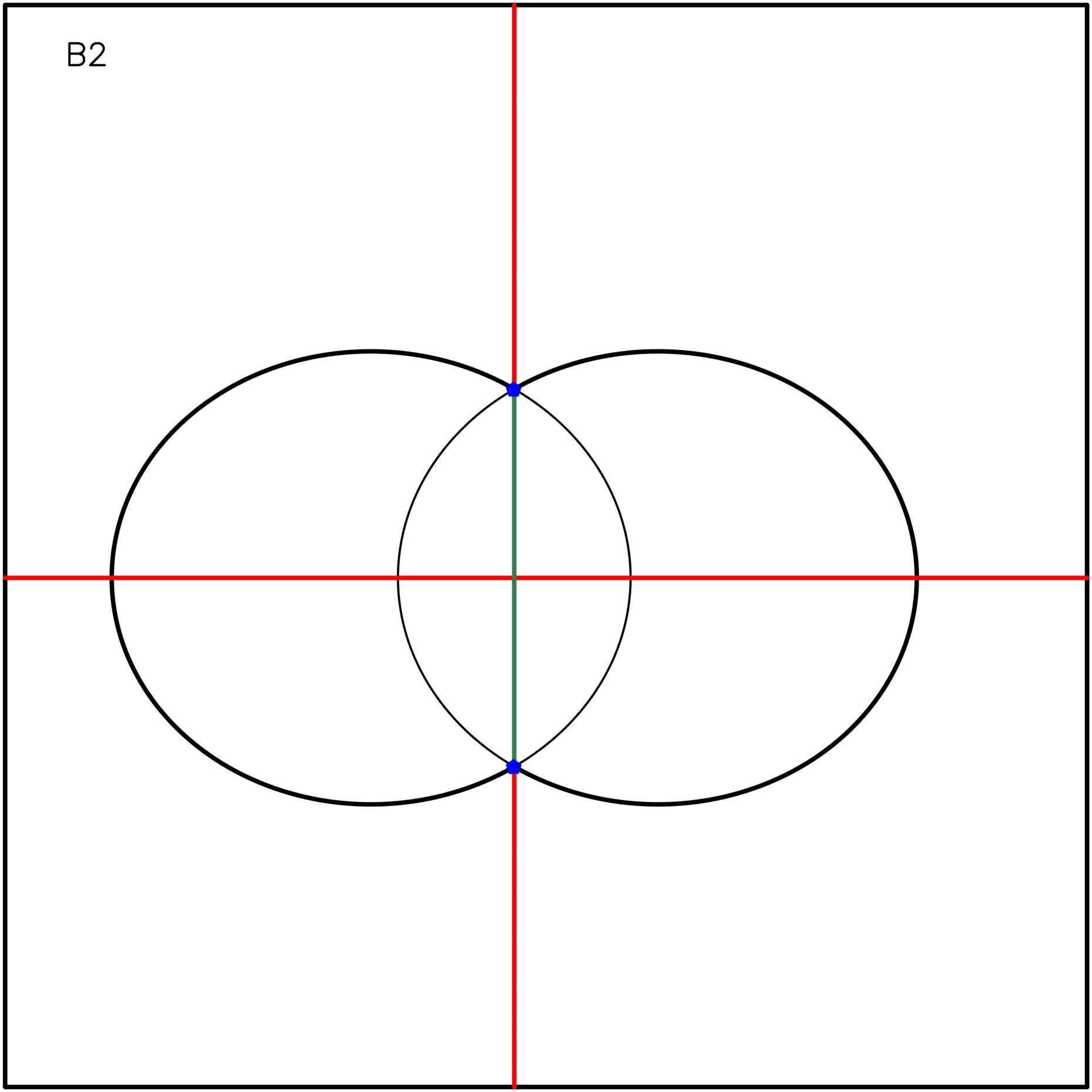}};
\node[anchor=south west,inner sep=0] (image) at (11.0,0) {\includegraphics[width=\textwidth,height=5.5cm,width=5.5cm]{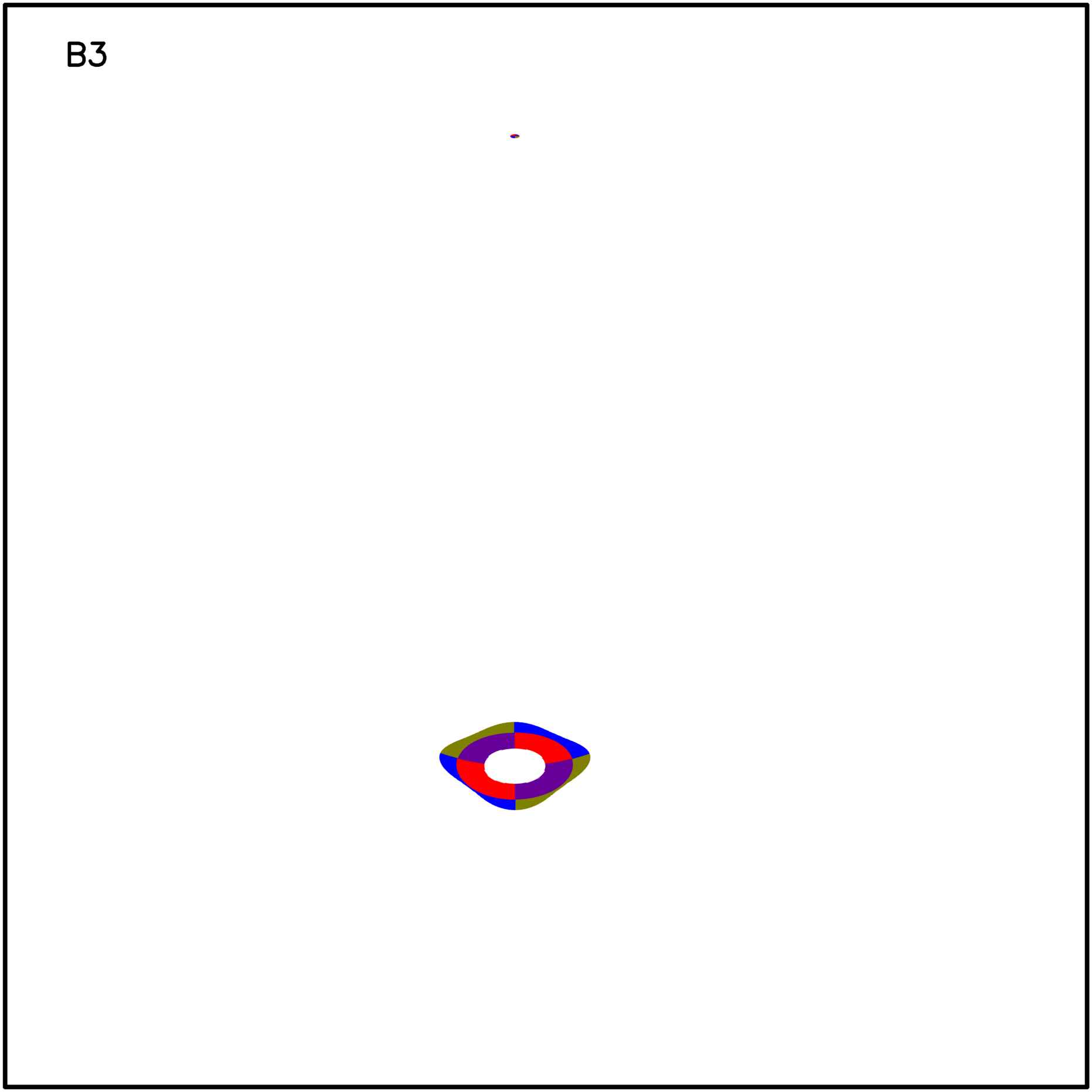}};
\end{tikzpicture}
\end{figure*}
\begin{figure*}
\centering
\begin{tikzpicture}
\node[anchor=south west,inner sep=0] (image) at (0,0) {\includegraphics[width=\textwidth,height=5.5cm,width=5.5cm]{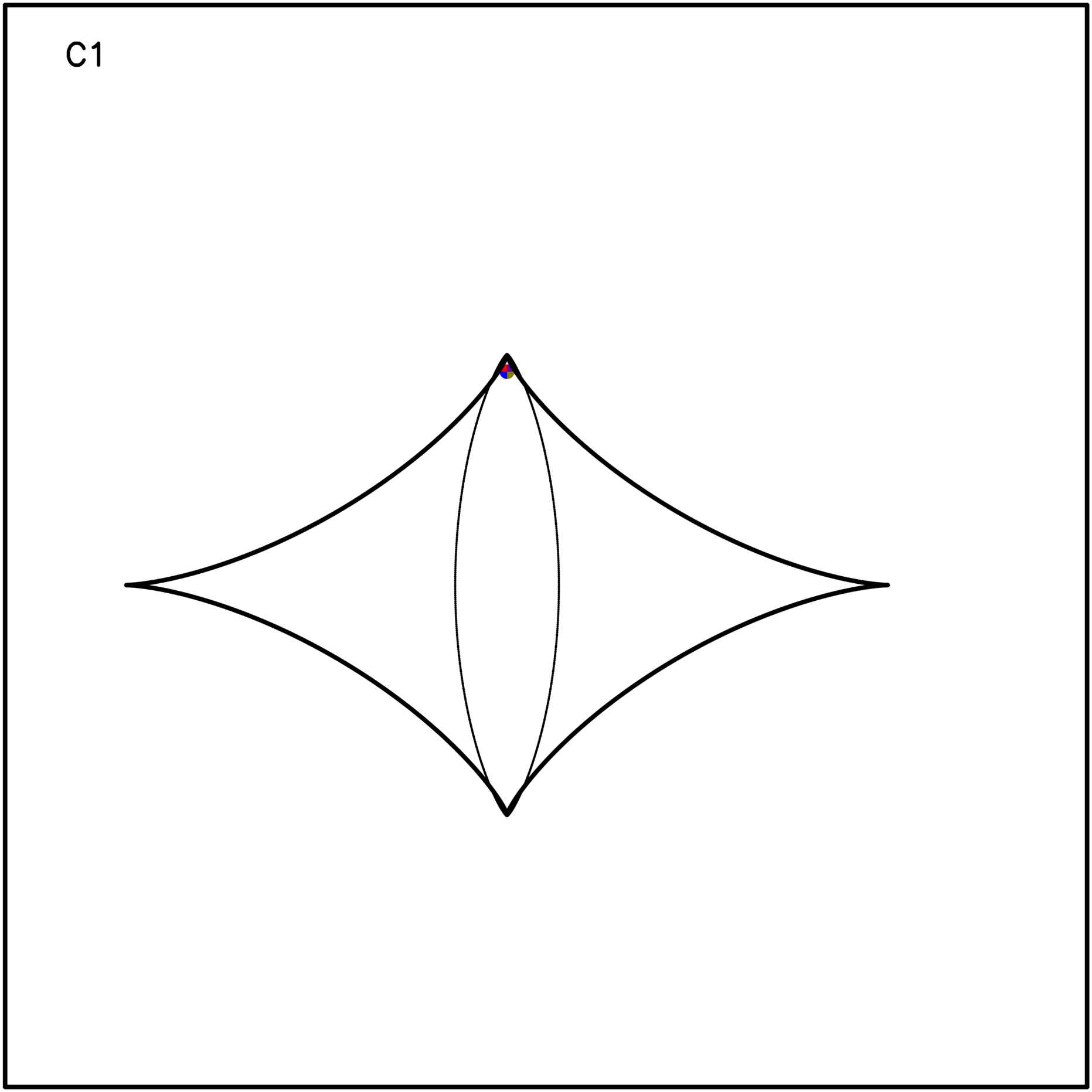}};
\begin{scope}[x={(image.south east)},y={(image.north west)}]
\node[anchor=south west,inner sep=0] (image) at (0.7,0.12) {\includegraphics[width=\textwidth,height=1cm,width=1cm]{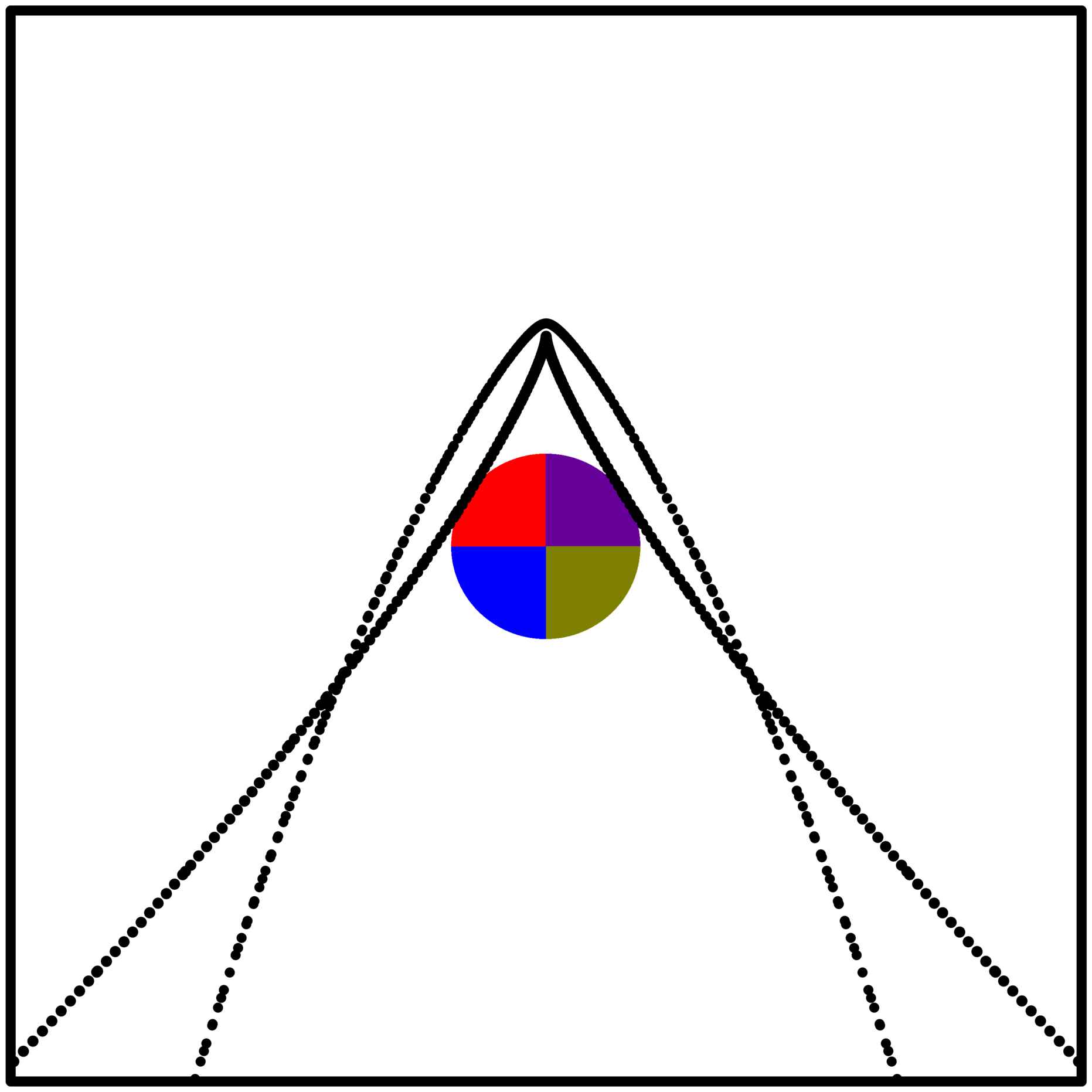}};
\end{scope}
\node[anchor=south west,inner sep=0] (image) at (5.5,0) {\includegraphics[width=\textwidth,height=5.5cm,width=5.5cm]{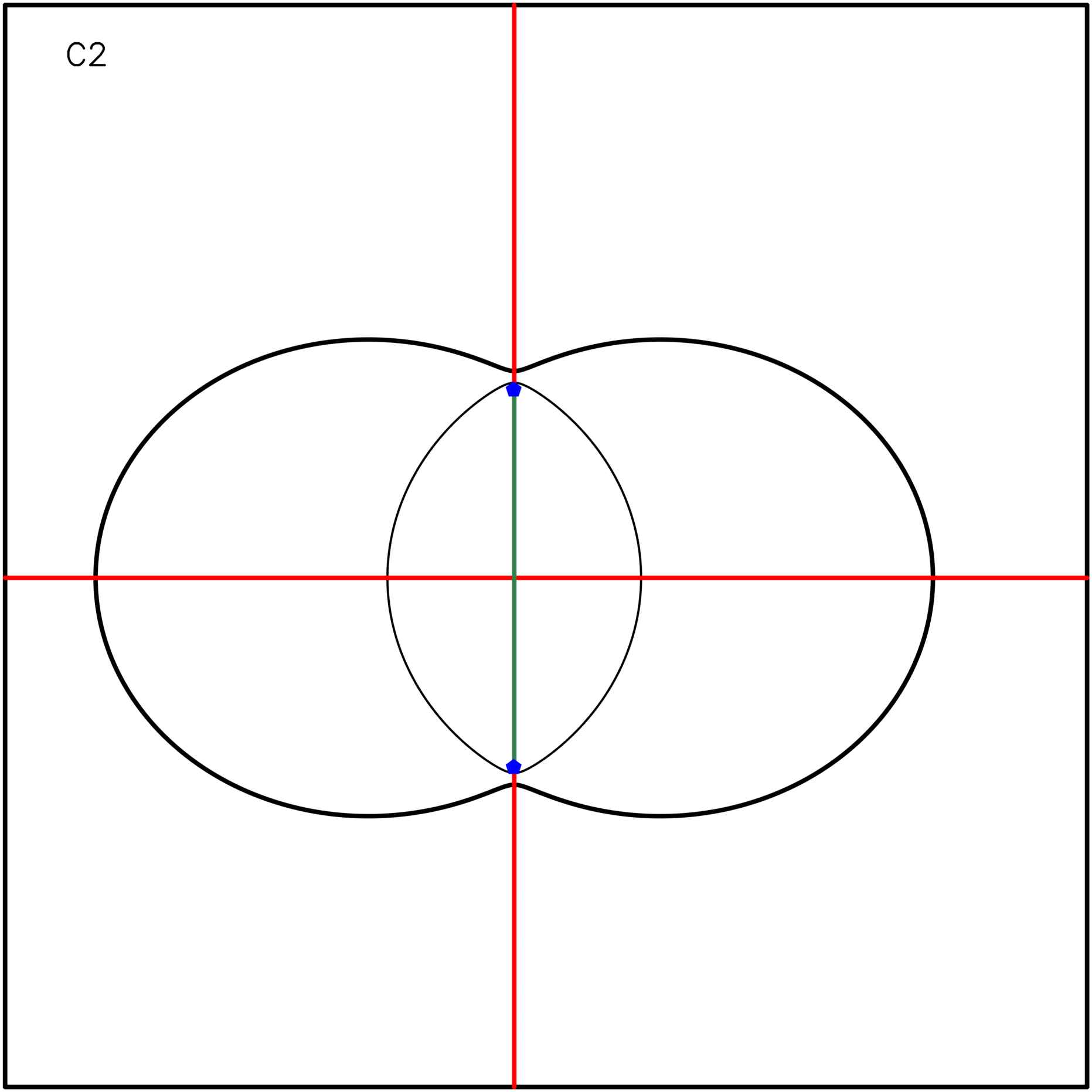}};
\node[anchor=south west,inner sep=0] (image) at (11.0,0) {\includegraphics[width=\textwidth,height=5.5cm,width=5.5cm]{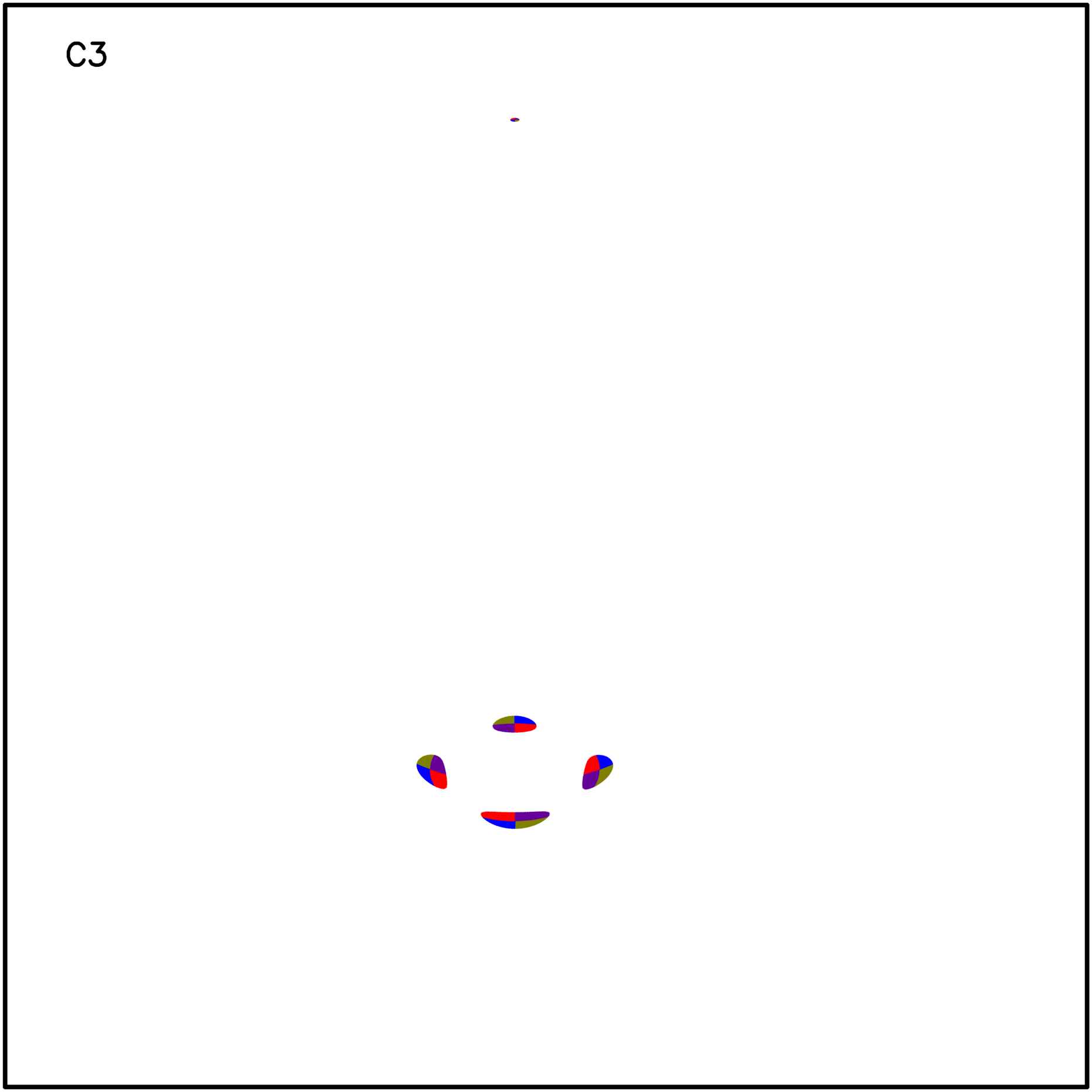}};
\end{tikzpicture}
\caption{Evolution of caustics and critical lines around a
    hyperbolic umbilic (purse). The left column (A1, B1, C1) shows the
    caustics in source plane for three different redshifts including
    redshift at which purse singularity becomes critical (panel
    B1). The middle column (A2, B2, C2) shows the corresponding
    critical lines and the singularity map including $A_{3}$-lines
    (red and dark green lines) and purse (blue point). And the right column
    (A3, B3, C3) shows the image formation. One can notice the exchange of
    the cusp between radial and tangential caustics (panel B1) and the ring
    shaped image formation (panel B3) at hyperbolic umbilic.  Kindly
    note that as the umbilics are symmetric, the image formation about
either one will be the same apart from a reflection.  Here we show
images corresponding to one of the umbilics, as marked by the source
position in the left column.}
\label{fig:purse_figure} 
\end{figure*}

\begin{figure*}
\centering
\begin{tikzpicture}
\node[anchor=south west,inner sep=0] (image) at (0,0) {\includegraphics[width=\textwidth,height=5.5cm,width=5.5cm]{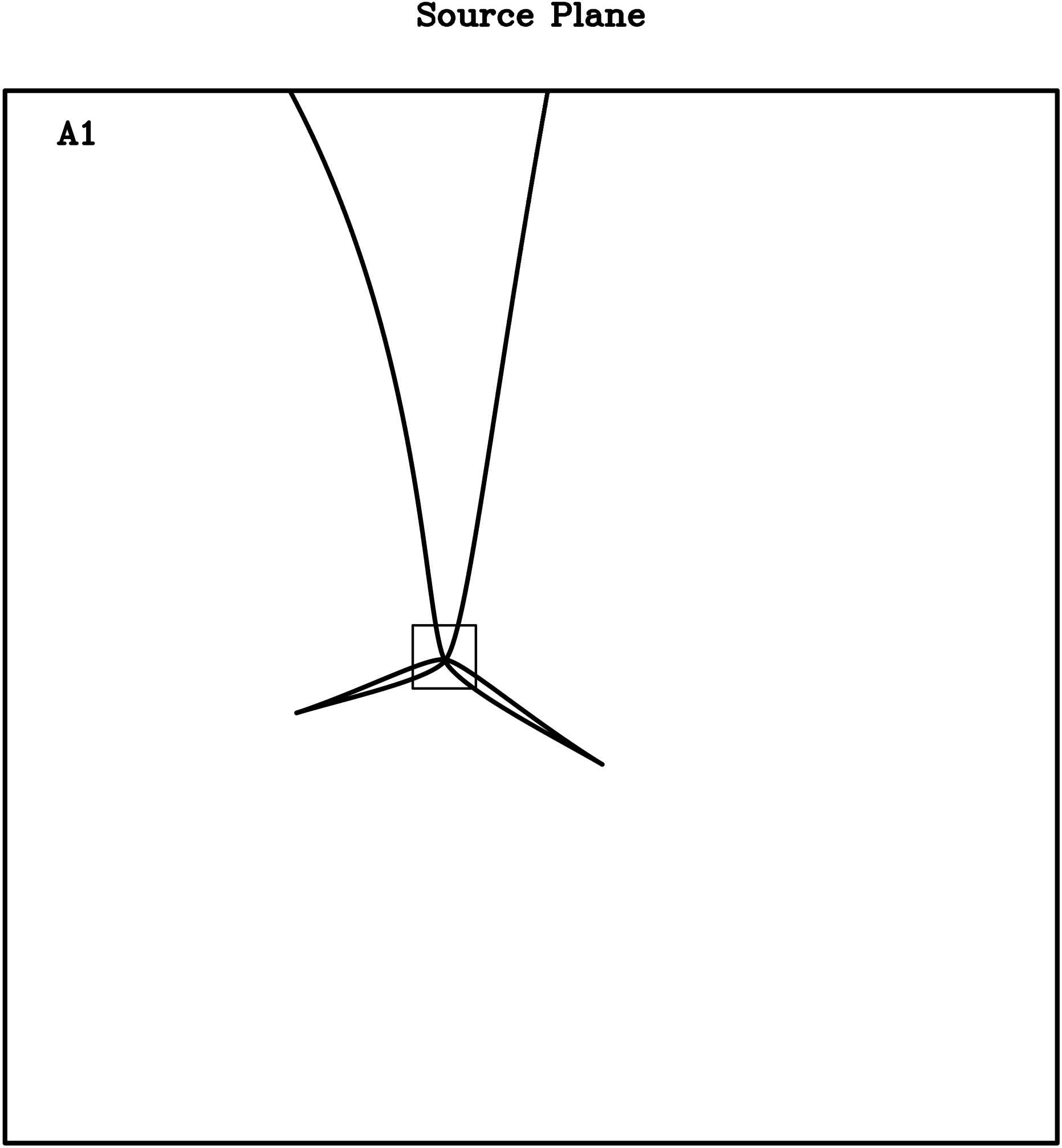}};
\begin{scope}[x={(image.south east)},y={(image.north west)}]
\node[anchor=south west,inner sep=0] (image) at (0.68,0.07) {\includegraphics[width=\textwidth,height=1.5cm,width=1.5cm]{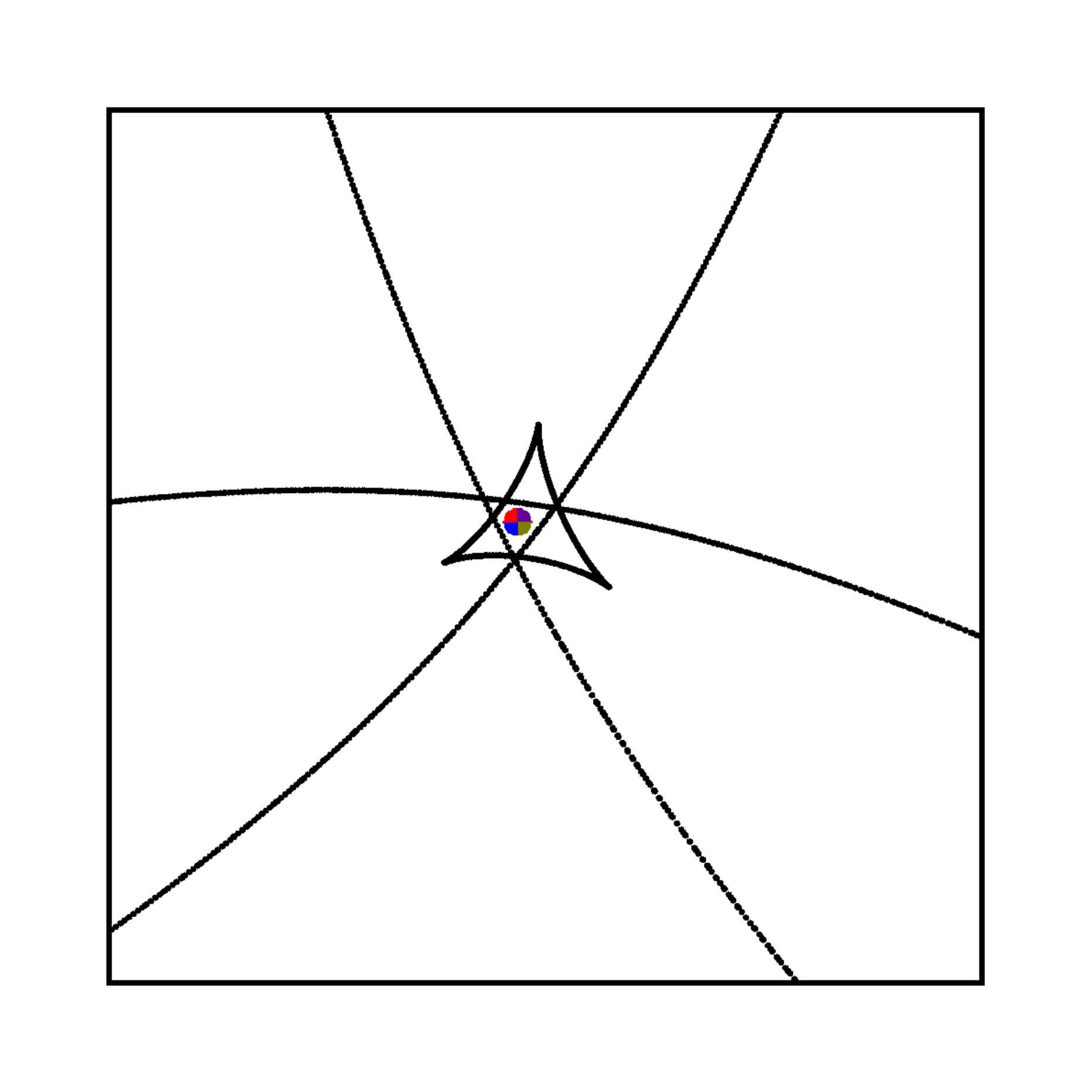}};
\end{scope}
\node[anchor=south west,inner sep=0] (image) at (5.5,0) {\includegraphics[width=\textwidth,height=5.5cm,width=5.5cm]{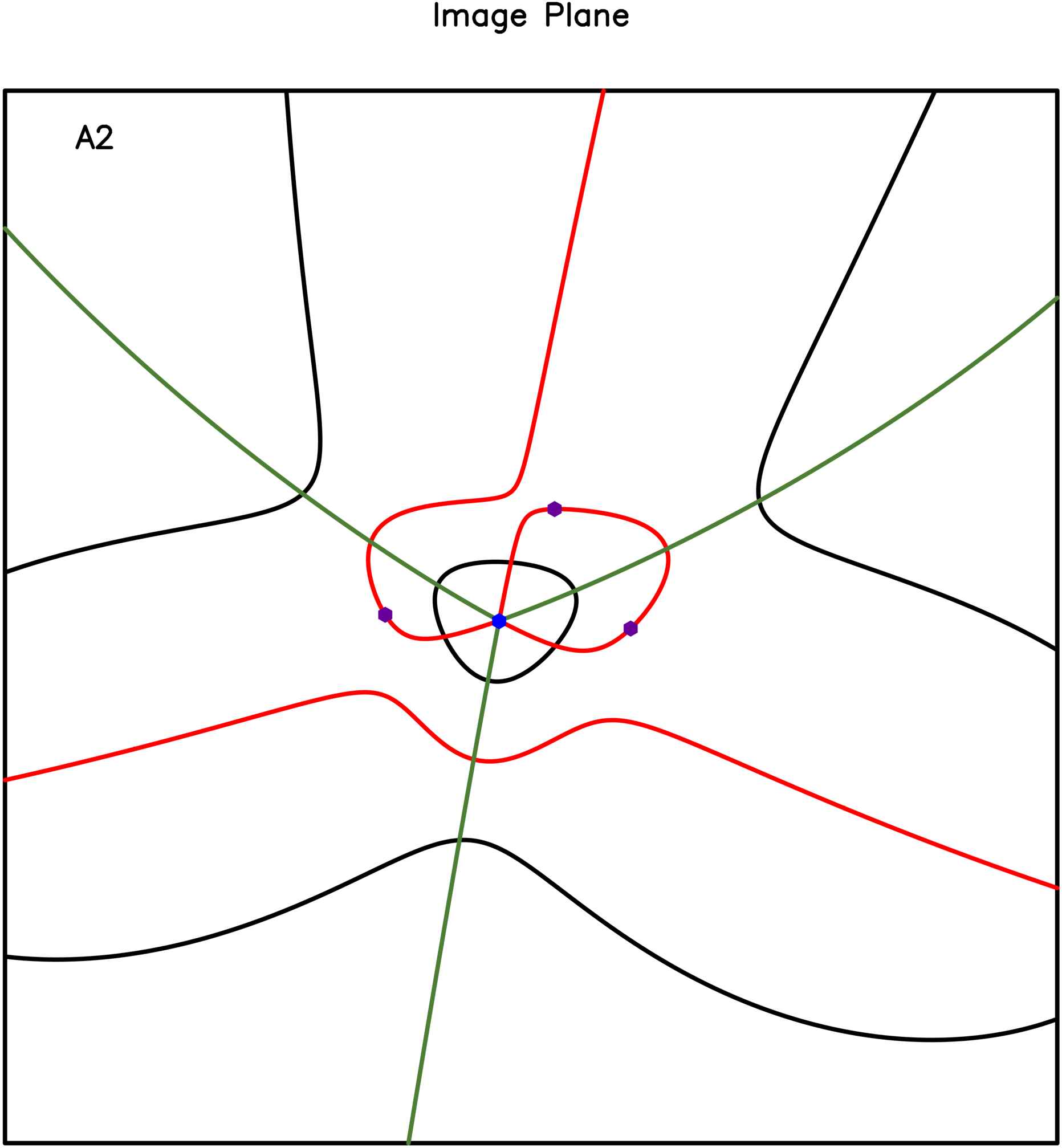}};
\node[anchor=south west,inner sep=0] (image) at (11.0,0) {\includegraphics[width=\textwidth,height=5.5cm,width=5.5cm]{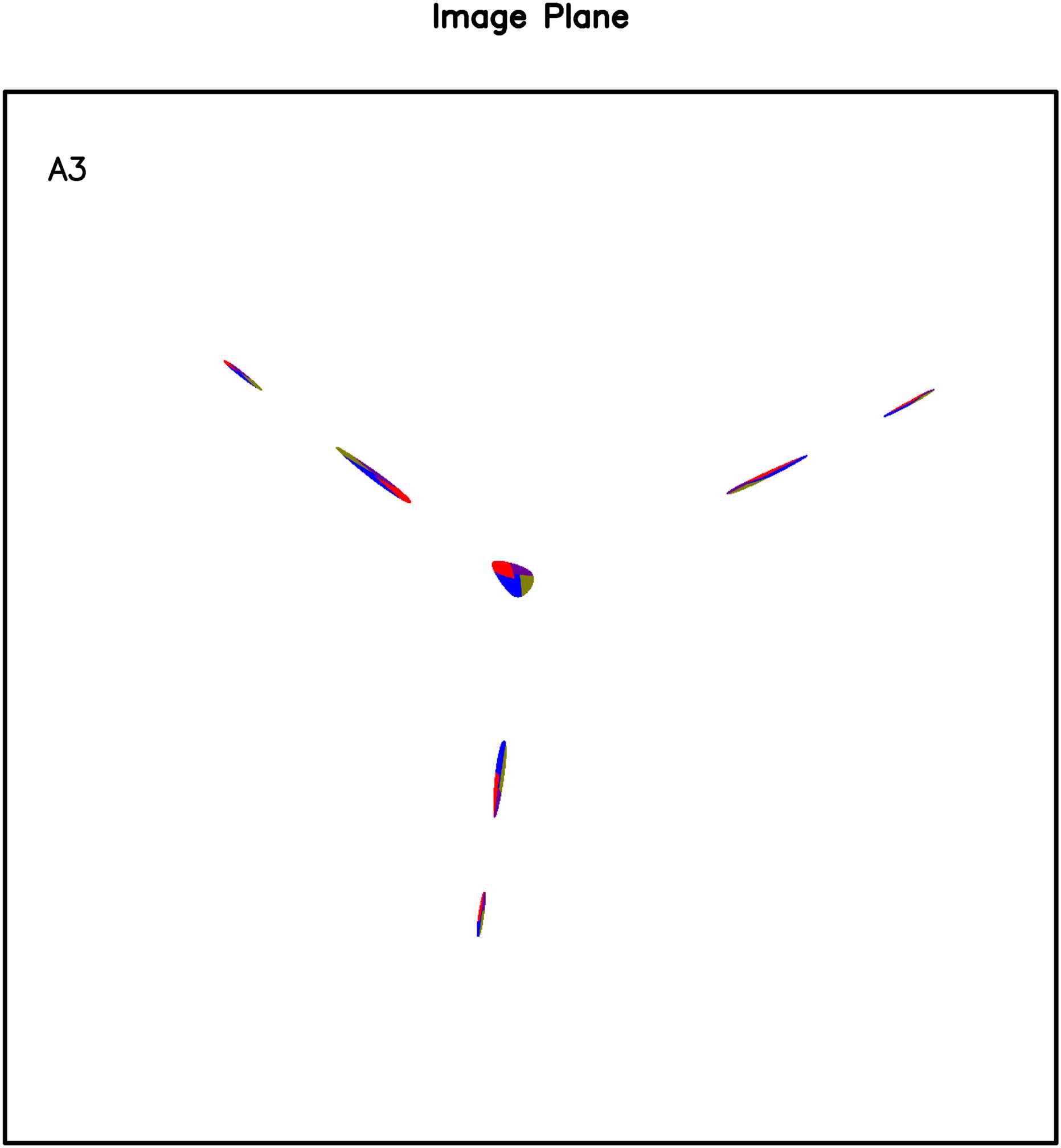}};
\end{tikzpicture}
\end{figure*}
\begin{figure*}
\centering
\begin{tikzpicture}
\node[anchor=south west,inner sep=0] (image) at (0,0) {\includegraphics[width=\textwidth,height=5.5cm,width=5.5cm]{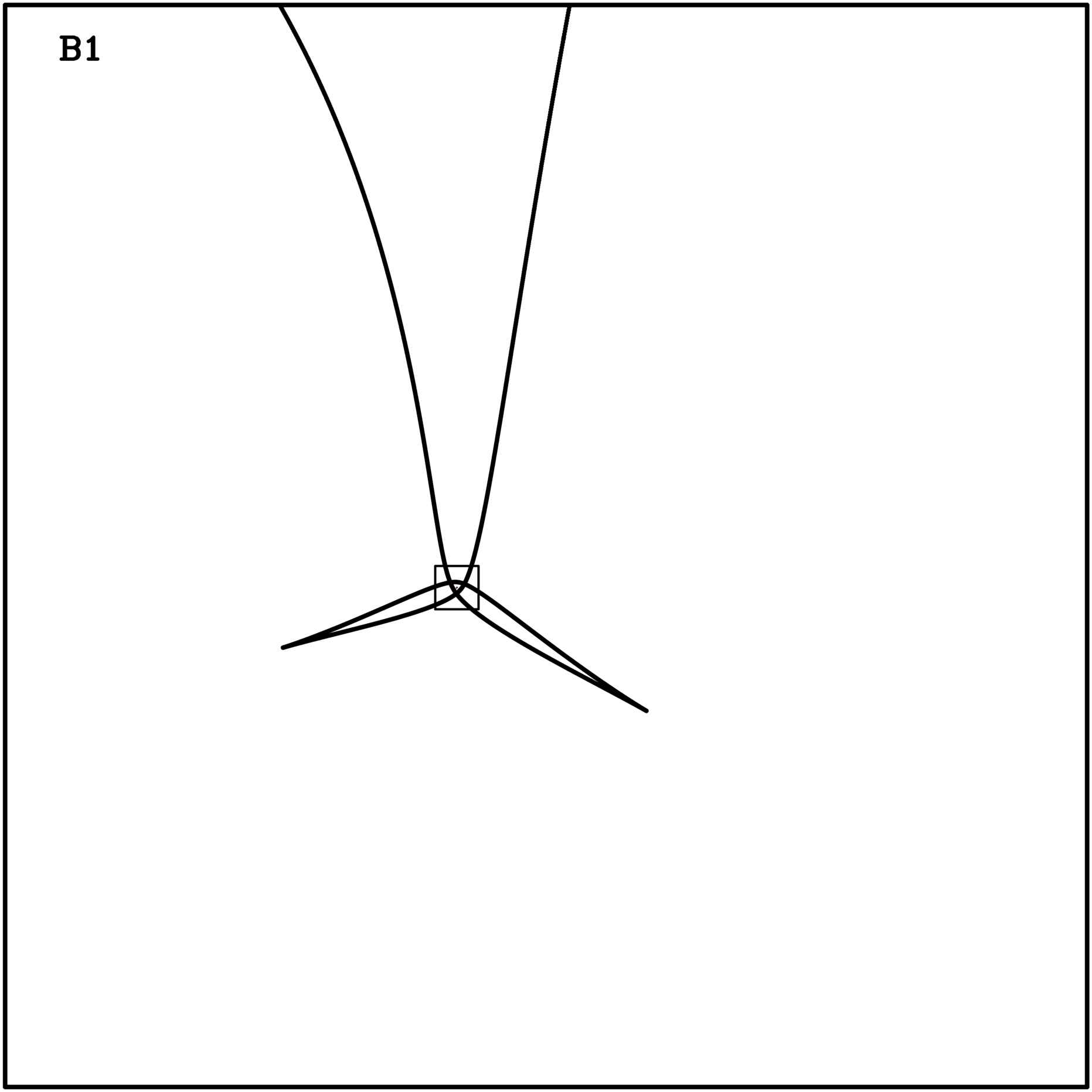}};
\begin{scope}[x={(image.south east)},y={(image.north west)}]
\node[anchor=south west,inner sep=0] (image) at (0.68,0.07) {\includegraphics[width=\textwidth,height=1.5cm,width=1.5cm]{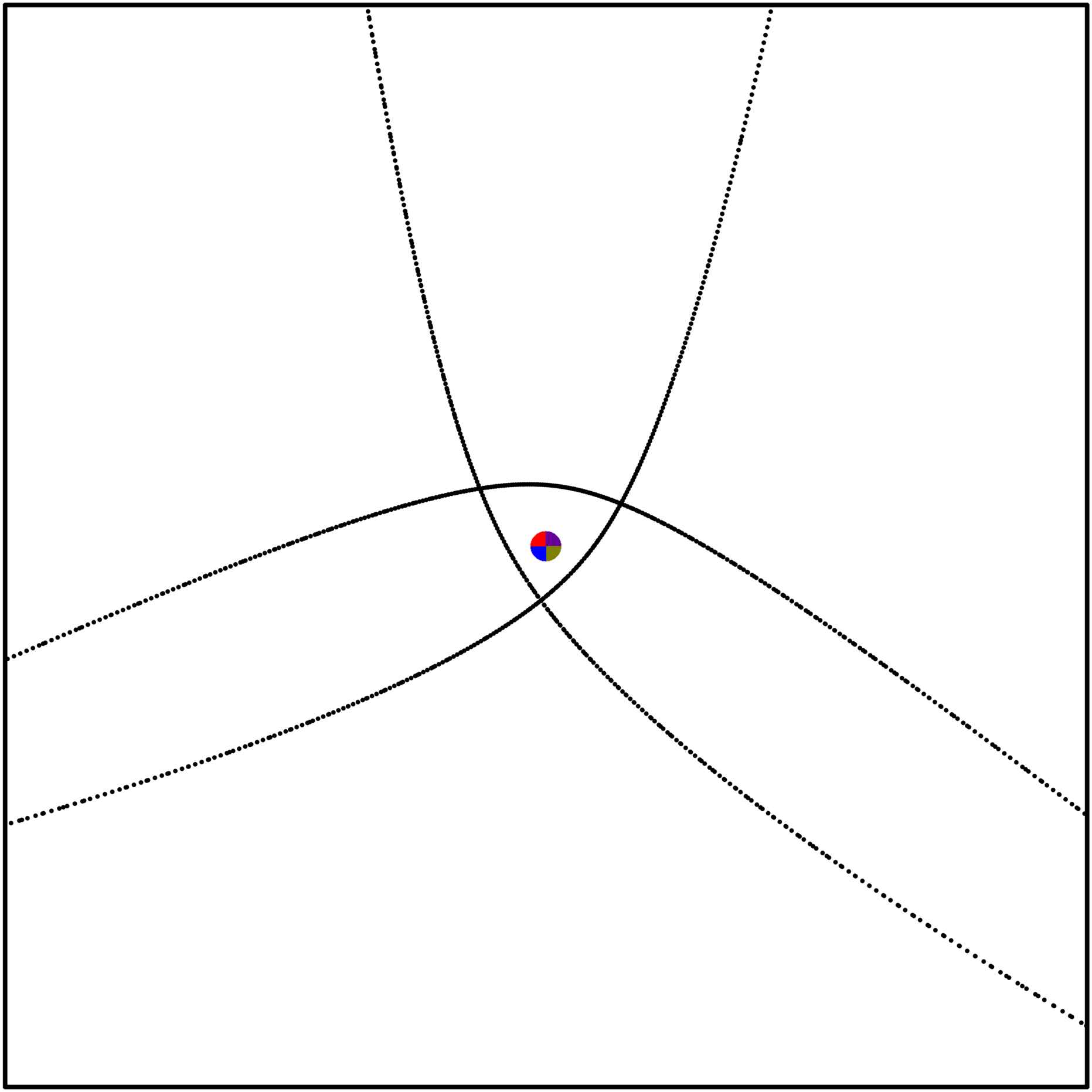}};
\end{scope}
\node[anchor=south west,inner sep=0] (image) at (5.5,0) {\includegraphics[width=\textwidth,height=5.5cm,width=5.5cm]{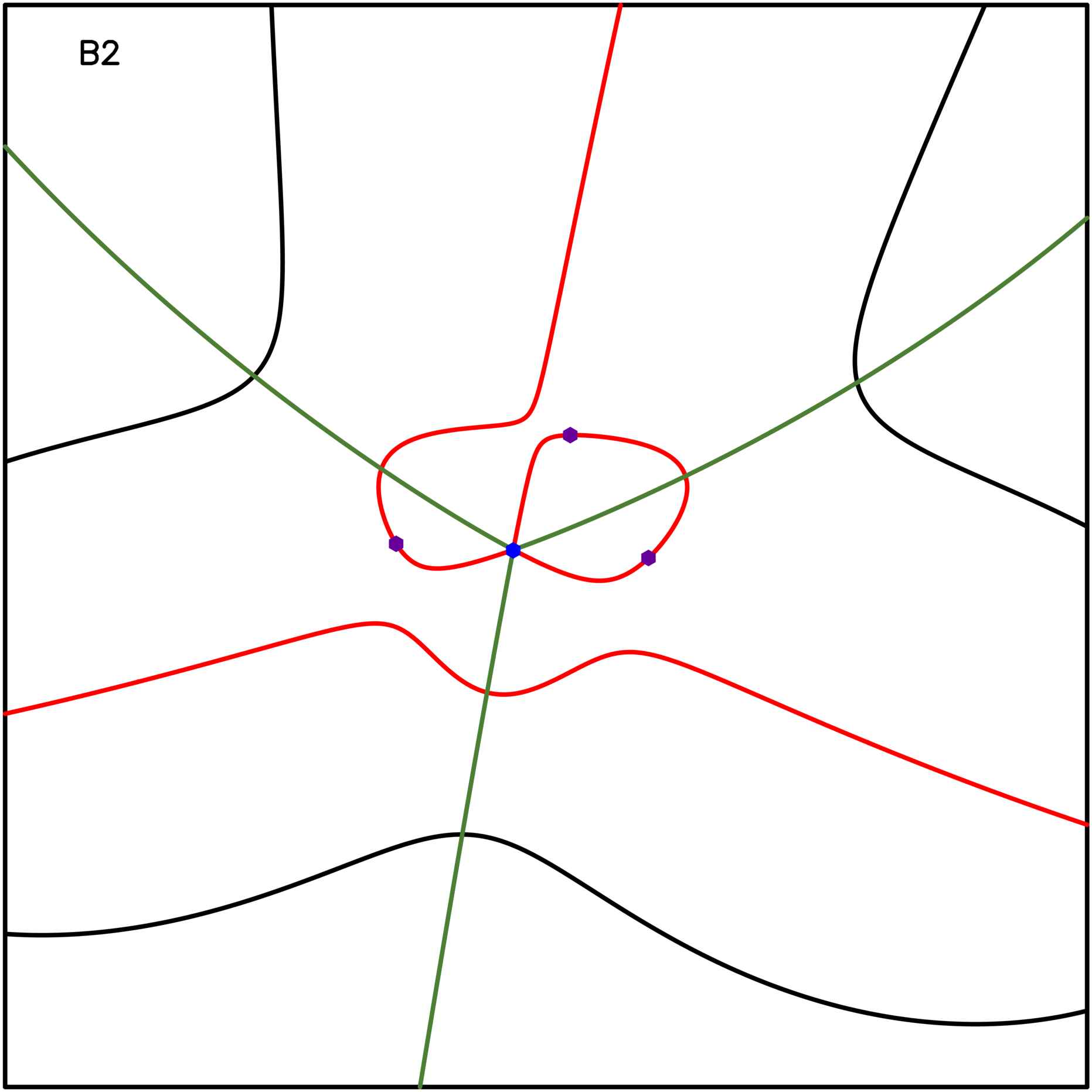}};
\node[anchor=south west,inner sep=0] (image) at (11.0,0) {\includegraphics[width=\textwidth,height=5.5cm,width=5.5cm]{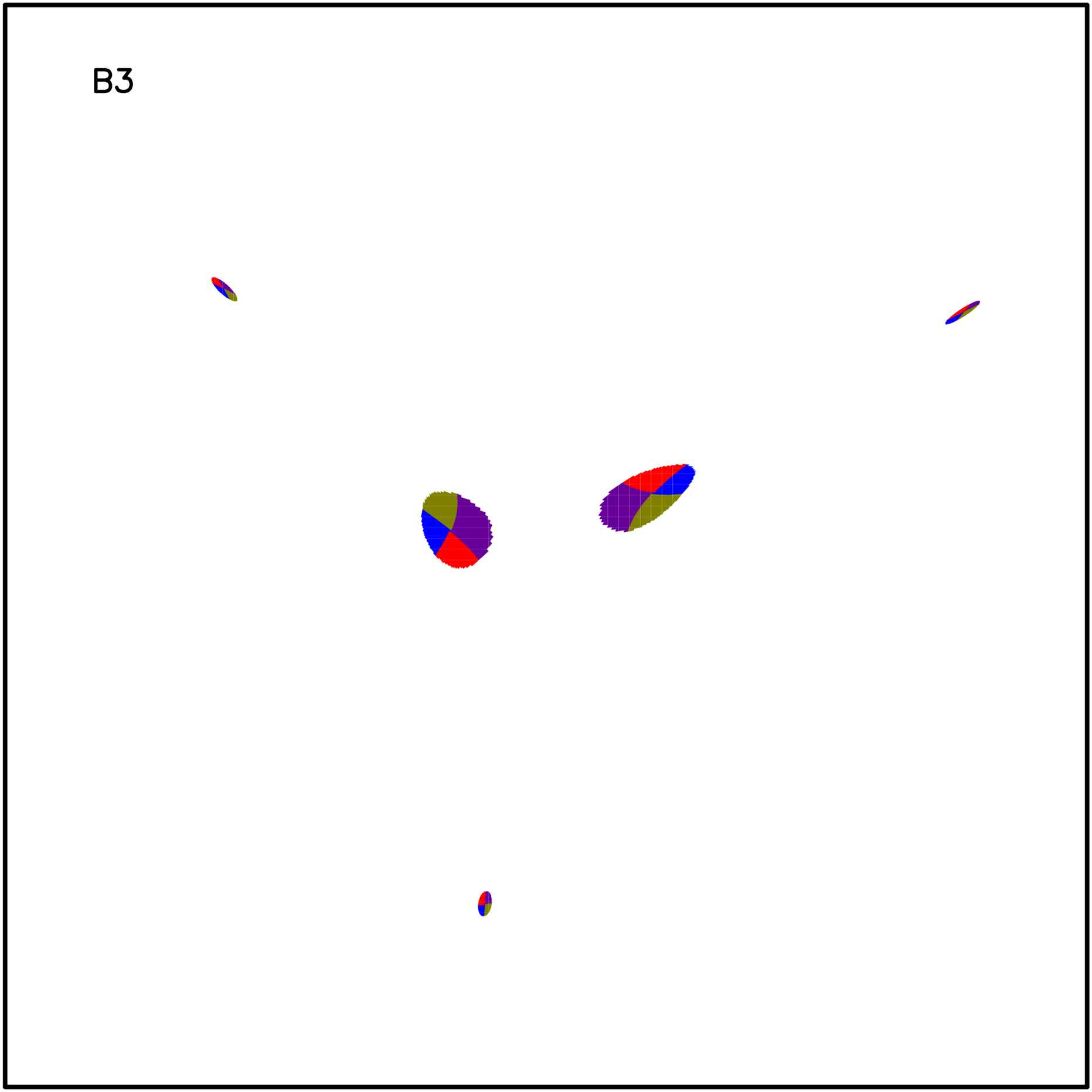}};
\end{tikzpicture}
\end{figure*}
\begin{figure*}
\centering
\begin{tikzpicture}
\node[anchor=south west,inner sep=0] (image) at (0,0) {\includegraphics[width=\textwidth,height=5.5cm,width=5.5cm]{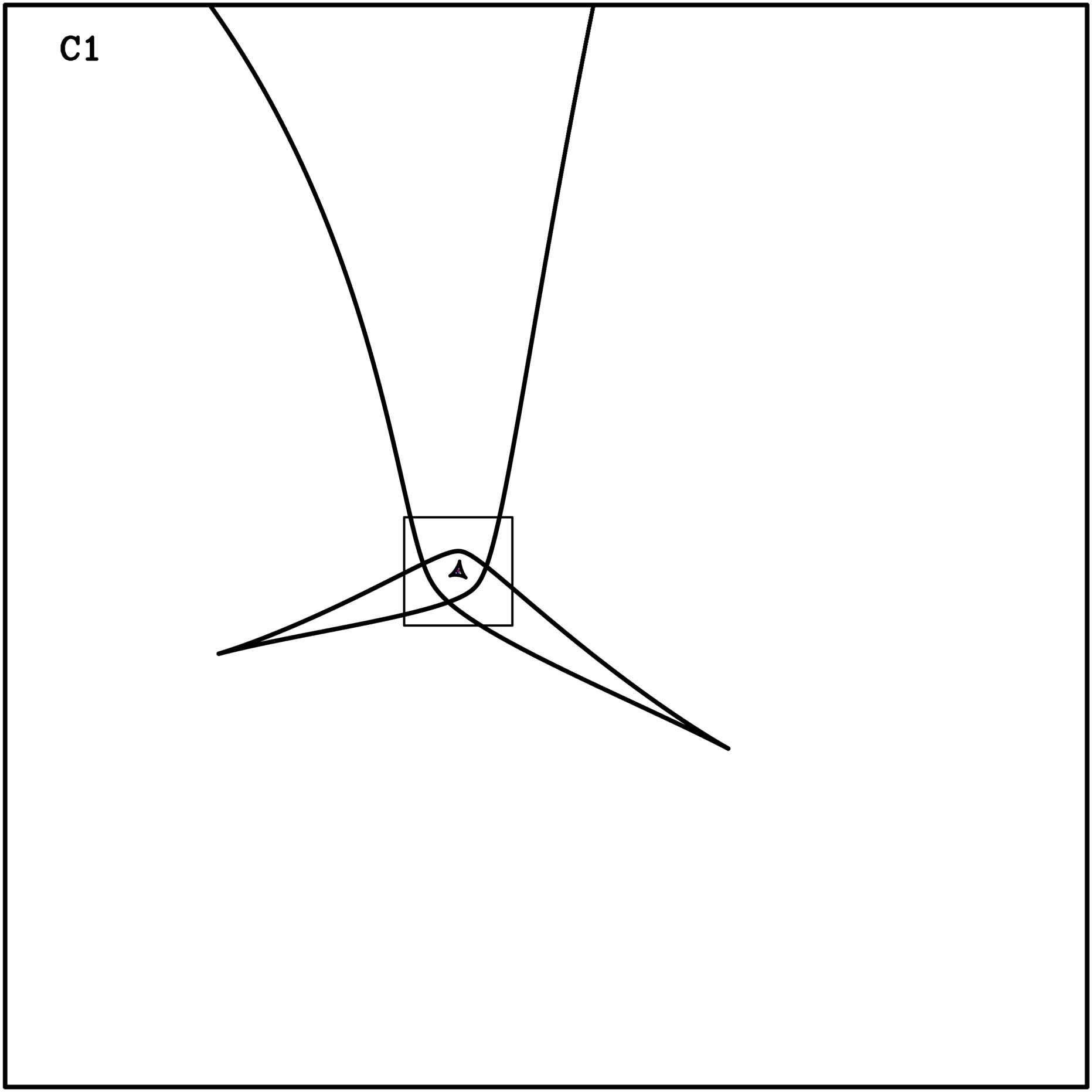}};
\begin{scope}[x={(image.south east)},y={(image.north west)}]
\node[anchor=south west,inner sep=0] (image) at (0.68,0.07) {\includegraphics[width=\textwidth,height=1.5cm,width=1.5cm]{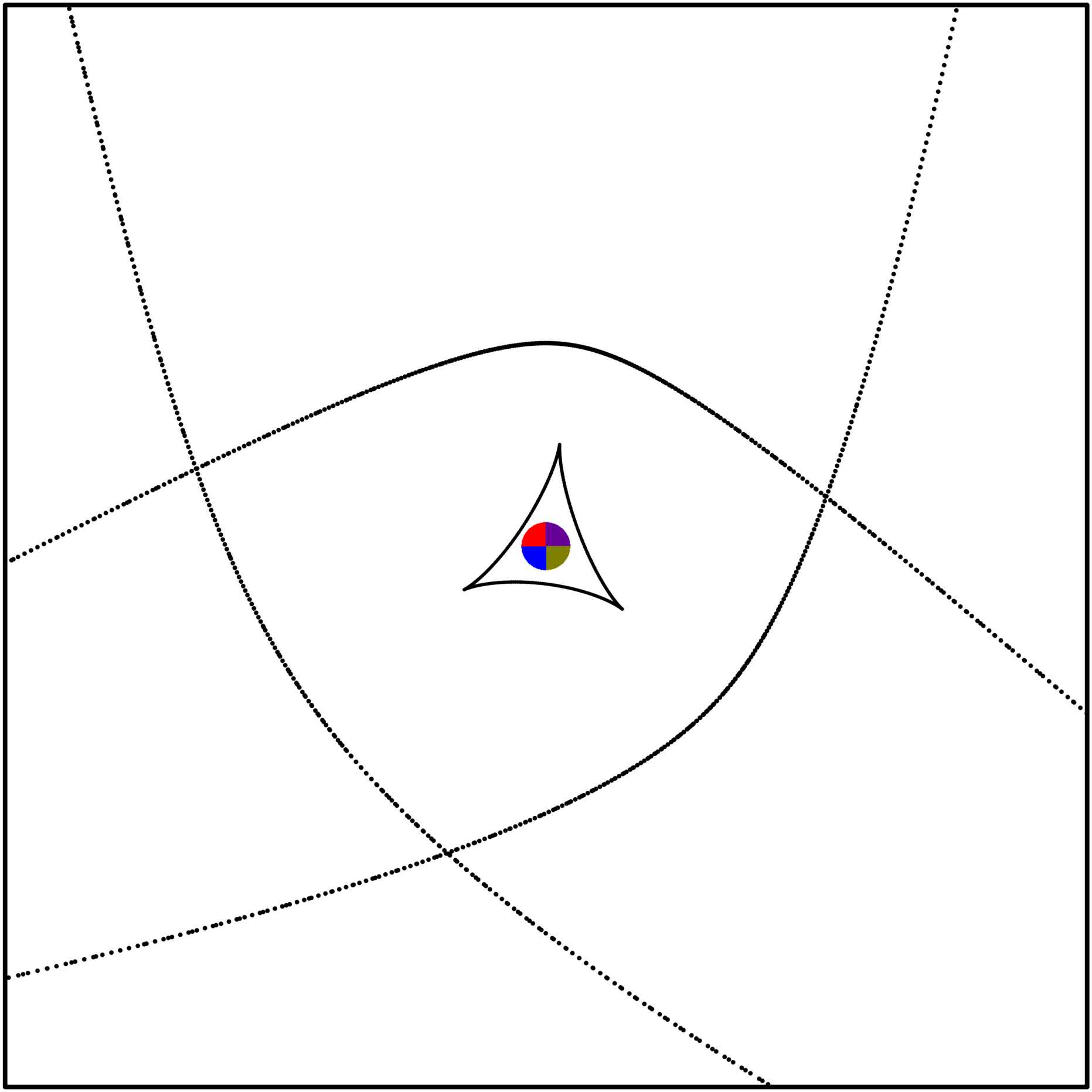}};
\end{scope}
\node[anchor=south west,inner sep=0] (image) at (5.5,0) {\includegraphics[width=\textwidth,height=5.5cm,width=5.5cm]{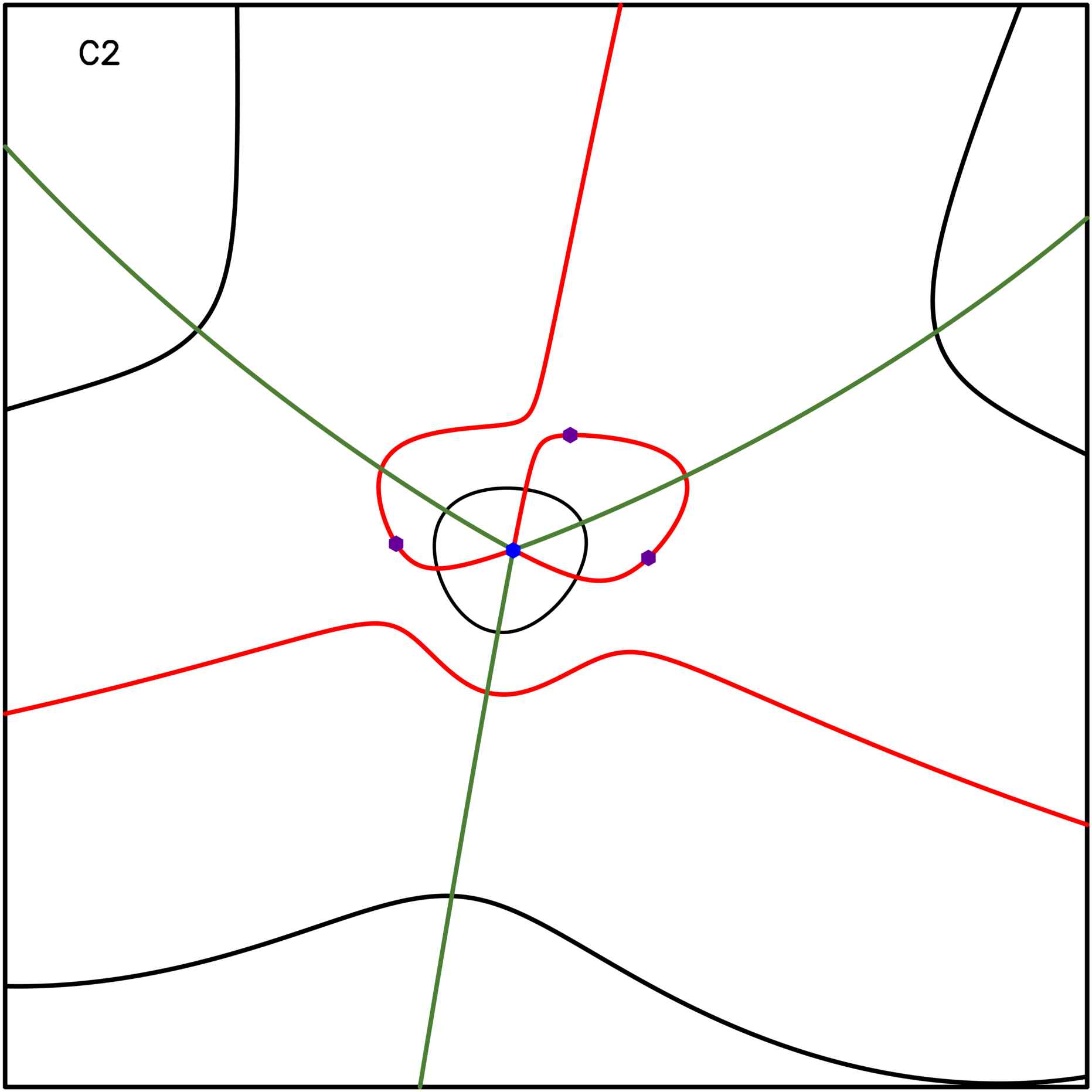}};
\node[anchor=south west,inner sep=0] (image) at (11.0,0) {\includegraphics[width=\textwidth,height=5.5cm,width=5.5cm]{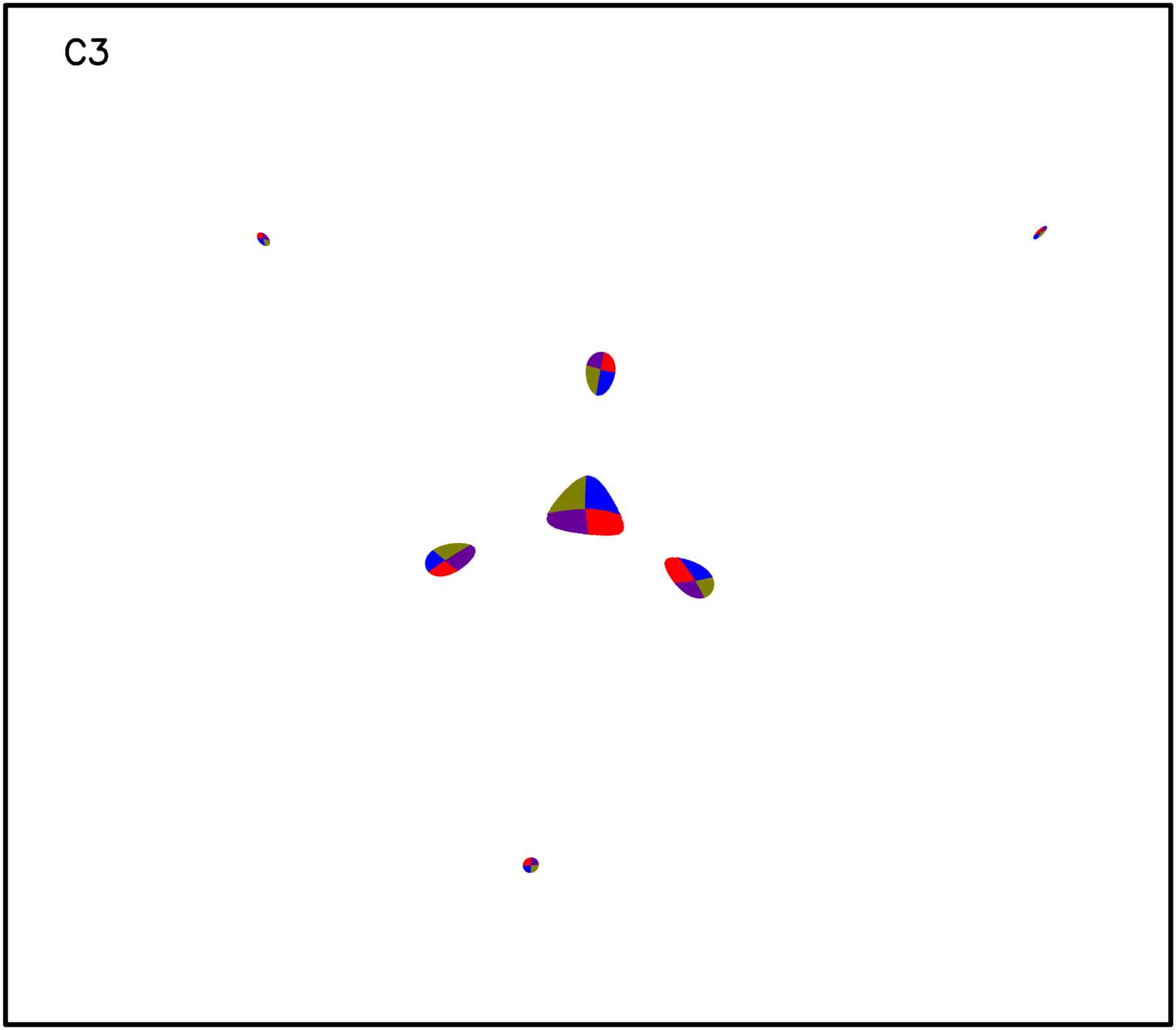}};
\end{tikzpicture}
\caption{Evolution of caustics and critical lines near a elliptic
  umbilic (denoted by blue point in singularity map). At elliptic
  umbilic triangular shaped caustic corresponding to the tangential
  caustic (panel A1) goes to a point caustic (panel B1) and emerge as
  a triangular shaped caustic corresponding to radial caustic (panel
  C1). The corresponding image formation shows a Y-shaped seven image
  configuration.} 
\label{fig:pyramid_figure}
\end{figure*}

\begin{figure*}
  \includegraphics[width=\textwidth,height=5.5cm,width=5.5cm]{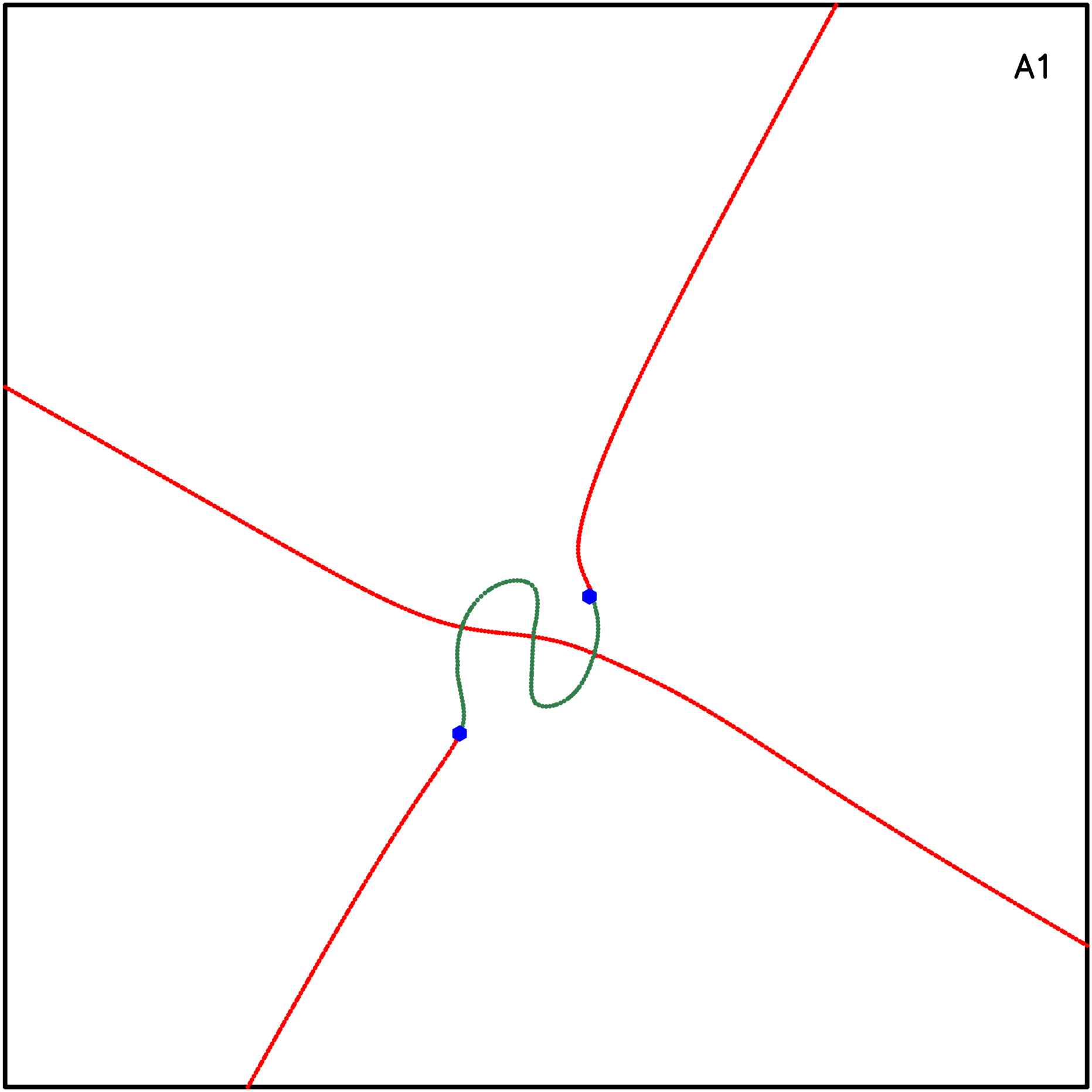}
  \includegraphics[width=\textwidth,height=5.5cm,width=5.5cm]{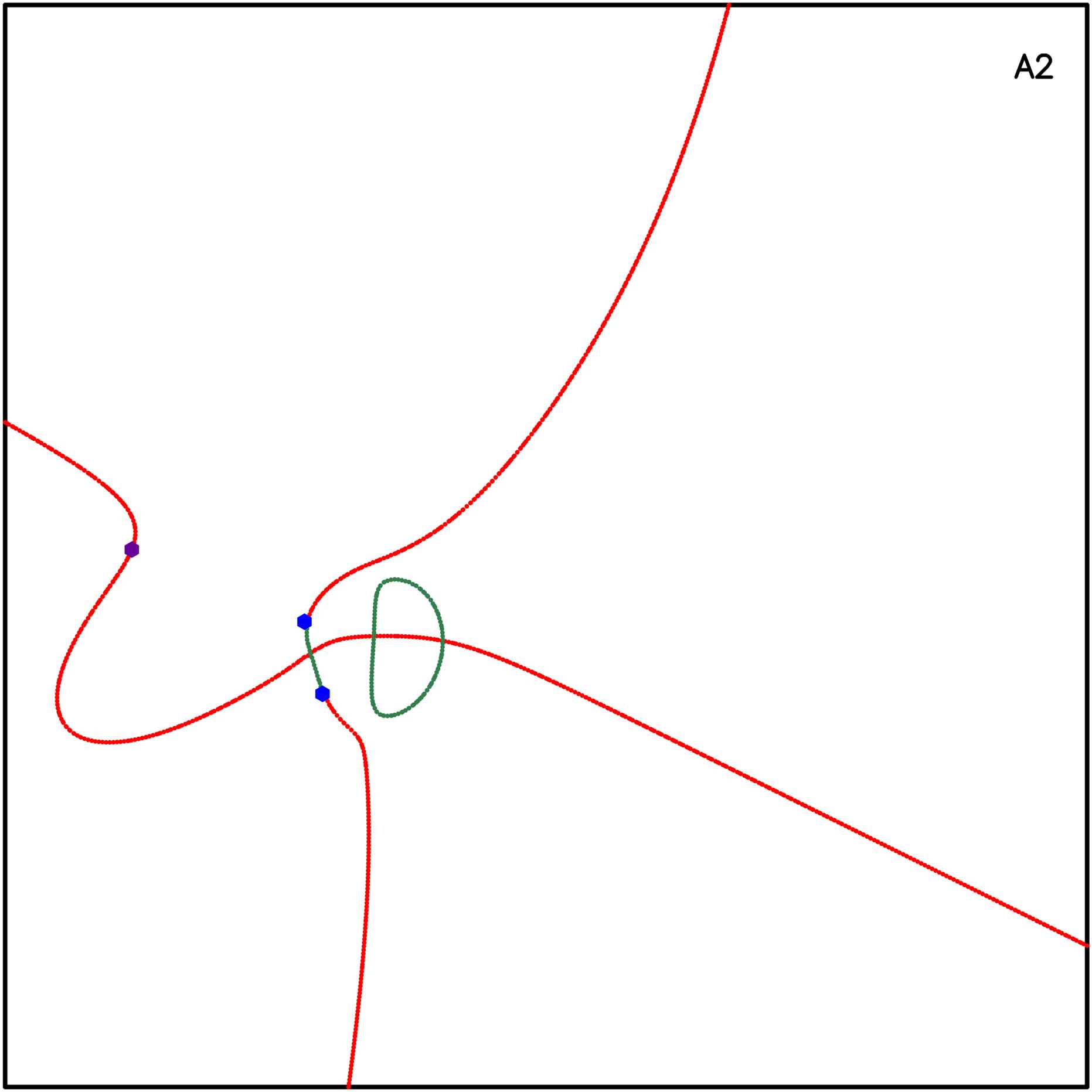} 
  \includegraphics[width=\textwidth,height=5.5cm,width=5.5cm]{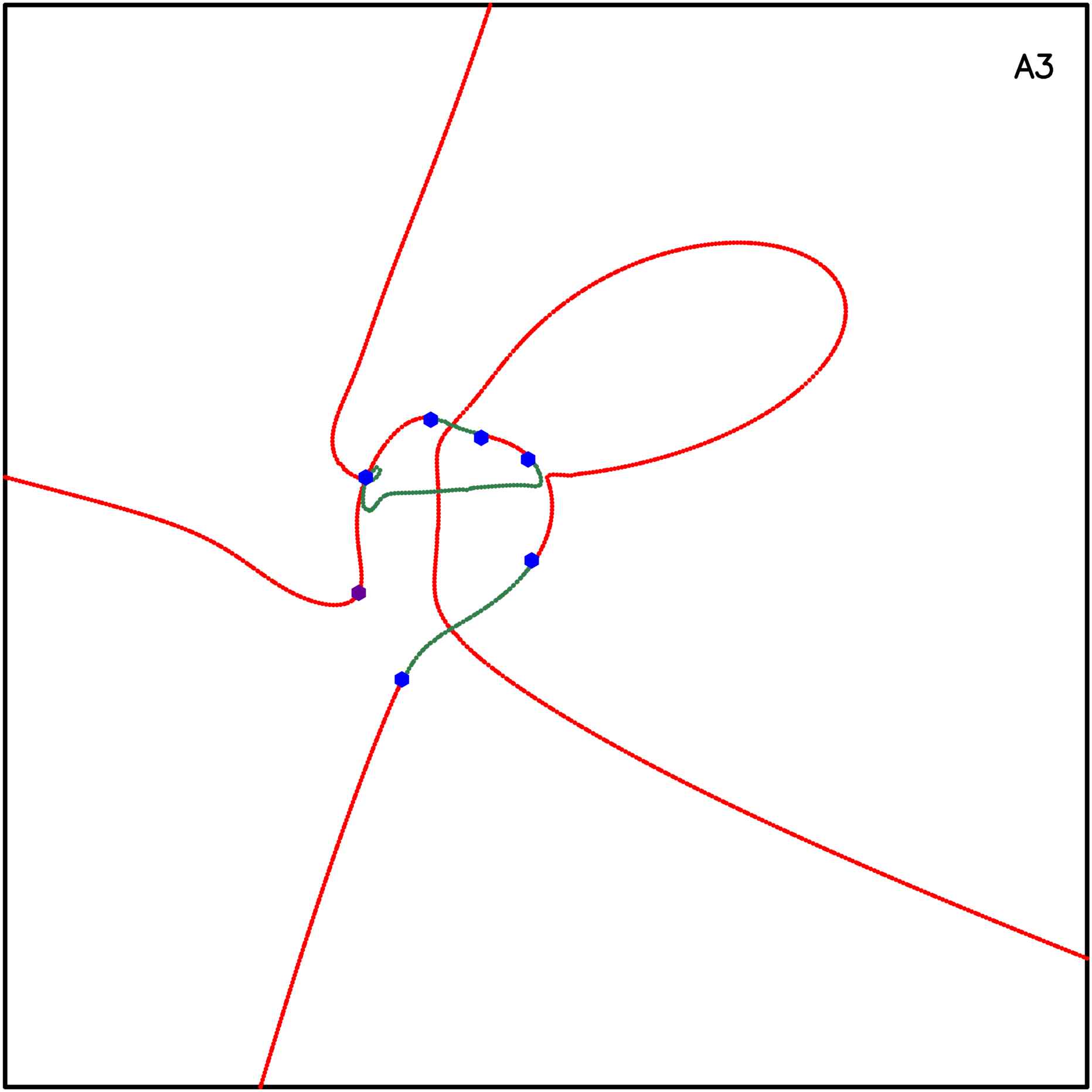}
  \includegraphics[width=\textwidth,height=5.5cm,width=5.5cm]{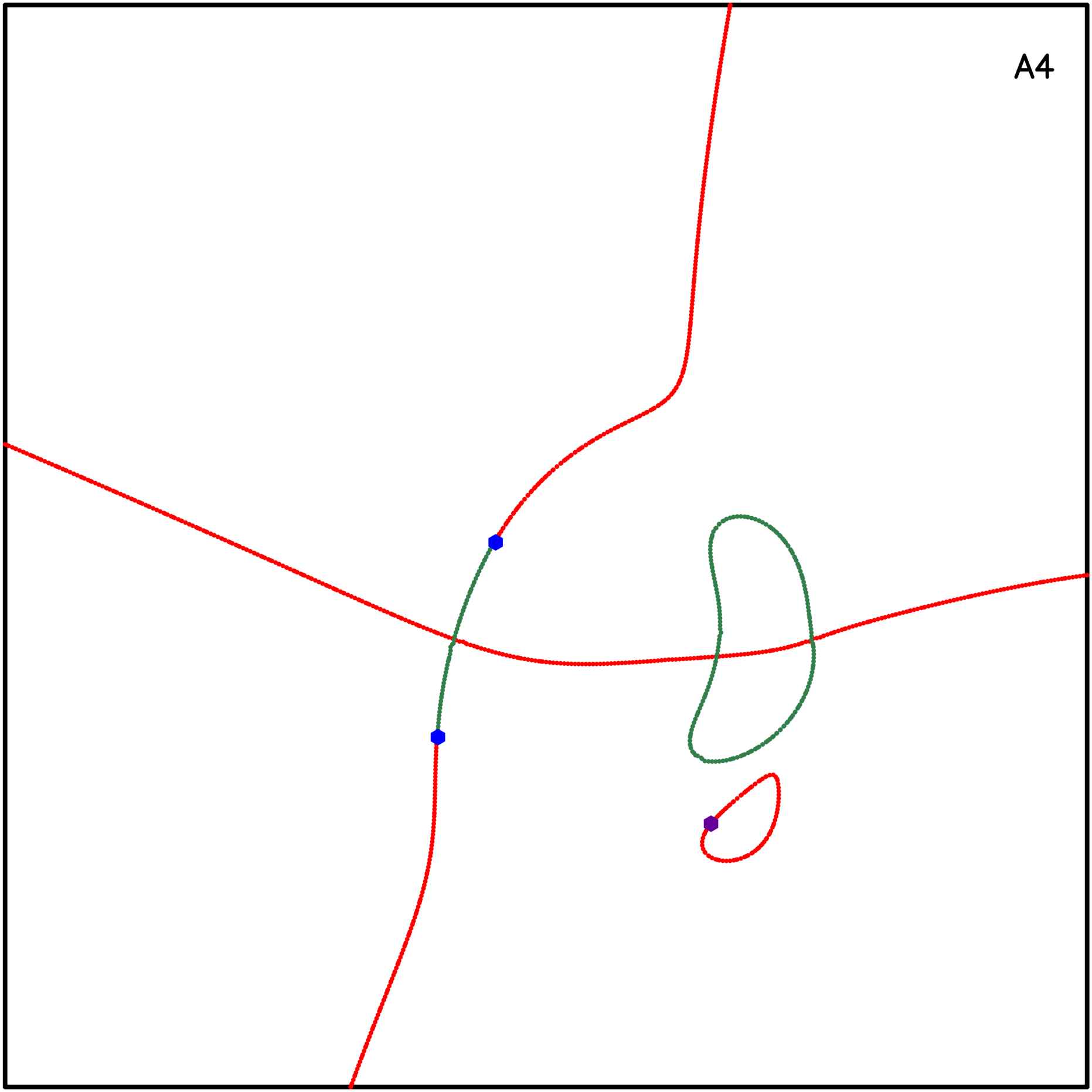}
  \includegraphics[width=\textwidth,height=5.5cm,width=5.5cm]{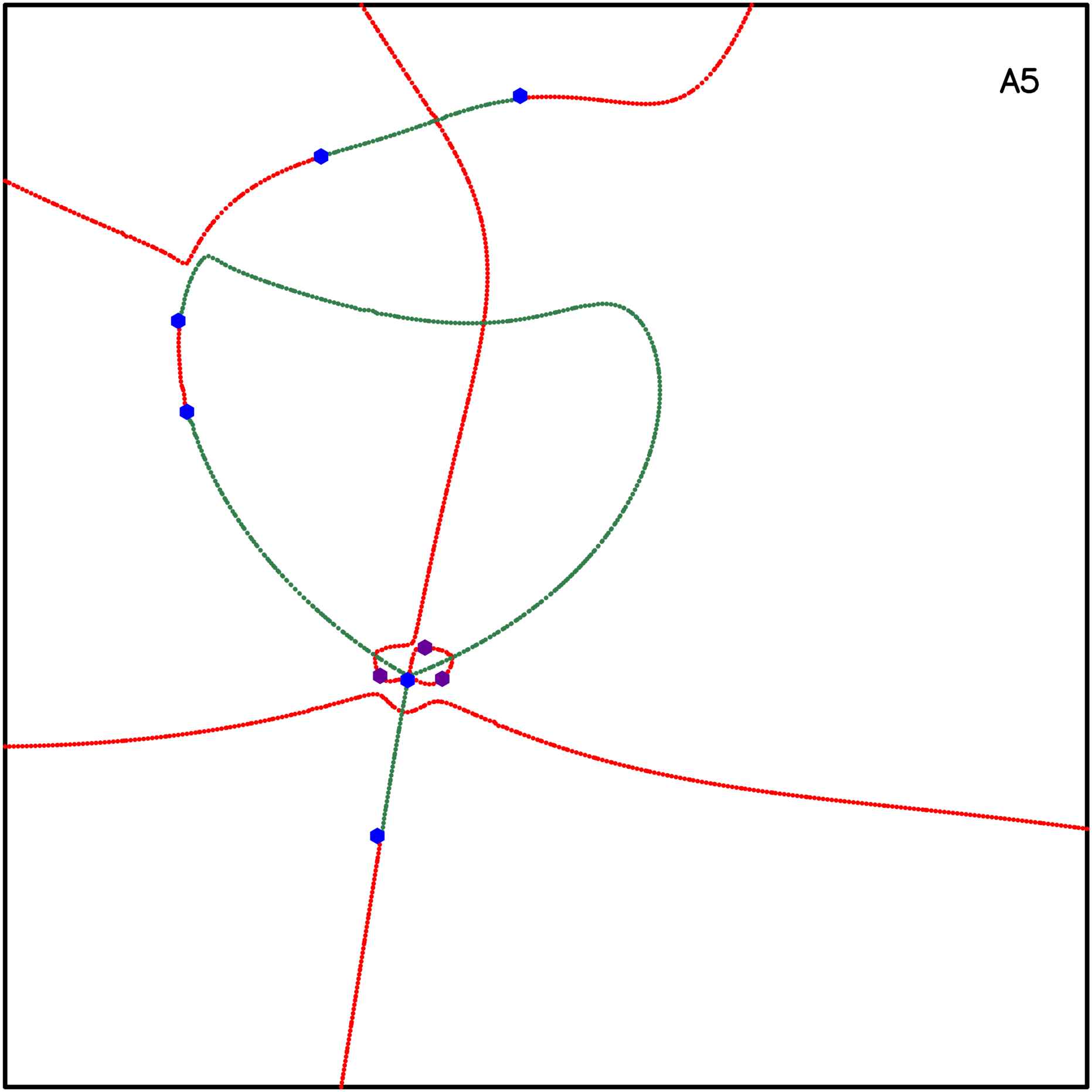} 
  \includegraphics[width=\textwidth,height=5.5cm,width=5.5cm]{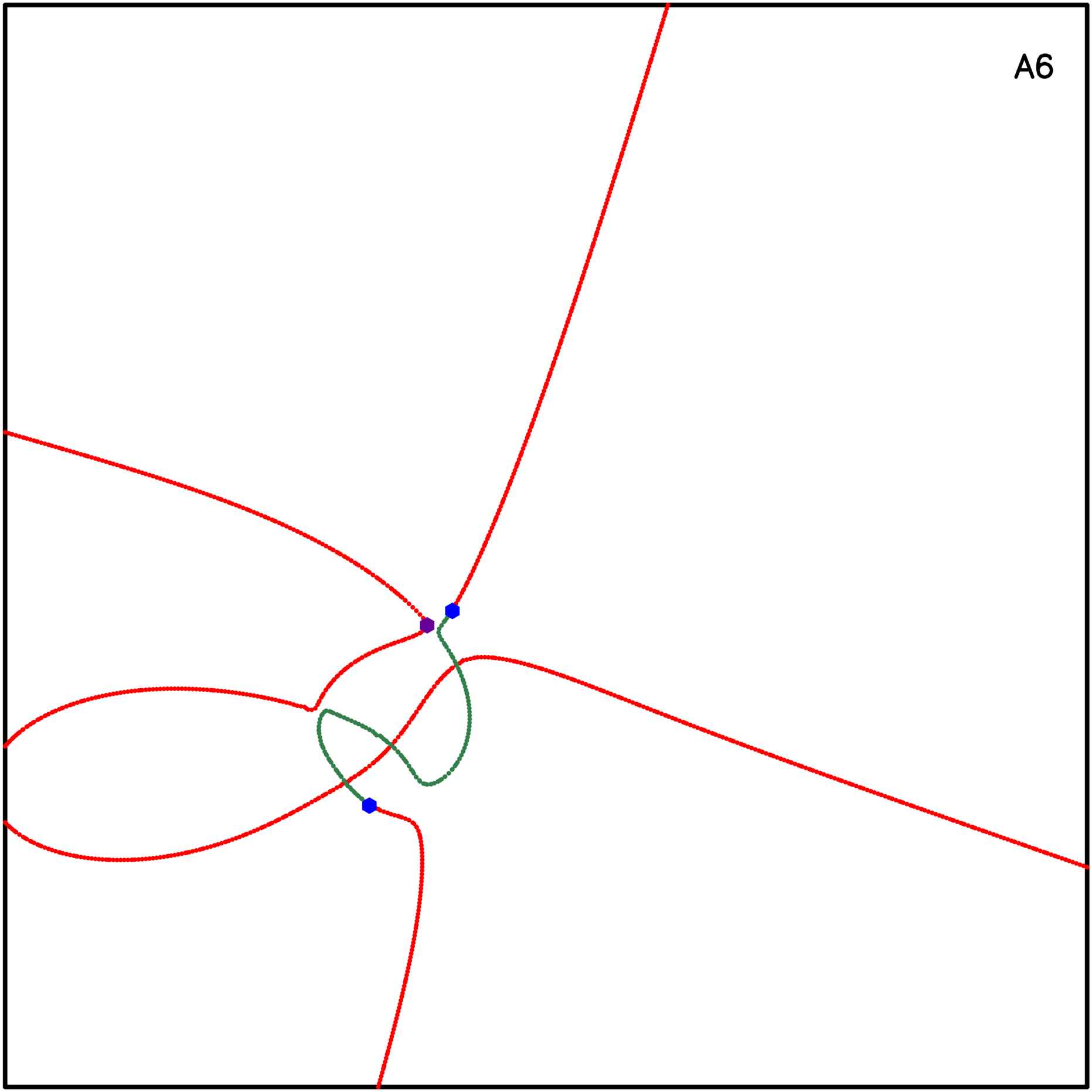}
  \includegraphics[width=\textwidth,height=5.5cm,width=5.5cm]{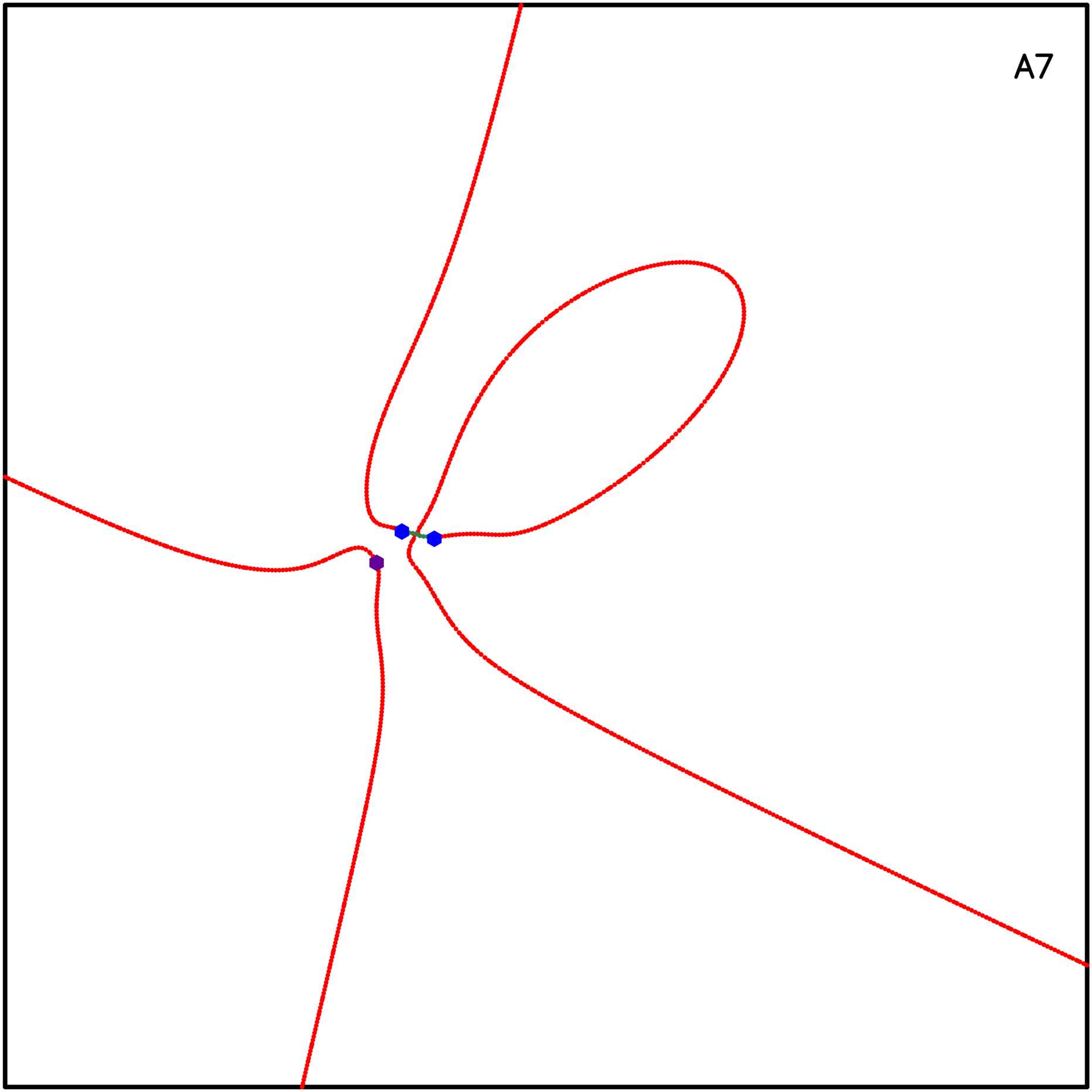}
  \includegraphics[width=\textwidth,height=5.5cm,width=5.5cm]{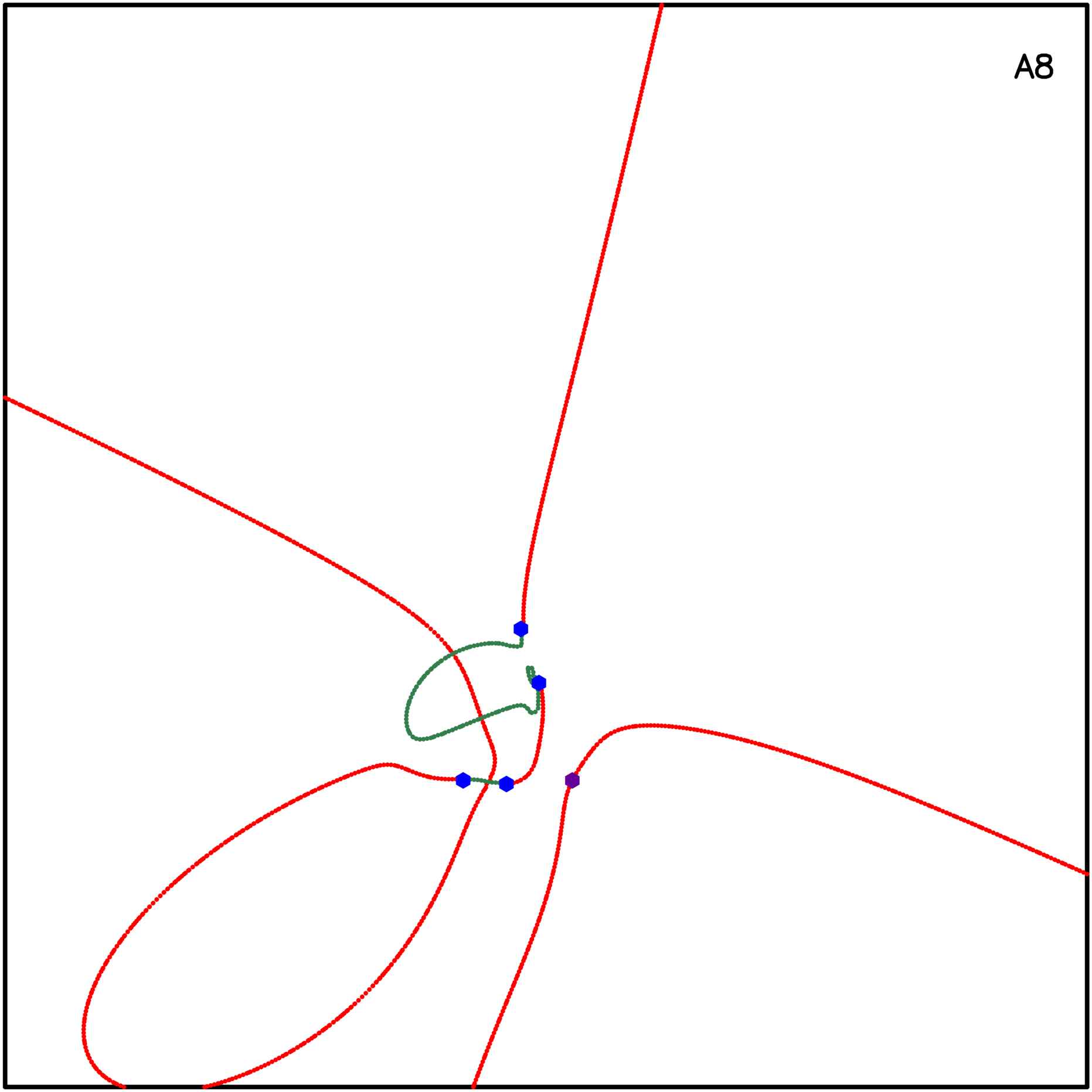} 
  \includegraphics[width=\textwidth,height=5.5cm,width=5.5cm]{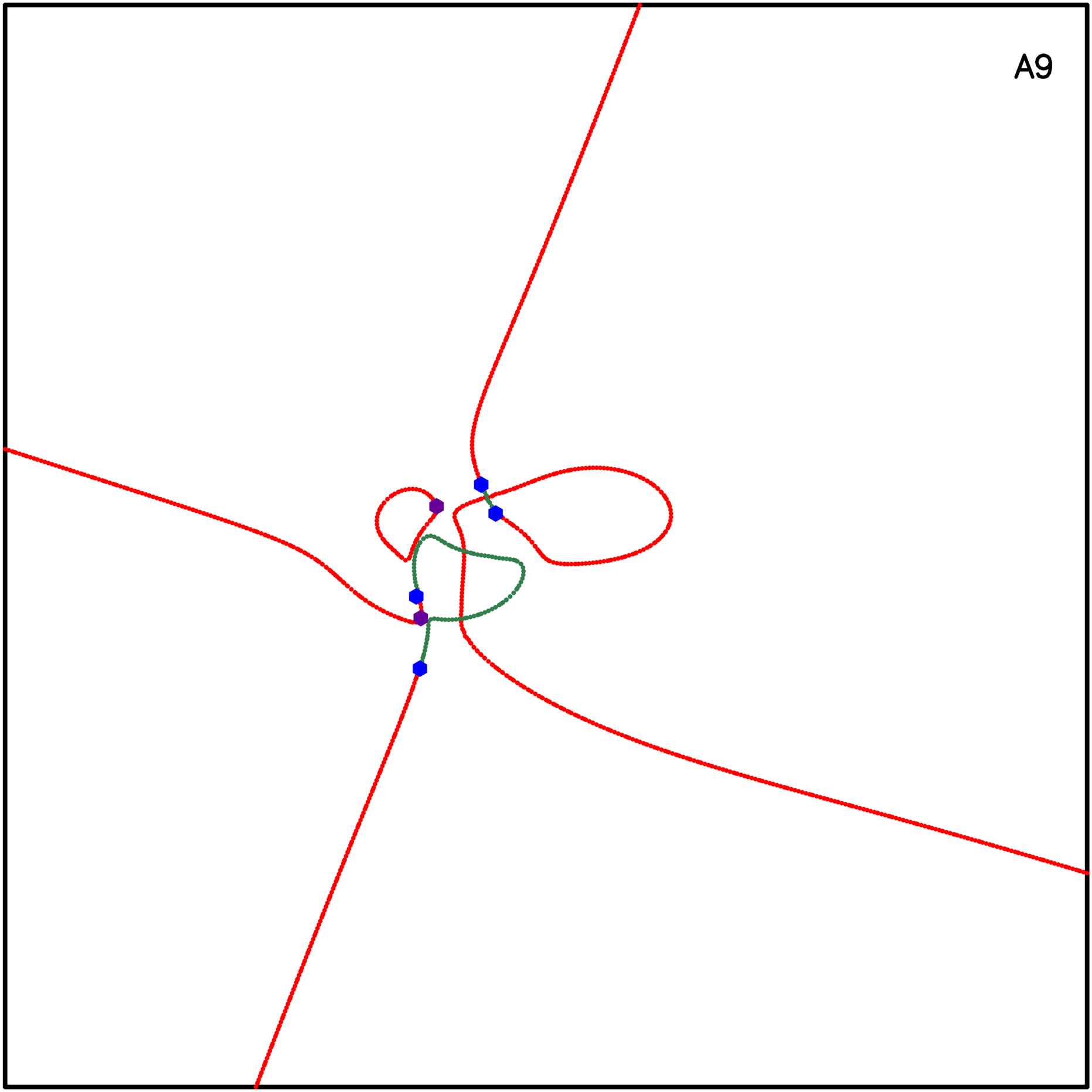}
  \caption{Singularity maps for different positions and orientations
    of the secondary lens in case of two-component elliptical lens
    with a fixed position and orientation of primary lens. The red and
    dark green lines are the $A_{3}$-lines with swallowtail and umbilics
    denoted by violet and blue points, respectively. The position
    and number of unstable singularities change with the change in
    lens parameters. Which shows the strong dependency of singularity
    map on the lens parameters.} 
  \label{fig:map_figure}
\end{figure*}

\begin{figure*}
  \includegraphics[width=\textwidth,height=5.5cm,width=5.5cm]{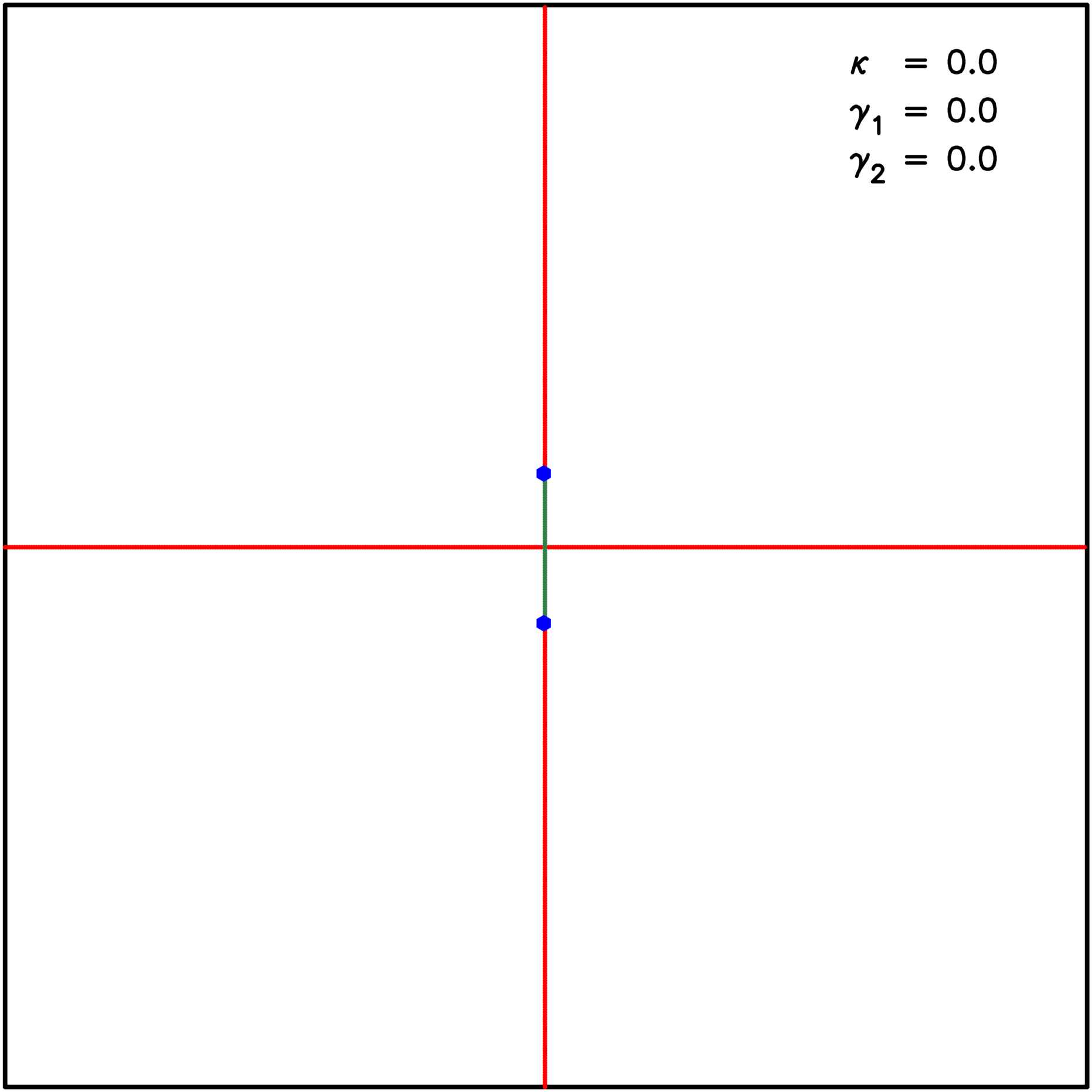}
  \includegraphics[width=\textwidth,height=5.5cm,width=5.5cm]{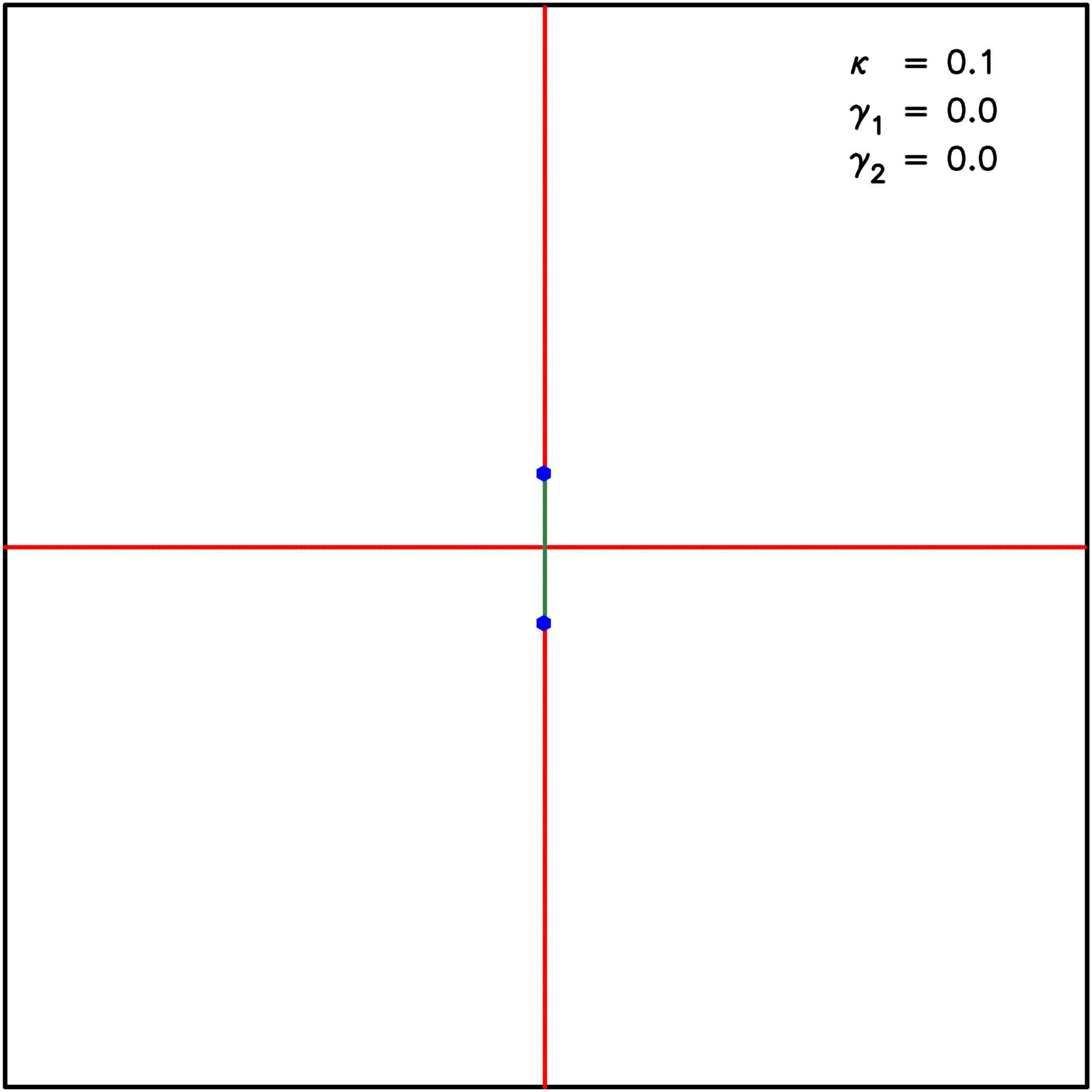} 
  \includegraphics[width=\textwidth,height=5.5cm,width=5.5cm]{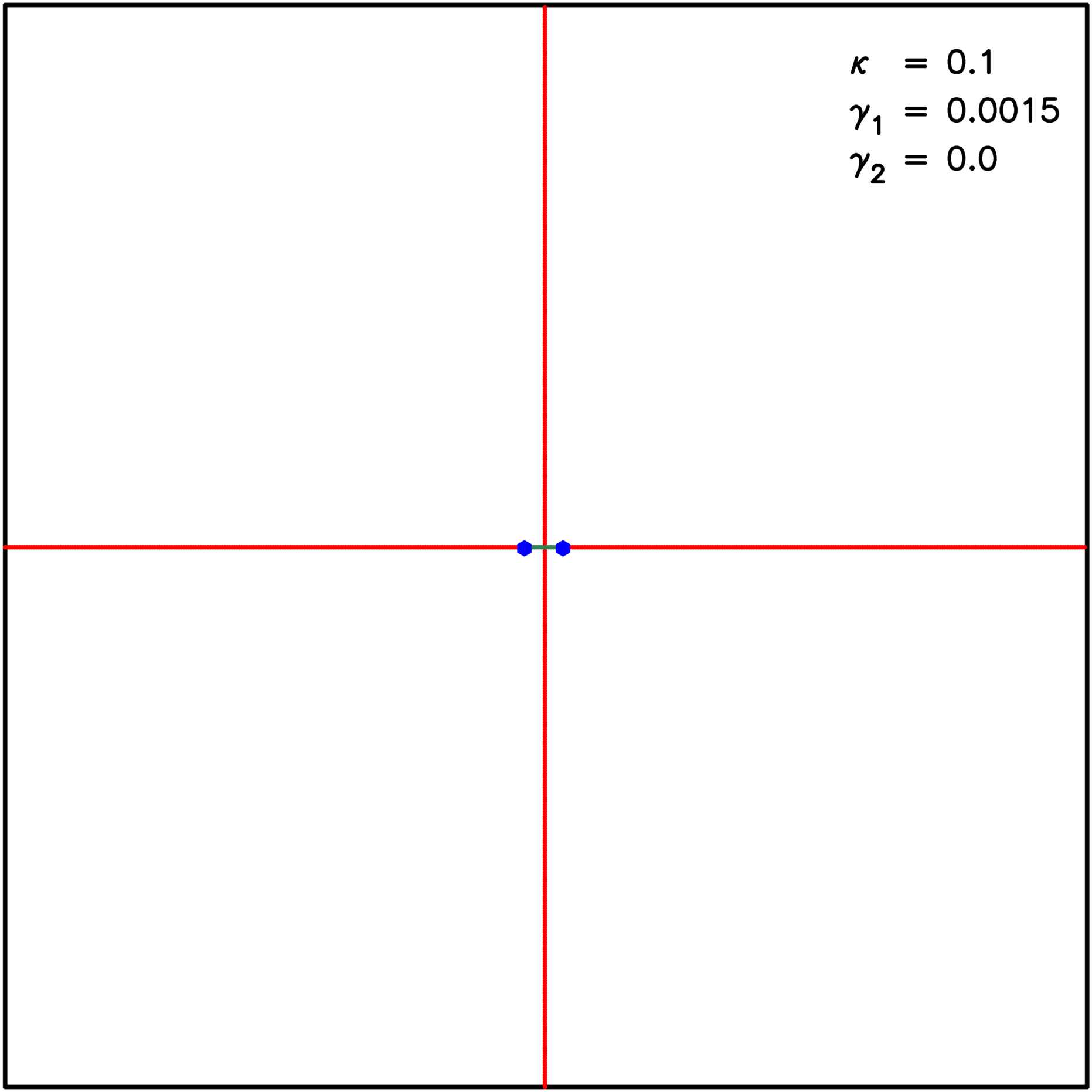}
  \includegraphics[width=\textwidth,height=5.5cm,width=5.5cm]{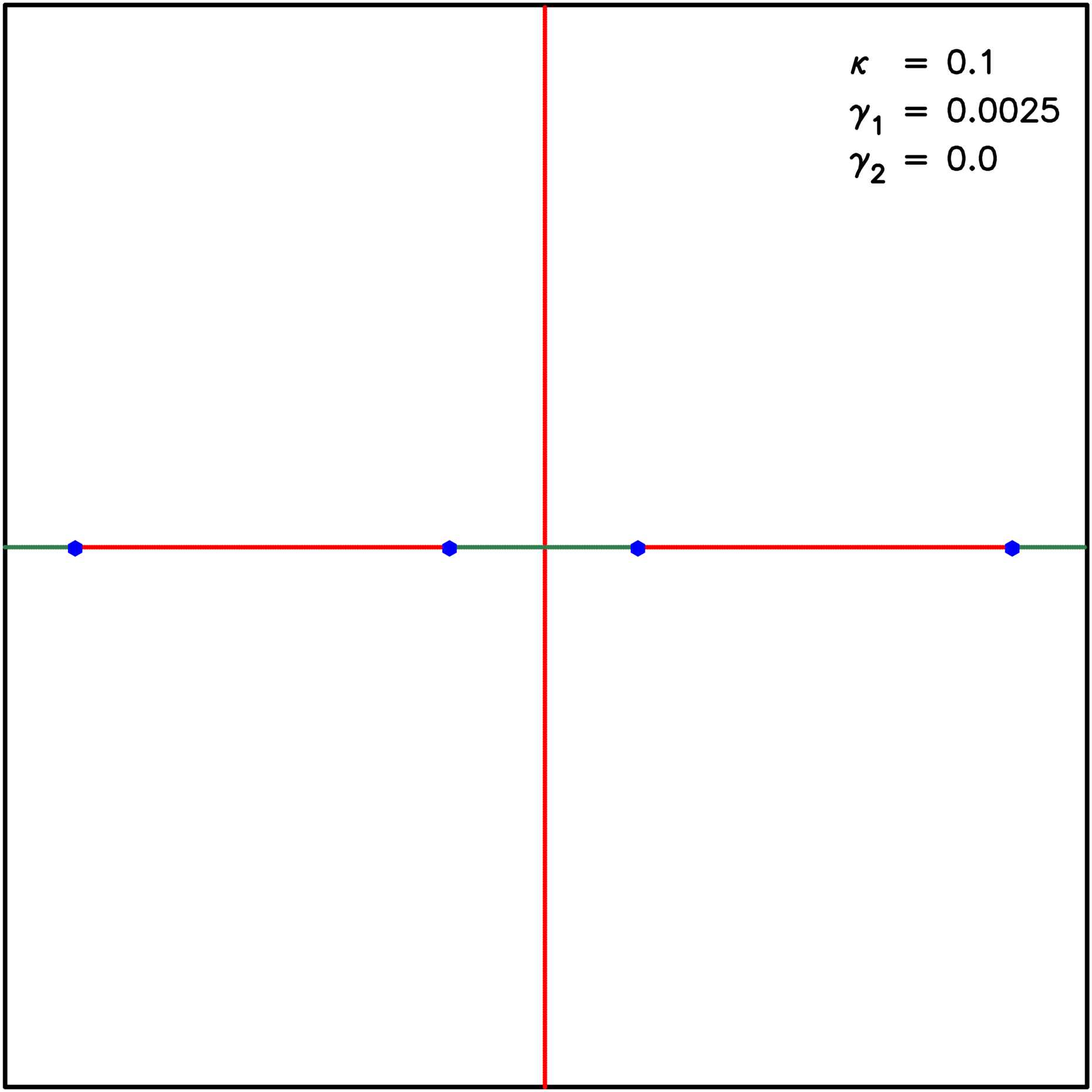}
  \includegraphics[width=\textwidth,height=5.5cm,width=5.5cm]{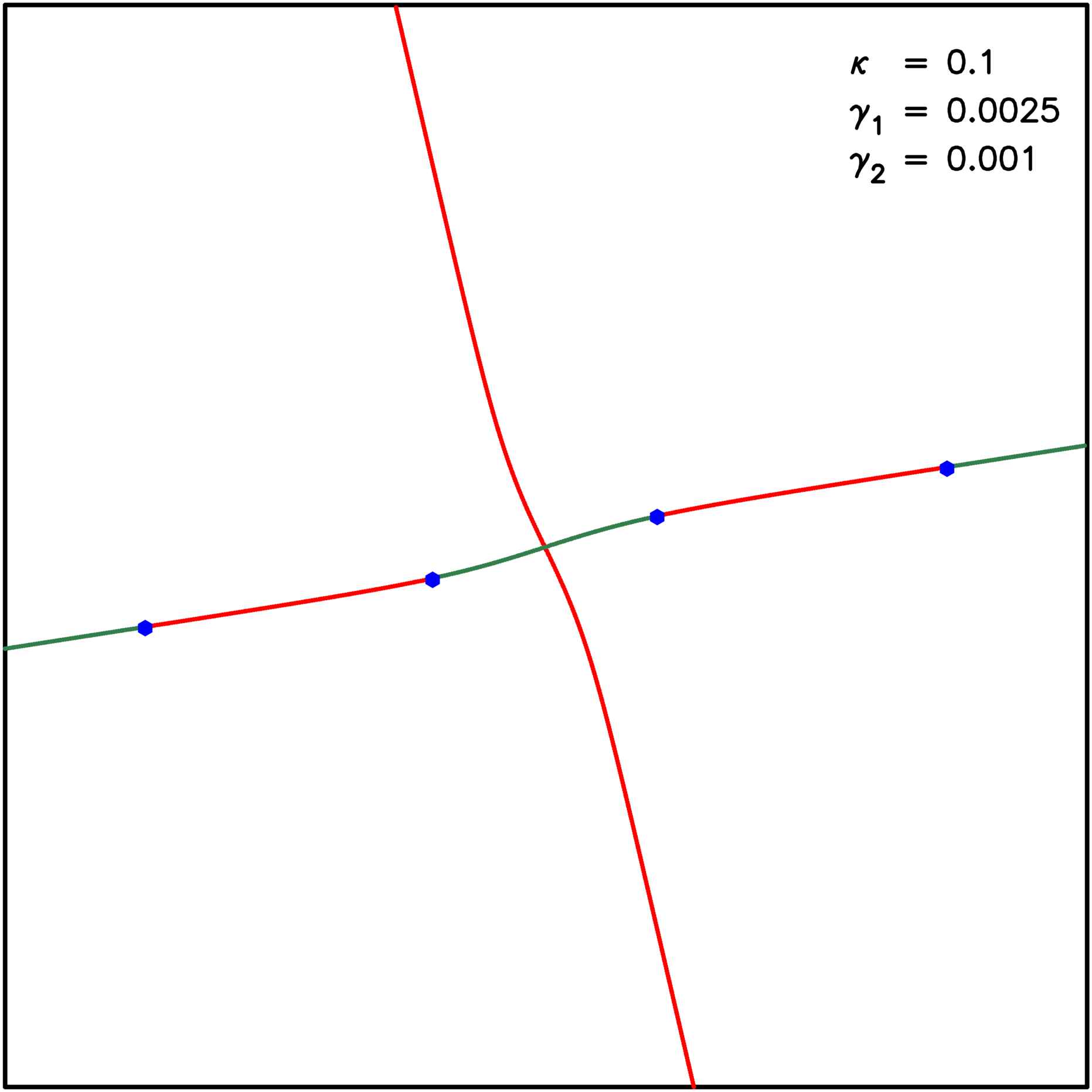} 
  \includegraphics[width=\textwidth,height=5.5cm,width=5.5cm]{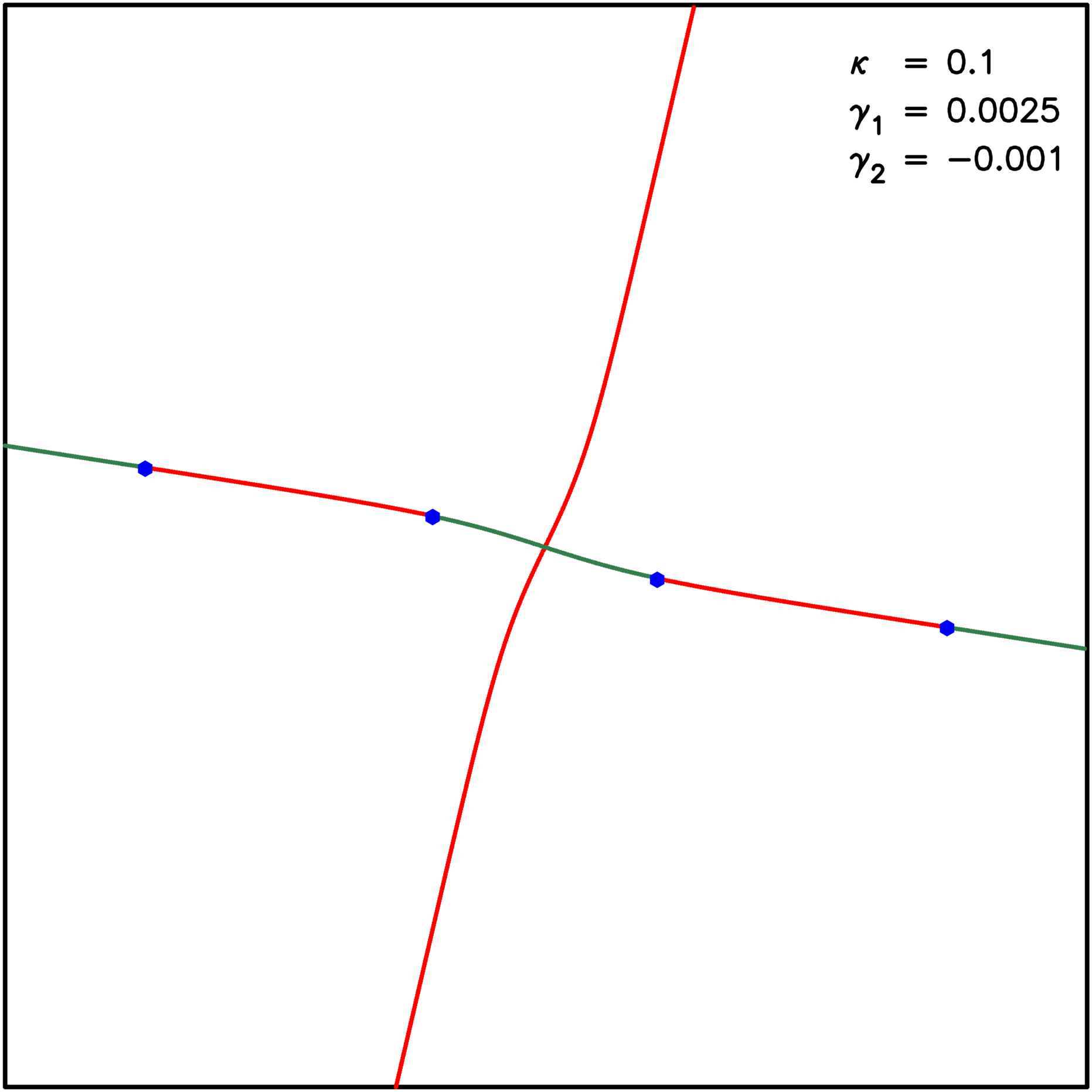}
  \includegraphics[width=\textwidth,height=5.5cm,width=5.5cm]{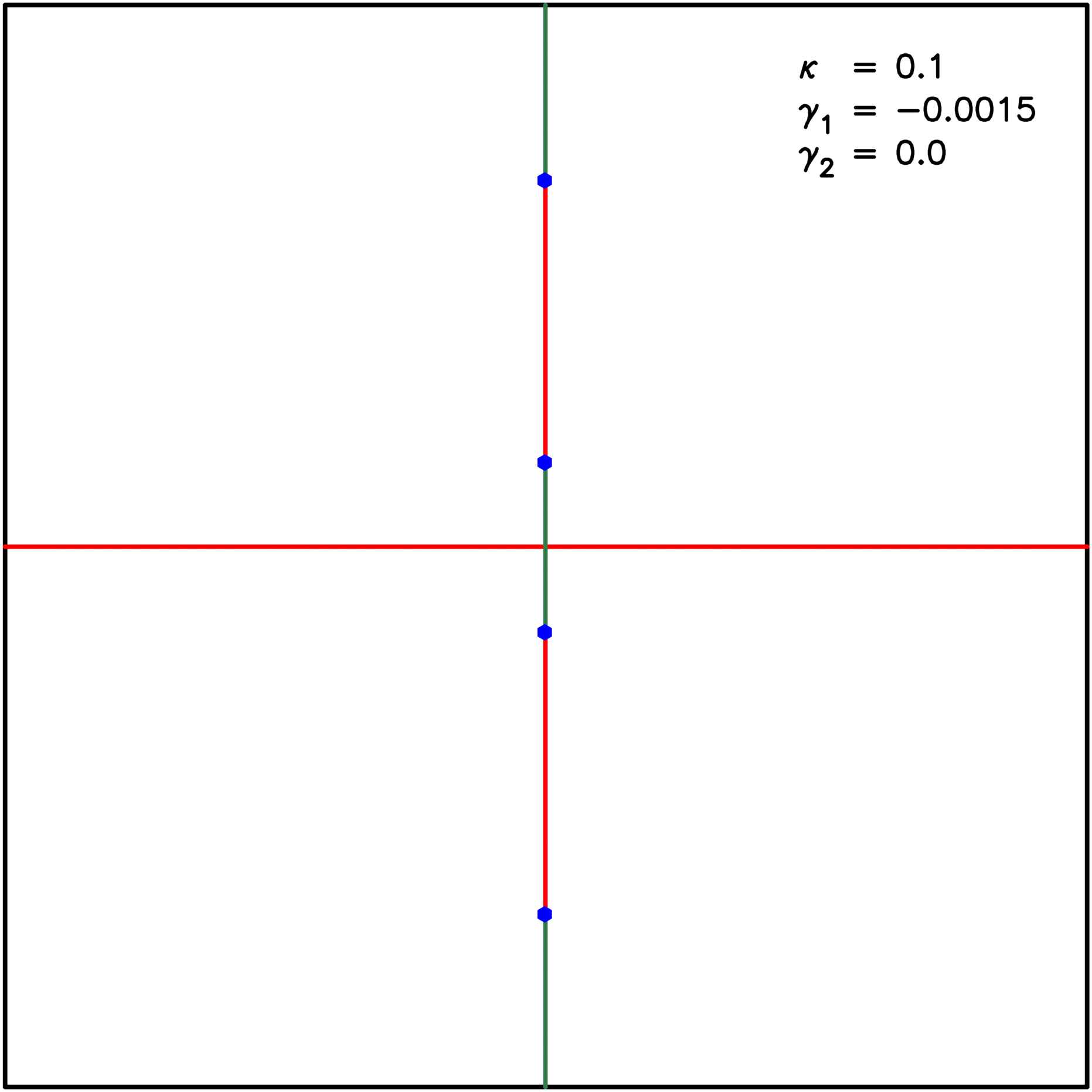}
  \includegraphics[width=\textwidth,height=5.5cm,width=5.5cm]{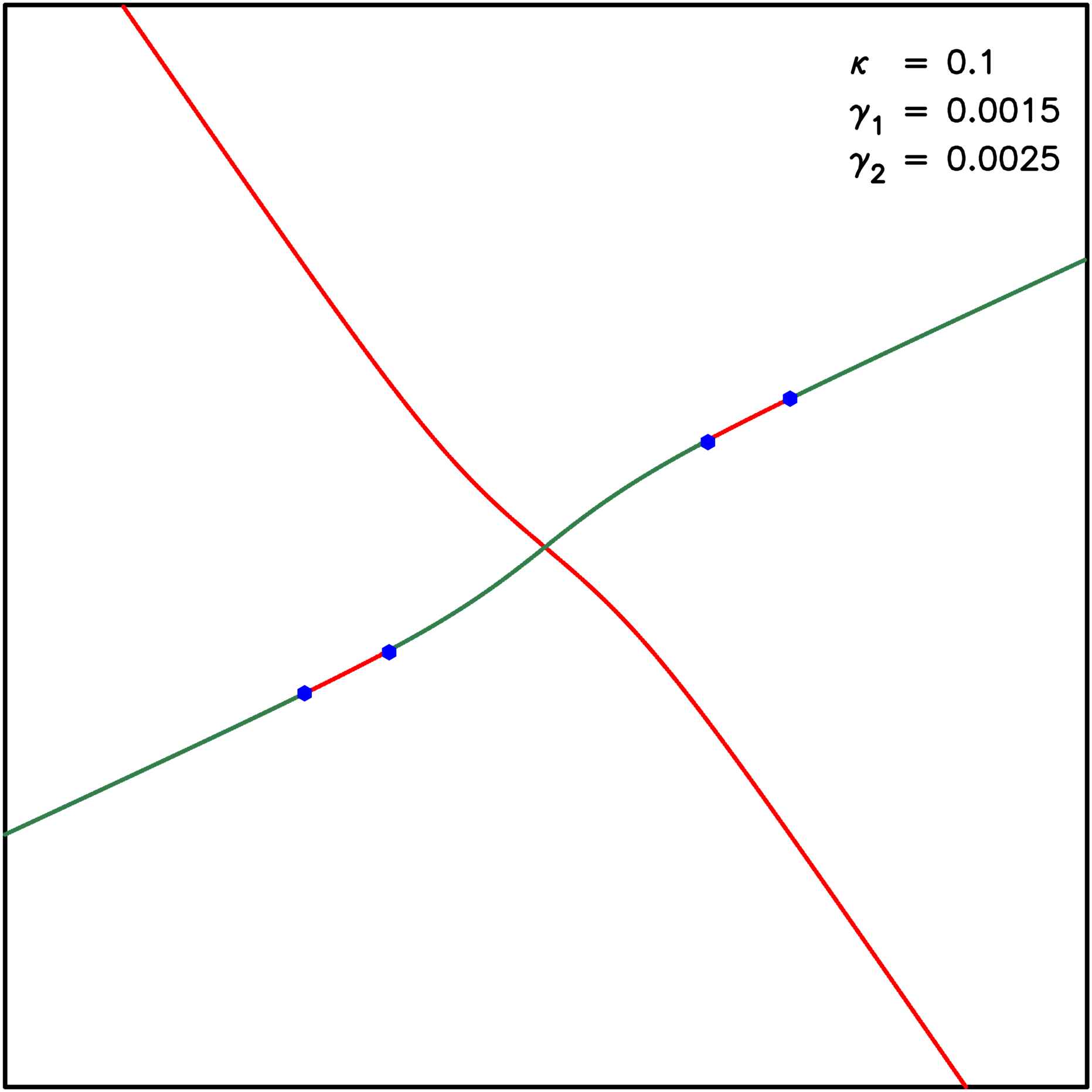} 
  \includegraphics[width=\textwidth,height=5.5cm,width=5.5cm]{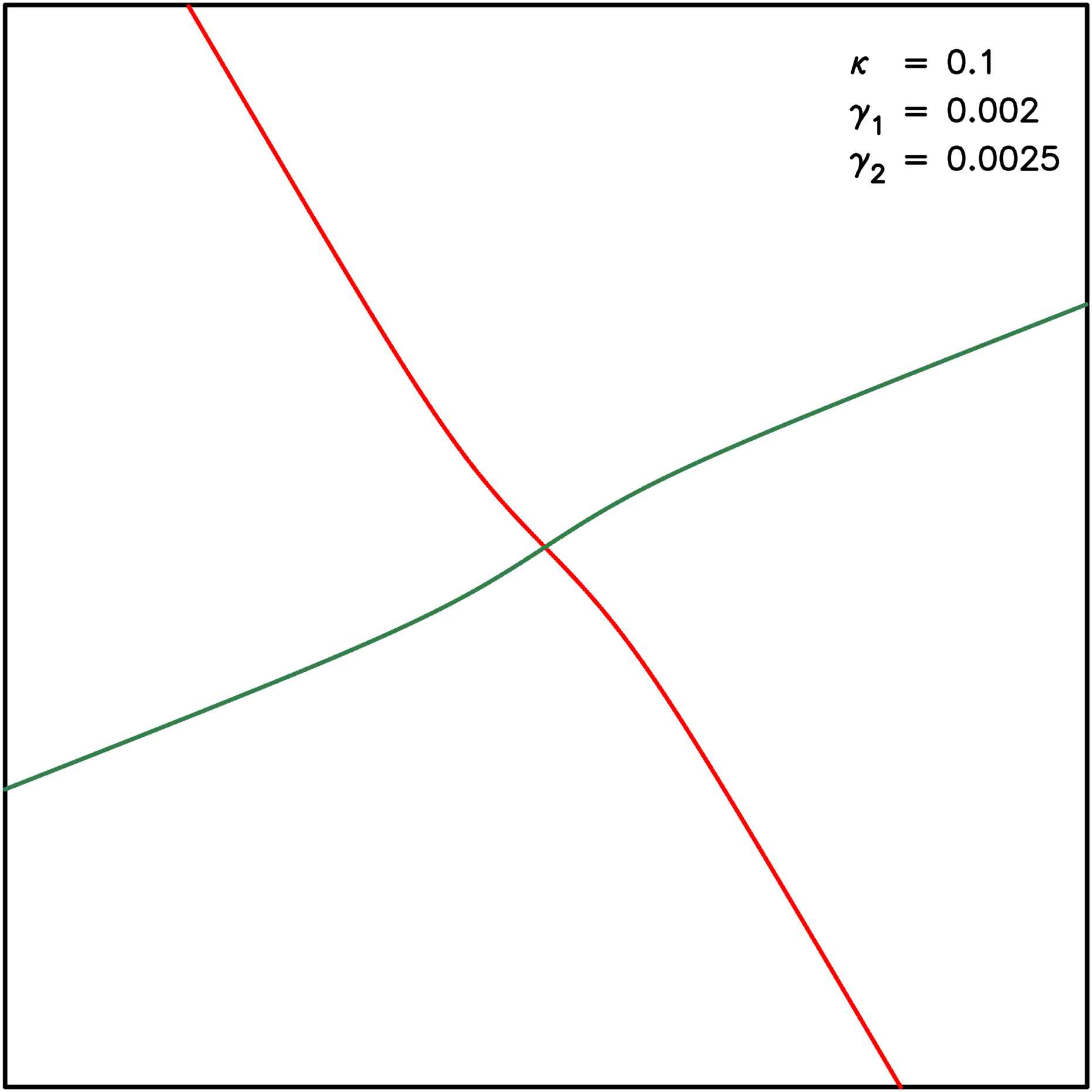}
  \caption{Effect of external convergence and shear on singularity map
    in case of a one-component elliptical lens. One can see the motion
    of extra pair of umbilics towards the already existing pair of
    umbilics. After a certain amount of external shear all point
    singularities disappear from the singularity map (bottom-right
    panel).} 
  \label{fig:stability1}
\end{figure*}

\section{Classification of Singularities}
\label{sec:Classification}

In this section we discuss classification of singularities 
presented in the lens mapping.
The singularities refer to situations where the map from the image
plane to the source plane is no longer one to one, and an
infinitesimal solid angle in the image plane maps to zero solid angle
in the source plane.
There are two stable singularities: fold and cusp, and these are
present in all situations when we have formation of multiple images.
Thus, in general, a caustic in source plane represents cusps connected
by folds.
The corresponding curve in the image plane is called the critical
curve and is expected to be smooth.
In the following discussion, we will encounter other singularities,
e.g., beak-to-beak, swallowtail, elliptic and hyperbolic umbilics, but
these are not stable.
These occur only for specific source redshifts with specific lens
parameters \citet{Schneider1992,Bagla 2001}.  
At these unstable singularities, cusps are either created or destroyed
or there is an exchange of cusp between radial and tangential caustics
in such a way that the total number of the cusps in source plane always
remains even.
All these unstable singularities are point singularities and have
characteristic image formations.

The classification of all these singularities is based on 
catastrophe theory.
In the context of lensing, catastrophe theory describes the
singularities in terms of derivatives of the Fermat potential,
$\phi(\mathbf{x},\mathbf{y})$ (e.g. SEF, \citet{Petters 2001}), which
is related to the lensing potential $\psi(\mathbf{x})$ as, 
\begin{equation}
\phi(\mathbf{x},\mathbf{y}) =
Const. \left(\frac{1}{2}\left(\mathbf{x}-\mathbf{y}\right)^2 -
  \psi(\mathbf{x})\right).   
\end{equation}
Instead of using Fermat potential one can also use deformation tensor
(which is completely determined by its eigenvalues and eigenvectors)
to discuss different singularities that occur in gravitational
lensing.
In this way, one does not have to worry about the source parameters,
which affect the Fermat potential.  
The benefit of using deformation tensor instead of Fermat potential is
that one does not have to draw critical lines and caustics for all
possible source redshift in order to find highly magnified regions in
the image plane.  
And the study of deformation tensor gives a singularity map
(in the lens plane) of all possible singularities that can occur for a
given lens model for all source redshifts. 
In our analysis we do not focus on the source redshift but list all the
singularities of the map.
This approach enables us to focus on these singularities and attempt
statistical analysis in typical lens models. 
This also brings in its own limitations: we are unable to discuss
folds as these have an explicit source redshift dependence.  
This however can be recovered without much work after the
singularities have been mapped. 

\subsection{$A_{3}$-lines}
\label{ssec:A3 lines}

$A_{3}$-lines are the essential elements of the singularity map for a
given lens model.
In the image plane, these are the lines on which cusps form.
As all point singularities are associated with creation, destruction
or exchange of cusps, our first goal is to identify the
$A_{3}$-lines for a lensing potential. 

In the image plane $A_{3}$-lines pass through the points where the
gradient of the eigenvalue of the deformation tensor is orthogonal to
the corresponding eigenvector $n_{\lambda}$, 
\begin{equation}
n_{\lambda}.\nabla_{x}\lambda = 0.
\label{eq:(A3 lines)}
\end{equation}
Which implies that at $A_{3}$-lines the eigenvector $n_{\lambda}$ is
tangent to the corresponding eigenvalue contour
\citep{Arnold 1982,Hidding 2014}.
The reader may note that this is also true at points where the
eigenvalues have extrema, however such points are isolated.
At generic points along these lines an infinitesimal portion of the
critical curve, which essentially is a contour level for the
eigenvalue, is mapped onto itself as we go from the image plane to the
source plane. 

In case of a spherically symmetric lens, every point in the lens plane
satisfies equation~(\ref{eq:(A3 lines)}).
As a result, a spherical symmetric lens gives a formation of point
caustic in source plane (SEF) at any point.

In general, we observe two different sets of $A_{3}$-lines in lens
plane, one for each eigenvalue of the deformation tensor.
The points in the lens plane where $A_{3}$-lines and the corresponding
eigenvalue contour (with $\alpha$ or $\beta=1/a$) cross each other
correspond to the cusp singularities in source plane at that
redshift. 

These lines do not intersect each other though as we shall see, lines
corresponding to the two eigenvalues can meet at degenerate points
($\alpha = \beta$).  
The presence of $A_{3}$-lines itself proves the stability of cusp
singularities in lens mapping: changing the redshift of the source
plane merely shifts the cusp to a neighbouring point. 

\subsection{Swallowtail Singularities}
\label{ssec:Swallowtail}

In the image/lens plane, swallowtail singularities mark the points where
eigenvector $n_{\lambda}$ of deformation tensor is tangent to the
corresponding $A_{3}$-line.  
Which implies that at a swallowtail singularity the corresponding
eigenvalue ${\lambda}$ reaches a local maxima along $A_{3}$-lines, but
this is not a true local maxima \citep{Arnold 1982,Hidding 2014}.
We use this method to identify swallowtail singularity in lens maps. 

The characteristic image formation for a swallowtail singularity is an
elongated arc.
This {\sl arc} is made up of four images.
As we move away from swallowtail singularity the arc changes into
distinct multiple images.  
At a swallowtail singularity, the number of cusps in source plane
change by two: a section of fold bifurcates into two cusps at this
point.   

Figure~(\ref{fig:swallowtail_figure}) illustrates the caustics and
critical curves in source and lens plane around a redshift at which a
swallowtail singularity becomes critical.
The lens model used here is a two-component softened elliptical
isothermal lens. 
The first column shows the formation of tangential (radial) caustics,
denoted by thick (thin) lines, in the source plane for three different
redshifts including redshift $z_{s}$, at which the swallowtail
singularity becomes critical (panel B1).
The second column shows the corresponding critical curves and the
singularity map consisting of $A_{3}$-lines (red for $\alpha$ and
dark green for $\beta$ eigenvalue) and other singularities in the lens
plane.
Position of the swallowtail singularity is denoted by a violet point on
the $A_{3}$-line: this is the point where the $A_{3}$ line is
tangential to the critical curve.
The blue points denote the position of hyperbolic umbilics,
discussed in the following subsection.  
The third column shows the corresponding image formation in lens plane
for a given source position in source plane. 
To see the multiple image formation, we take a circular source: a
different color in each quadrant.
Such a multi-color source is helpful to recognize positive and negative
parity images. 
The source is shown in the source plane in panels in the left column.
A circle is plotted around the source for easy localisation, this
circle is not used in the lensing map.
The top-left panel (A1) shows the caustics for a redshift smaller than
the $z_{s}$ with a circular source lying outside to both caustics.  
In the lens plane (top-right panel (A3)), we observe a single
distorted image.  
As the source redshift is set to $z_{s}$ (panel (B1)), we can see a
kink (origin of two extra cusps) in the tangential caustic near the
source position.
In panel (B1), the centre of the source in source plane lies on this kink. 
In the corresponding lens plane (B2) at swallowtail singularity three
vectors: tangent to the $A_{3}$-line, tangent to the eigenvalue
contour and the local eigenvector are parallel to each other.
As we can see in panel (B2), an $A_{3}$ line coming in 
from outside the critical curve just touches
the critical curve at the $A_{4}$ point and falls back to the outer 
region. 
The corresponding image formation (B3) shows formation of a tangential
arc made of four images.
The magnification factor ($|\mu\left(r\right)|$) around a swallowtail
singularity is proportional to $r^{-3/4}$, where $r$ is the distance
from the singular point.
Whereas in the case of fold (cusp), the magnification factor
proportional to the $r^{-1/2}$ ($r^{-2/3}$).
Hence, the slope of the magnification factor around the swallowtail
singularities is steeper than fold and cusp \citep{Arnold 1982}. 

As we further increase the source redshift (C1) newly formed cusps in
source plane move away from each other and the corresponding arc in
lens plane (C3) become more stretched. 
One can see that the arc in the lens plane is made of four images, two
of them have positive parity and two of them have negative parity.
Due to the finite size of the source, the images shown here are
merging into one another.
And the image on the upper left corner has positive parity. 
Eventually, the gradual increment in the source redshift changes
the arc into four individual images.
Formation of such giant arcs around swallowtail singularities has been
already encountered in investigations of strong lens systems, see,
e.g., \cite{Saha 1998,Suyu 2010}. 

\subsection{Umbilics}
\label{ssec:Umbilics}

For a given lens model, the presence of umbilics in the corresponding
singularity map denote the points with zero shear
$\left(\gamma\right)$ in lens plane.  
At these points, both of the eigenvalues of the deformation tensor are
equal to each other ($\alpha=\beta$).
The dependence on both eigenvalues simultaneously separates these
singularities from the $A_{3}$-lines and swallowtail singularities,
which have dependency on one eigenvalue in their definitions.  
The equality of both eigenvalues implies that at umbilics,
eigenvectors of the deformation tensor are degenerate.  
As a result, any vector at these points can behave as an eigenvector.  
We can always choose the eigenvector in such a way that $A_{3}$-line
condition is always satisfied (for a quantitative analysis see
\citet{Hidding 2014}).  
At these points $A_{3}$-lines corresponding to different eigenvalues
meet with each other.  
There are two types of umbilics present in gravitational lens mapping:
elliptic and hyperbolic umbilics.  
This division of the umbilics depends on the sign of the quantity $s_{D}$,
\begin{equation}
\begin{split}
s_{D} \equiv t_{111}^2t_{222}^2-3t_{112}^2t_{122}^2-
6t_{111}t_{112}t_{122}t_{222} \\
+4t_{111}t_{122}^3+4t_{222}t_{112}^3  \,\,  ,
\end{split}
\label{eq:(Hessian)}
\end{equation}
where $t_{ijk}=\frac{\partial\psi_{ij}}{\partial x_{k}}$. 
If $s_{D}$ is positive, the singularity is called hyperbolic umbilic
and if it is negative then the singularity is an elliptic umbilic.  
At umbilics, the number of cusps in the source plane remains unchanged
but there is an exchange of one or three cusps between tangential and
radial caustics depending on the type of the umbilic.
In case of a hyperbolic umbilic, one cusp is exchanged between the
tangential and the radial caustic: in the image plane an $A_{3}$-line
corresponding to each of the two eigenvalues meet at this point.
Whereas three $A_{3}$-lines of each of the two eigenvalues meet at the
elliptic umbilic in the image plane, and three cusps are exchanged
between the tangential and the radial caustic in the source plane.

In order to discuss the evolution of the caustics and critical curves
near a hyperbolic umbilic (because of the simplicity of its
singularity map) we use a one-component elliptical lens. 
The evolution of caustics and critical curves near a hyperbolic
umbilic is shown in figure~\ref{fig:purse_figure}.  
The $A_{3}$-lines in the singularity map (middle column) are
denoted by red and dark green lines for two different eigenvalues.  
The positions of hyperbolic umbilic in lens plane is denoted by blue
points, at which two $A_{3}$-lines (one for $\alpha$ and one for
$\beta$ eigenvalue) meet with each other. 
For a redshift smaller than the redshift at which hyperbolic umbilic
becomes critical, $z_{u}$ both (radial and tangential) caustics in
source plane each have two cusps (A1). 
As we increase the source redshift to $z_{u}$, there is an exchange
of cusp from radial caustic to tangential caustic (panel B1) (For the
single component elliptical lens model, because of the symmetry of the
lens model, both of the hyperbolic umbilics become critical at the
same redshift.  The symmetry is broken in presence of a second
component or shear.). 
The corresponding image formation (panel B3) shows a single
demagnified image with positive parity and a loop formed by four
images, two of them with positive parity and two of them with negative
parity.  
As we increase the source redshift further, source plane has a diamond
shaped tangential caustic and a smooth radial caustic (panel C1) and
in lens plane the highly magnified ring shaped image changes into four
individual less magnified images (panel C3).
The ring and the cross (for higher redshifts) is not centered at the
lens centre but is off centre.
We have studied the location of the ring by varying the mass profile
of the lens and we find that the ring is located where the projected
surface density begins to drop sharply.
The magnification factor $|\mu|$ falls as $r^{-1}$ around both
umbilics as one moves away from the singular point.  
Thus magnification factor falls much more rapidly around umbilics than
other singularities.

The characteristic image formation for the hyperbolic umbilic is a
ring or a cross like system centered away from the lens centre.
The curvature of the image formation is much stronger and the radius
of curvature is much smaller than the size of the lens. 
So far only one lens system (Abell 1703) with image formation near a
hyperbolic umbilic has been seen \citep{Orban 2009}. 

Unlike the hyperbolic umbilic, at an elliptic umbilic, there are six
$A_{3}$-lines (three each for each of the two eigenvalues of the
deformation tensor) meet with each other.   
For an illustration, formation of an elliptic umbilic in case of a 
two-component elliptical lens model is shown in the
figure~(\ref{fig:pyramid_figure}). 
We find that often, two of three $A_{3}$-lines of one or both
eigenvalues form a small closed loop.
This can be seen in examples shown in figure~\ref{fig:map_figure}.
In panel (A1), we only see tangential caustics, and the source lies
inside the triangular shaped caustic. 
Panel (A3) shows the characteristic image formation (seven images in a
shape of Y) near an elliptic umbilic.  
The central image has positive parity. The next three images from the
central image have negative parity.
And the three outer images again have positive parity. 
As we increase the source redshift, the size of the triangular shaped
caustic decreases and at the same time, it moves away from the source
position.  
At a redshift $z_{p}$, where the elliptic umbilic become critical it
become a points caustic (panel B1) and the source lies close to this
point caustic.  
The corresponding images still form a Y-shaped structure in lens plane
but with only five images.
As we further increase the source redshift, the point caustic turns
into a triangular shaped radial caustic (panel C3).  
Which implies that at the elliptic umbilic there is an exchange of three
cusps between tangential and radial caustic. 
In panel (C1), we moved the source inside the triangular caustic, to
see whether it still gives a Y-shaped image formation.
We get a different kind of image formation with central image rotated
by $\pi/2$.

Figure~\ref{fig:pyramid_figure}, shows the singularity map close to
the elliptic umbilic (shown by blue point).  
Swallowtail singularities are shown as violet points. 
The complete singularity map for figure~\ref{fig:pyramid_figure} is
given in figure~\ref{fig:map_figure} (panel A5).  
   
The characteristic image form is six images radiating out from the
singularity.
The singularity need not coincide with the centre of the lens,
The images do not have any tangential deformation. 

\section{Algorithm}
\label{sec:Algorithm}

We briefly discuss the algorithm used to find out the singularities
for a given lens model, we focus on singularities that are discussed
in above section.
We set up a uniform grid in the lens plane for calculations of
physical quantities in order to locate the singularities.
The grid-size depends on the resolution required for the lens model,
in general we require adequate resolution as we are dealing with
non-linear combinations of second derivatives of the lensing
potential, even the smallest features should be well resolved on the
grid.
We use finite difference methods to compute derivatives on the grid.
To calculate the position of the umbilics in the lens plane, we use
the fact that at umbilics, both components of the shear tensor vanish,
identically.
Our approach closely follows that of \citet{Hidding 2014}.
The flow of the code is as follows:  
\begin{itemize}
\item
  INPUT (Lens Potential)
\item
  CALCULATE first and second derivatives of the potential
\item
  CALCULATE eigenvalue and eigenvectors of the deformation tensor
\item
  CALCULATE gradient of eigenvalue
\item
  CALUCLATE extrema
  \begin{itemize}
  \item
    CALCULATE local maxima
  \item
    CALCULATE local minima
  \end{itemize}
\item
  Identify points on $A_{3}$-lines using Equation~\ref{eq:(A3 lines)}.
\item
  Identify $A_{4}$-points using the algorithm given in \S{3.2}
\item
  Identify $D_{4}$-points using the algorithm given in \S{3.3}
\end{itemize}
The potential of the given lens model is the input in this algorithm.
The potential can be computed from a mass model, or be provided
directly.
Given the potential, the deformation tensor is computed at each point
followed by calculation of its eigenvalues and corresponding
eigenvectors.  
This information along with gradient of eigenvalues is sufficient to
identify points on the $A_{3}$-lines (eq.~(\ref{eq:(A3 lines)})).
Note that points on the $A_{3}$-lines can be identified on the mesh
and need not coincide with the grid points.
The points need to be ordered to construct curves: this is required
for identifying $A_{4}$ points as we need to locate maxima of
eigenvalue along the $A_{3}$-lines. 
By rejecting such extrema that are also local maxima, we are left with
$A_{4}$ points. 

We find that it is simpler to identify umbilics by realizing that the
diagonal components of the deformation tensor are equal, and the
off-diagonal component of the deformation tensor (shear) vanishes.
Each of these conditions specifies curves in the image plane, and
intersections of these curves give us umbilics. 
We can classify the type of umbilics by counting the number of
$A_{3}$-lines that converge at this point.  

\section{Results}
\label{sec:Results}

We apply our algorithm by applying it to a single component lens
where the potential can be expressed in a closed form and the image
structure has been studied in detail.
Then we apply it to multi-component lenses and study a variety of
configurations.
Lastly we apply it to one real lens model. 
As mentioned above in  subsection~\ref{ssec:A3 lines}, this technique
does not work in case of isolated spherically symmetric lenses because
of the absence of cusp formation.   
We study elliptical lenses with one and two components.
In the case of one-component elliptical lens, there are two $A_{3}$-lines and
two hyperbolic umbilics.  
Whereas for two-component lens, the location of $A_{3}$-lines and other
singularities in lens plane depend on the lens configuration.

We also discuss the behaviour of swallowtail and umbilics in a lens
model under external perturbations in case of one and two component
elliptical lenses.   
This gives us the estimate about the amount of external shear under
which a singularity does not vanish from the lens plane and hence gives
us an idea about how robust these point singularities really are.     
In the following subsections, we will discuss the singularity maps for
elliptical lenses and Abell 697 in detail.  
After that we study the stability of these different singularities in
lens mapping.   

\begin{figure*}
\centering
\begin{tikzpicture}
\node[anchor=south west,inner sep=0] (image) at (0,0) {\includegraphics[width=\textwidth,height=5.2cm,width=5.2cm]{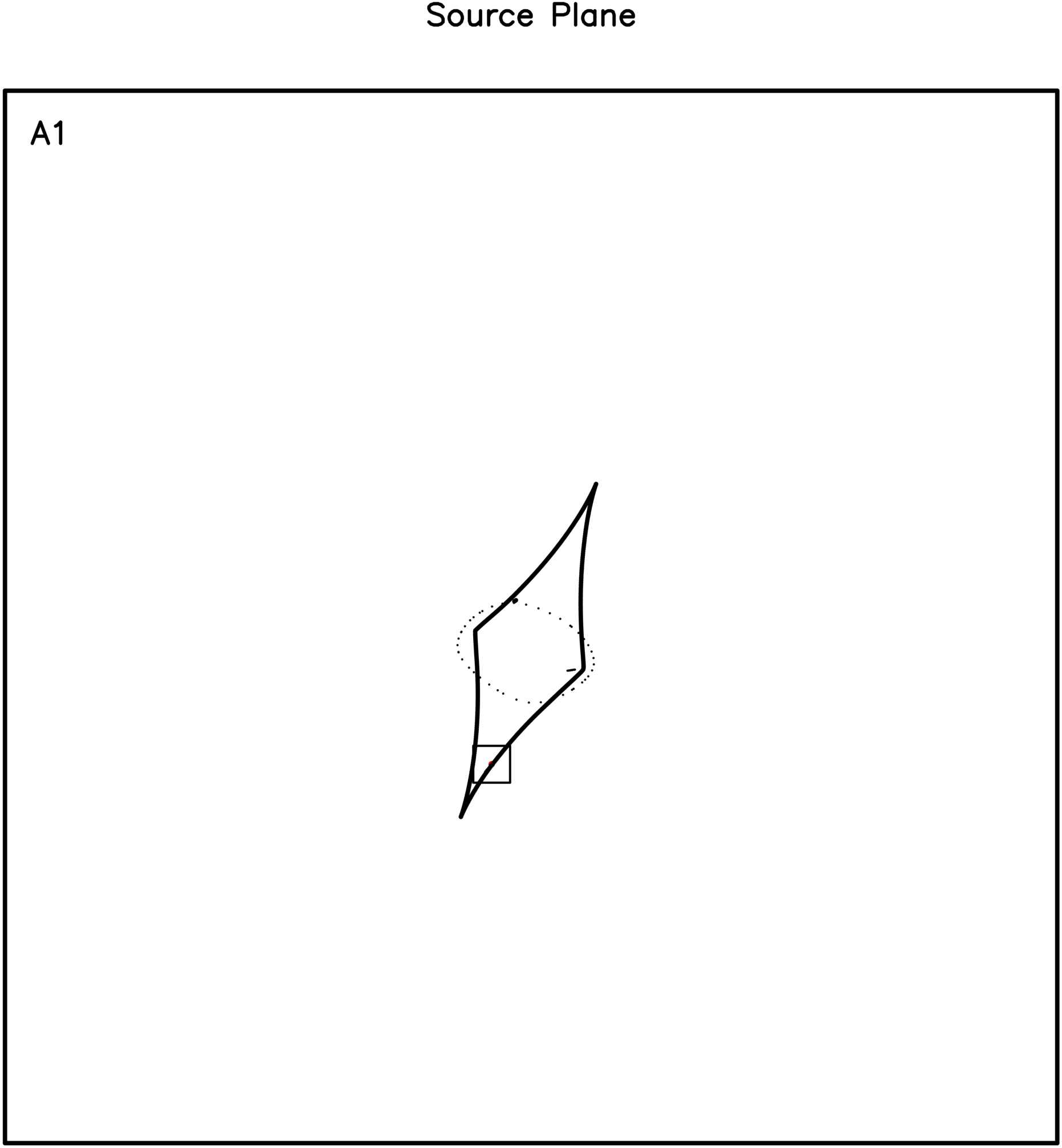}};
\begin{scope}[x={(image.south east)},y={(image.north west)}]
\node[anchor=south west,inner sep=0] (image) at (0.7,0.12) {\includegraphics[width=\textwidth,height=0.8cm,width=0.8cm]{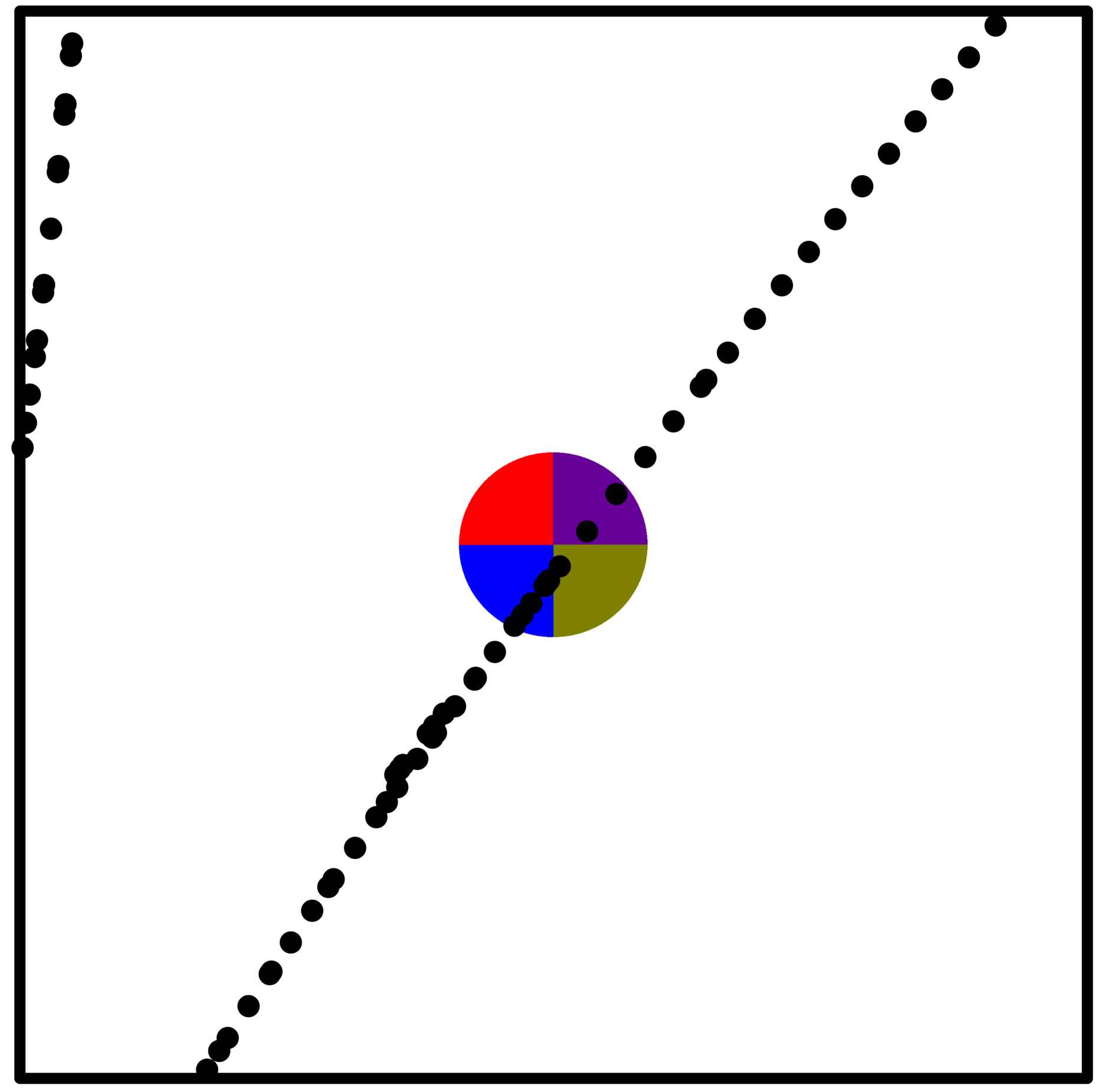}};
\end{scope}
\node[anchor=south west,inner sep=0] (image) at (5.5,0) {\includegraphics[width=\textwidth,height=5.2cm,width=5.2cm]{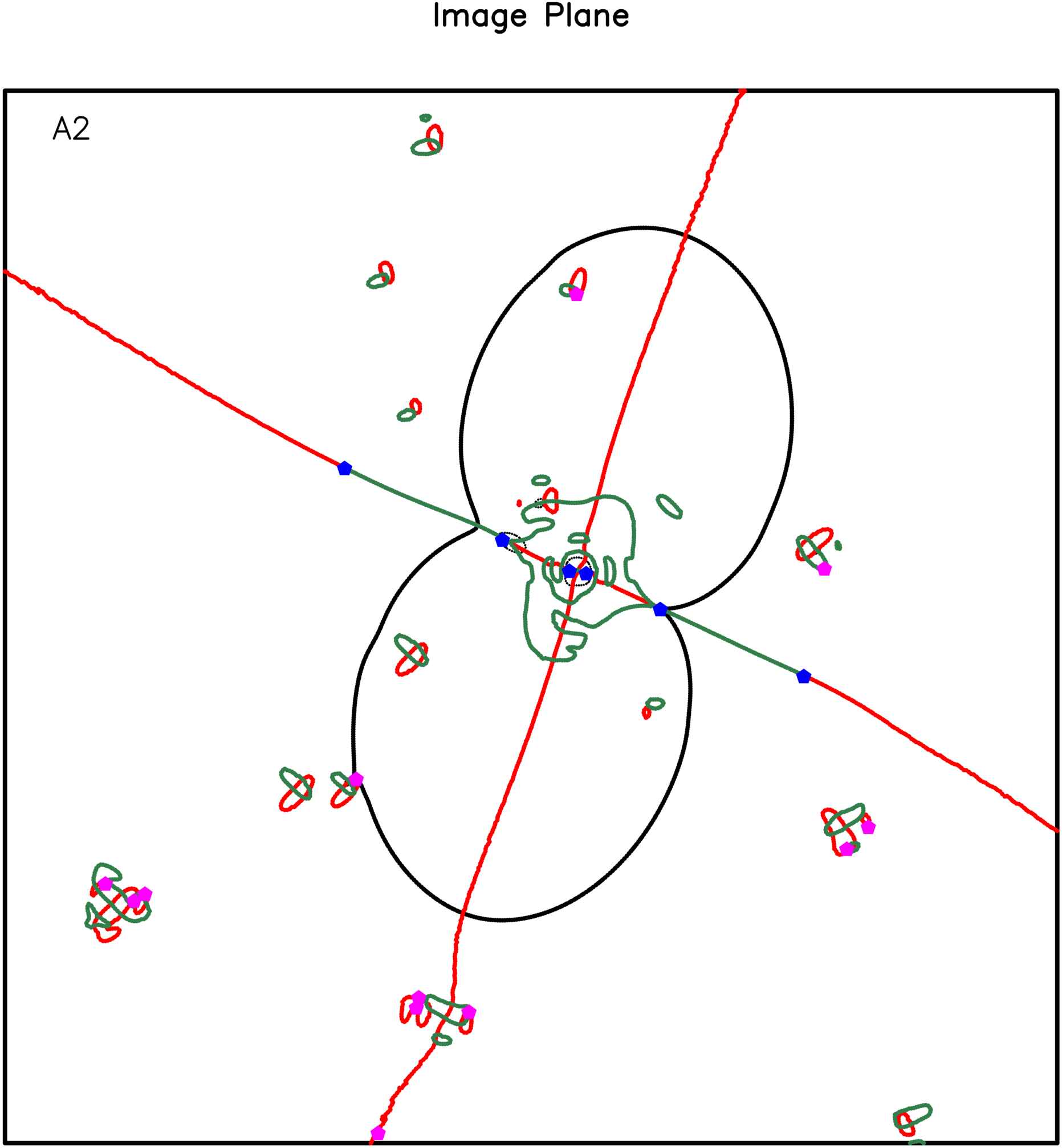}};
\node[anchor=south west,inner sep=0] (image) at (11.0,0) {\includegraphics[width=\textwidth,height=5.2cm,width=5.2cm]{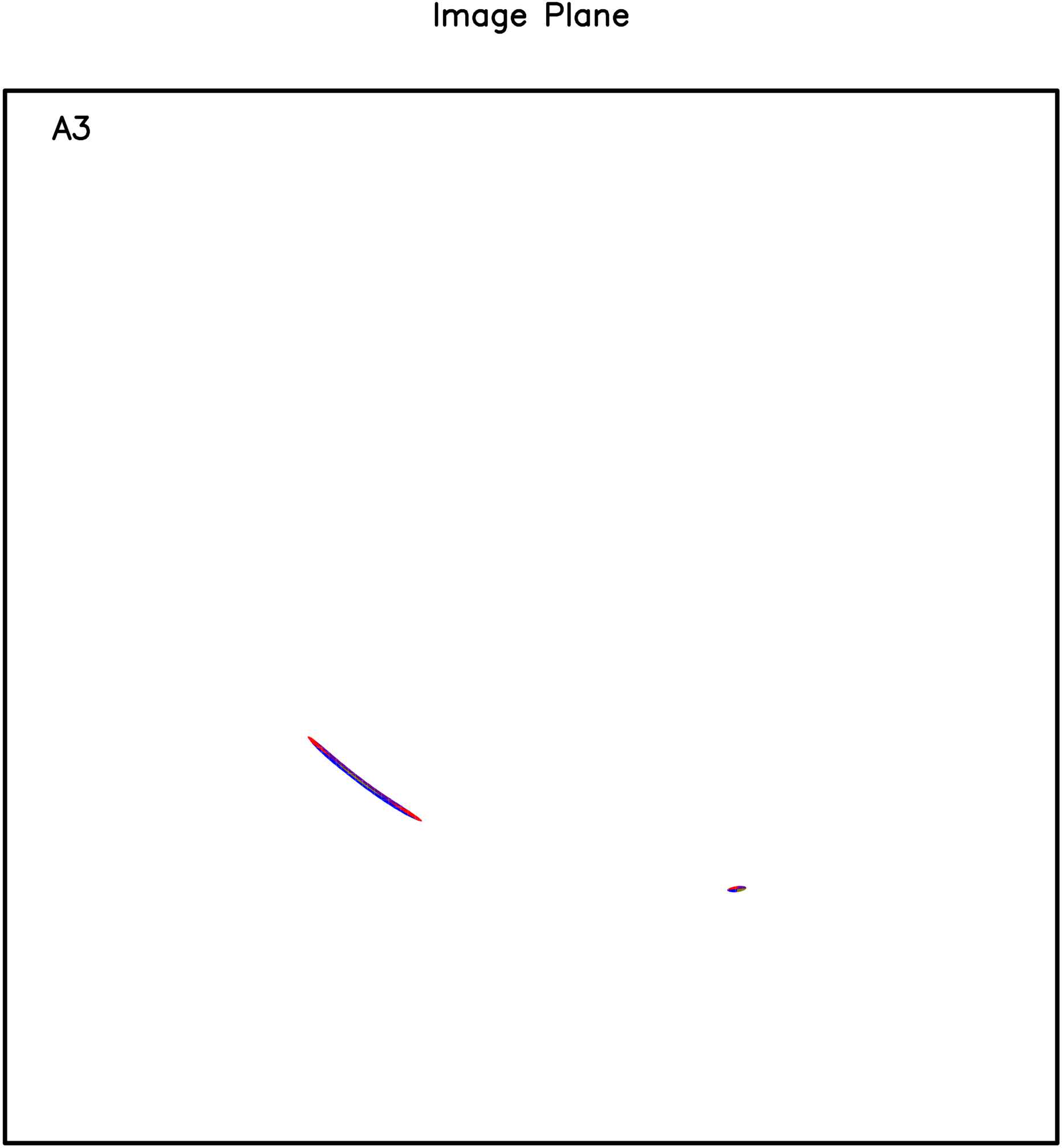}};
\end{tikzpicture}
\end{figure*}
\begin{figure*}
\centering
\begin{tikzpicture}
\node[anchor=south west,inner sep=0] (image) at (0,0) {\includegraphics[width=\textwidth,height=5.2cm,width=5.2cm]{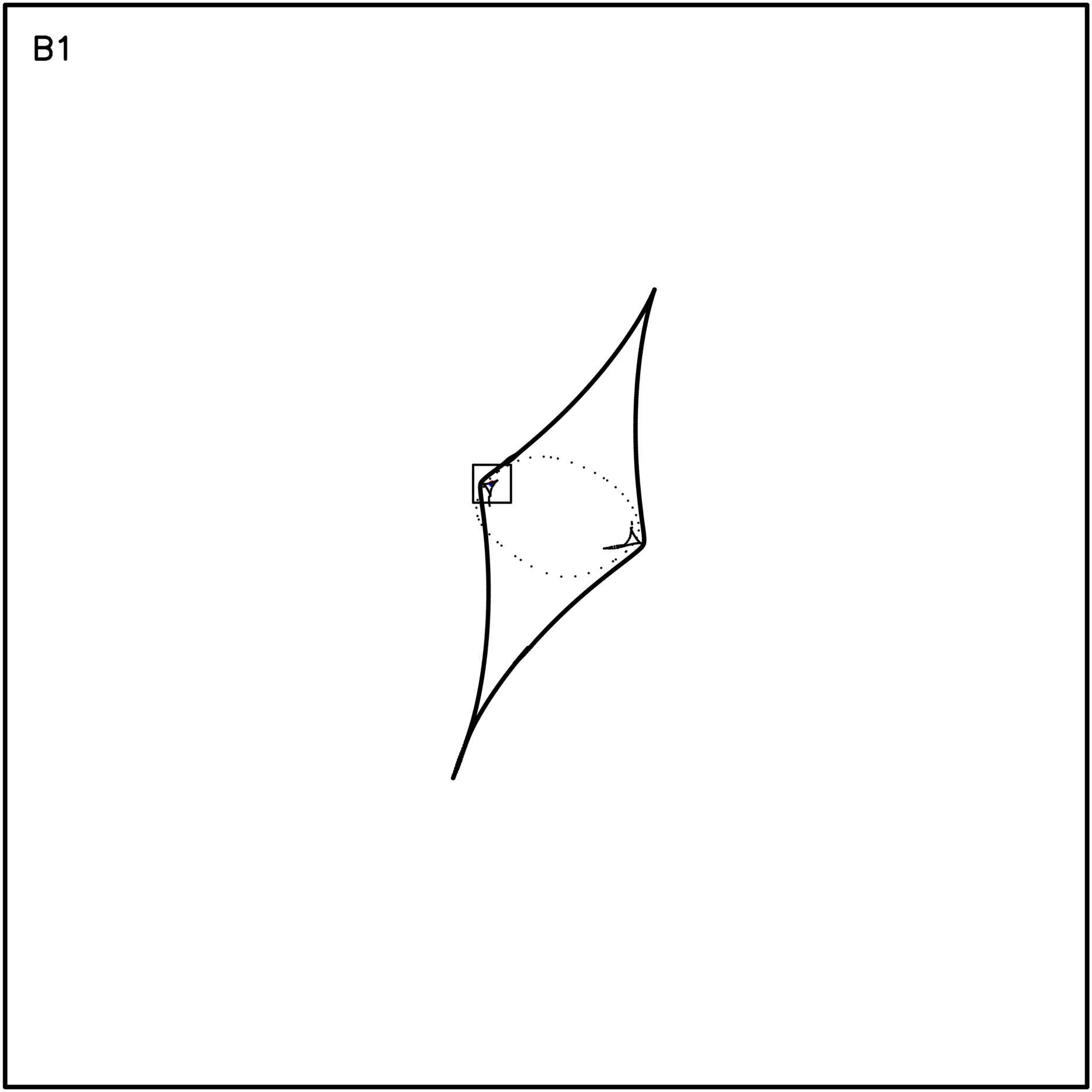}};
\begin{scope}[x={(image.south east)},y={(image.north west)}]
\node[anchor=south west,inner sep=0] (image) at (0.7,0.12) {\includegraphics[width=\textwidth,height=0.8cm,width=0.8cm]{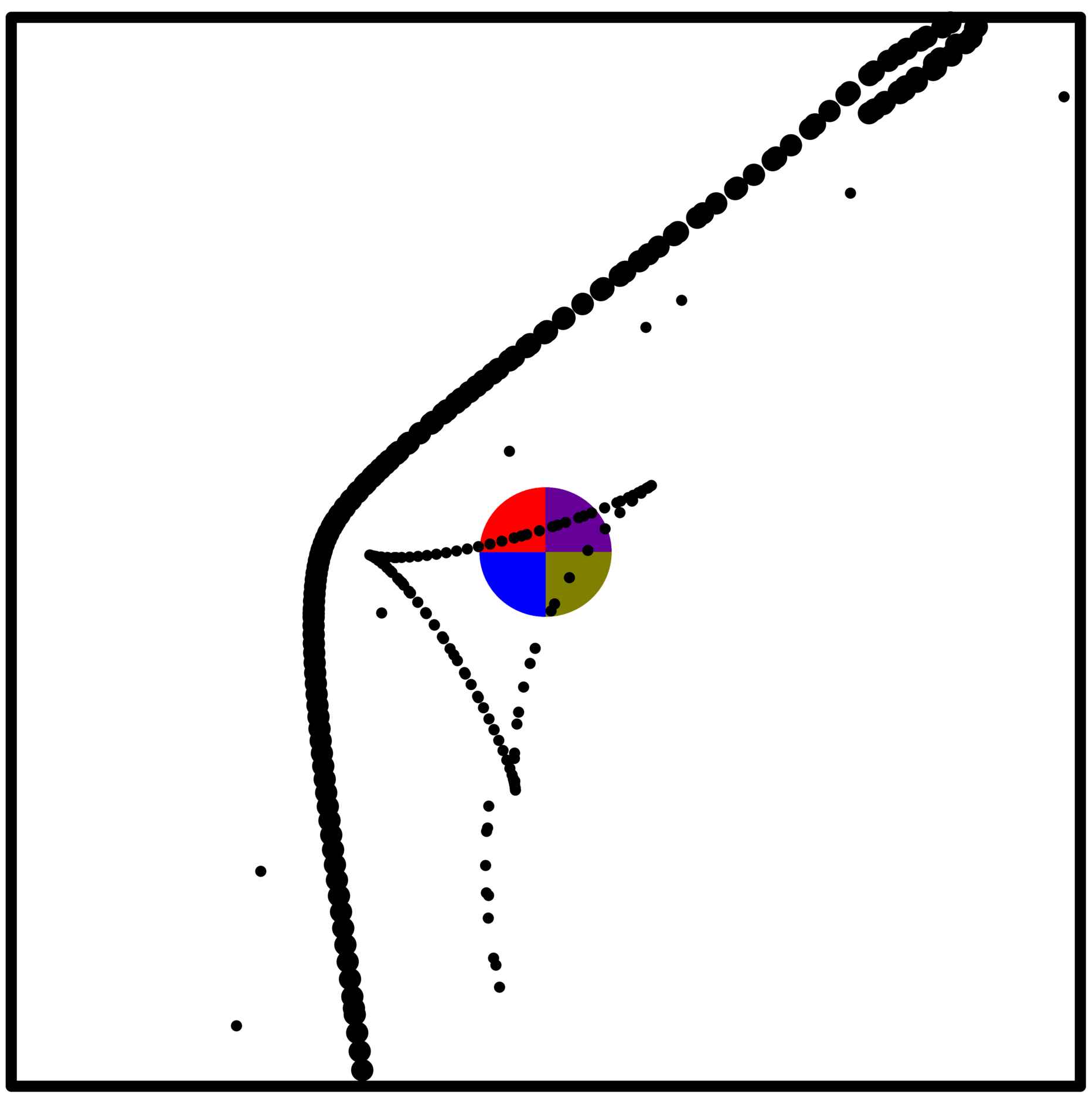}};
\end{scope}
\node[anchor=south west,inner sep=0] (image) at (5.5,0) {\includegraphics[width=\textwidth,height=5.2cm,width=5.2cm]{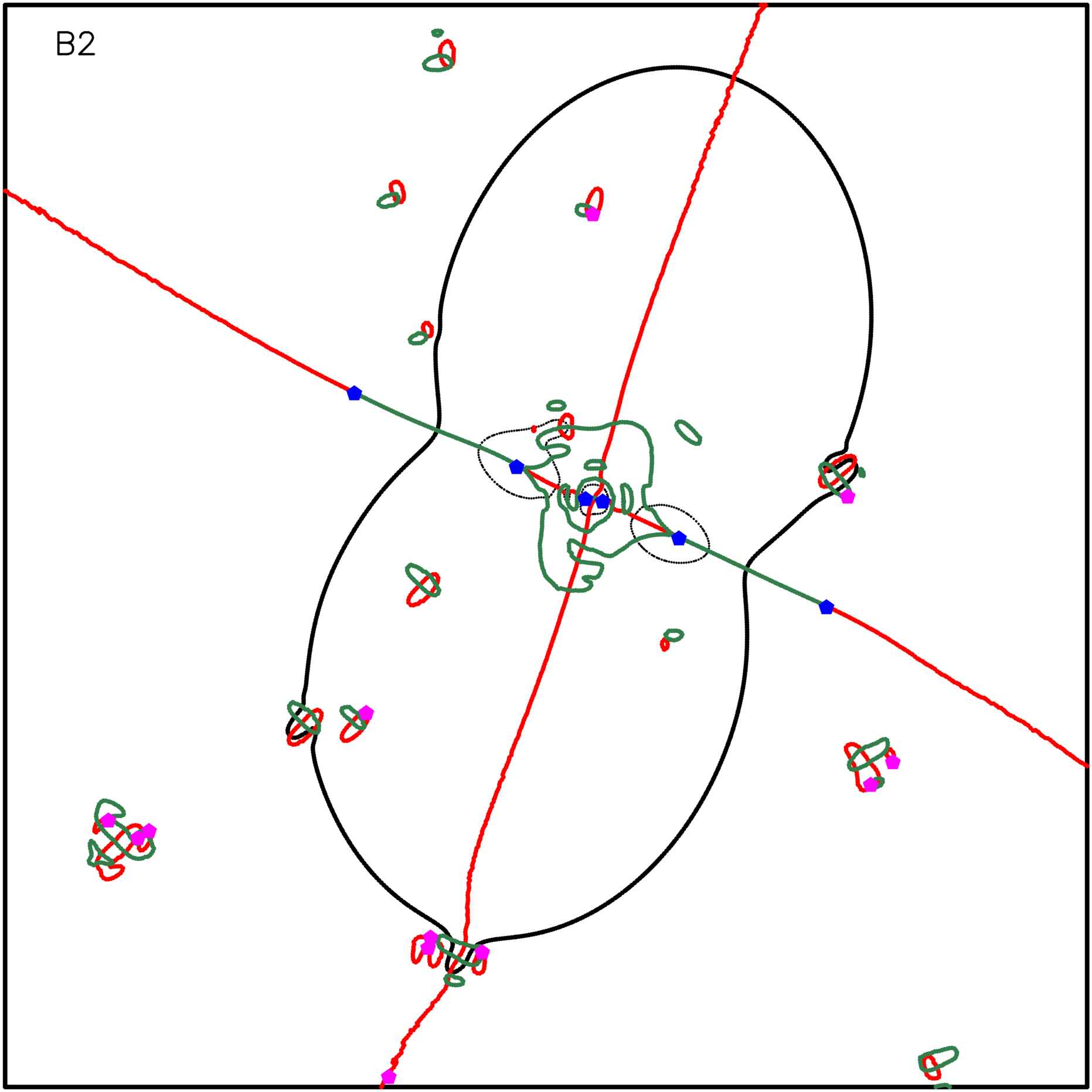}};
\node[anchor=south west,inner sep=0] (image) at (11.0,0) {\includegraphics[width=\textwidth,height=5.2cm,width=5.2cm]{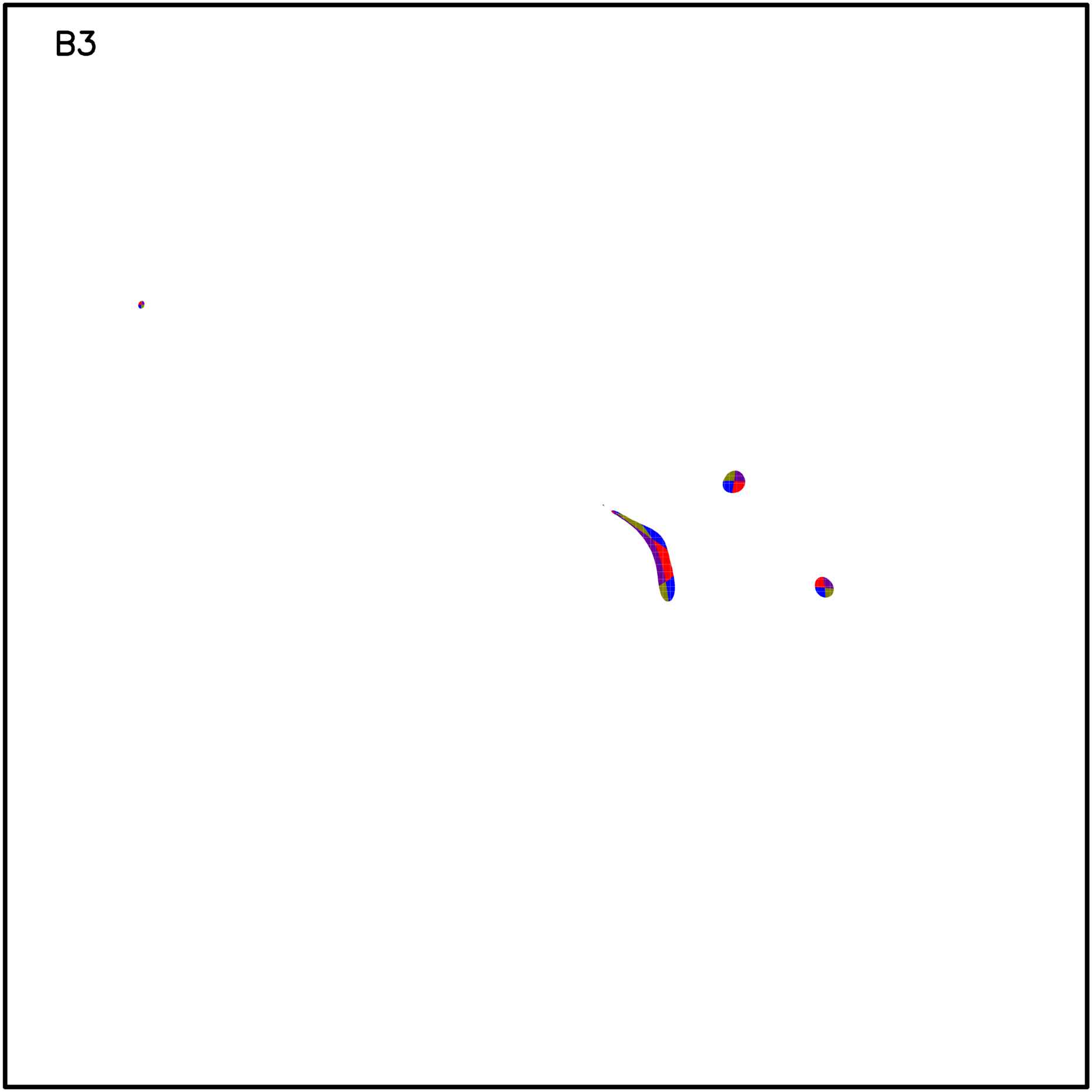}};
\end{tikzpicture}
\end{figure*}
\begin{figure*}
\centering
\begin{tikzpicture}
\node[anchor=south west,inner sep=0] (image) at (0,0) {\includegraphics[width=\textwidth,height=5.2cm,width=5.2cm]{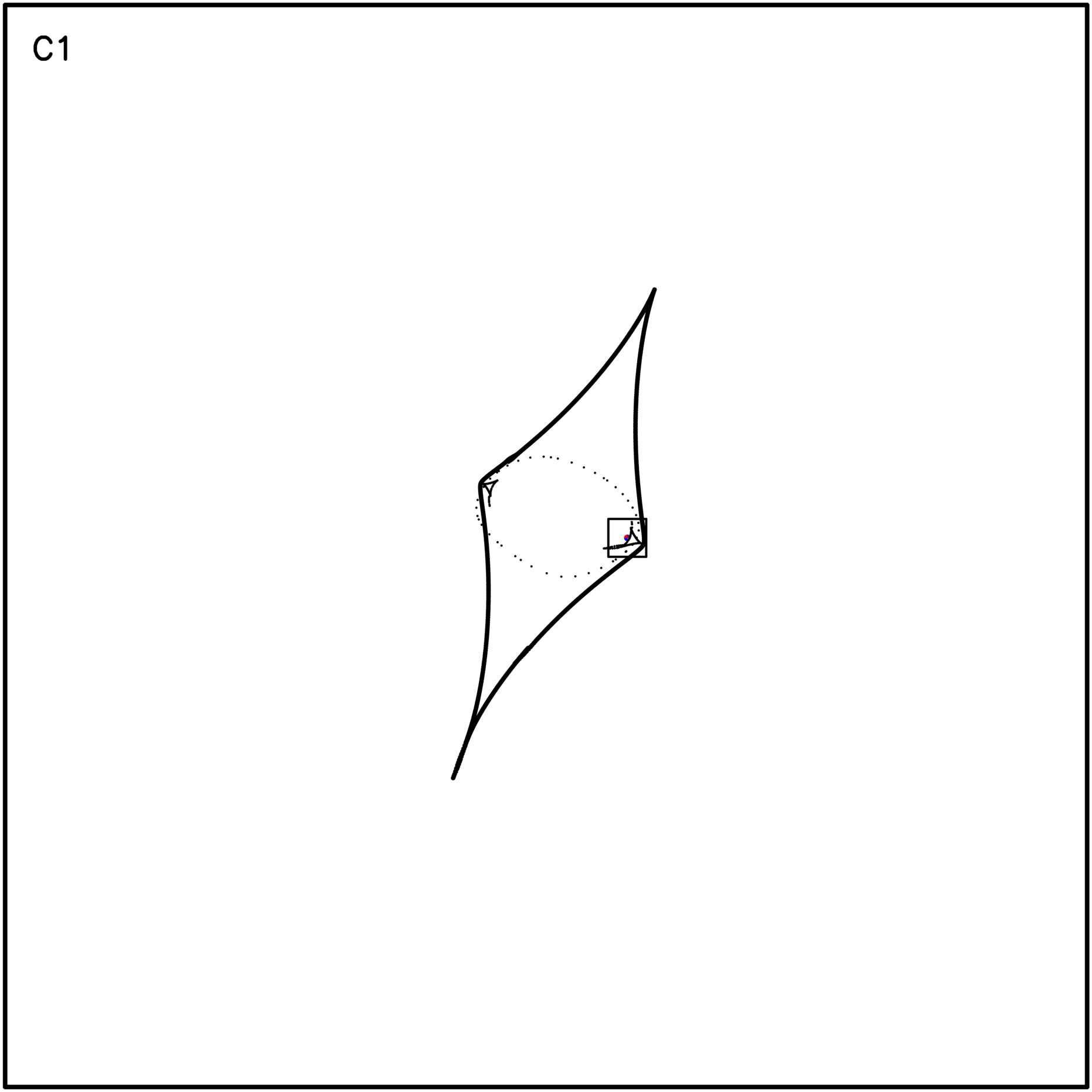}};
\begin{scope}[x={(image.south east)},y={(image.north west)}]
\node[anchor=south west,inner sep=0] (image) at (0.7,0.12) {\includegraphics[width=\textwidth,height=0.8cm,width=0.8cm]{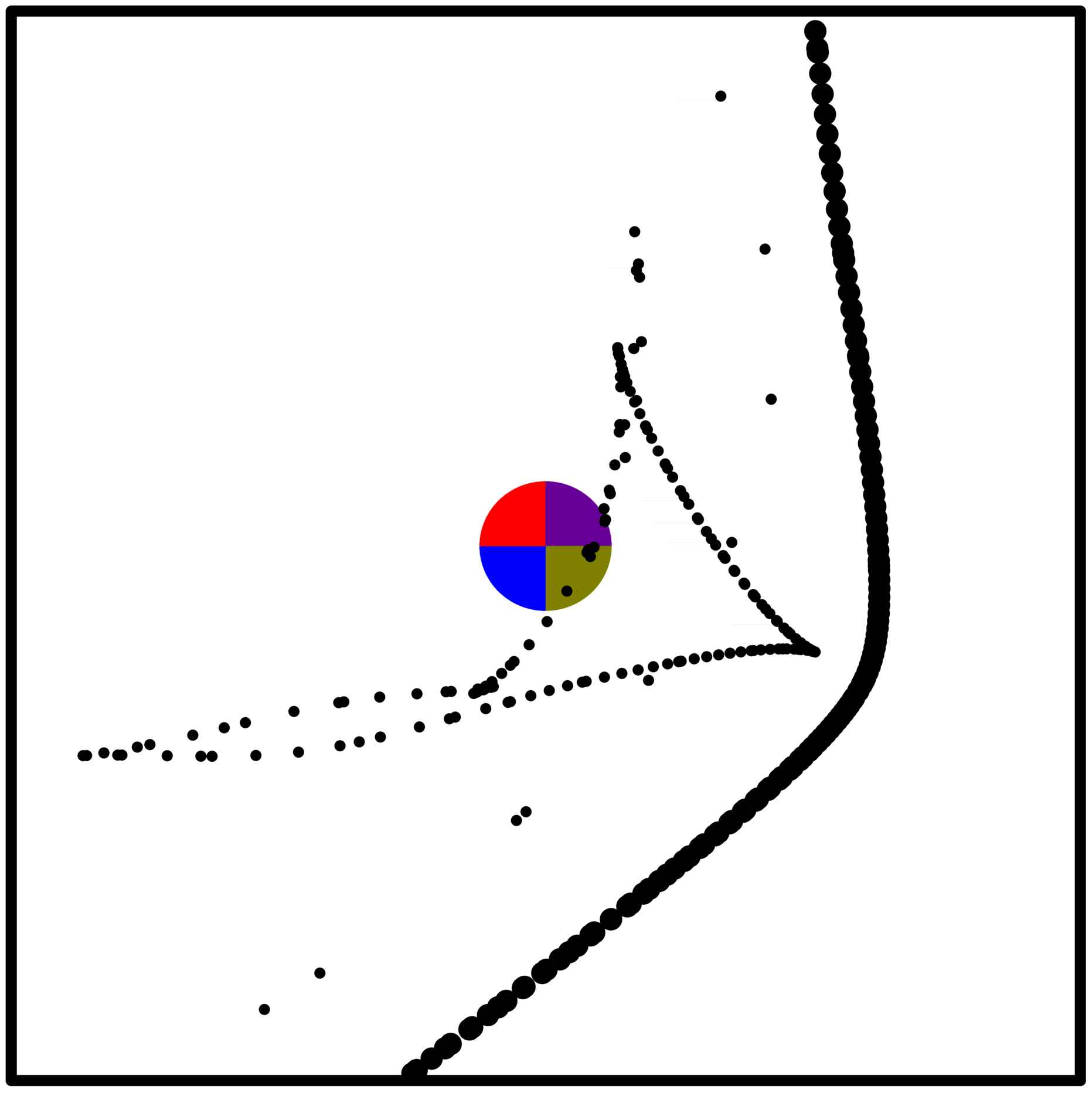}};
\end{scope}
\node[anchor=south west,inner sep=0] (image) at (5.5,0) {\includegraphics[width=\textwidth,height=5.2cm,width=5.2cm]{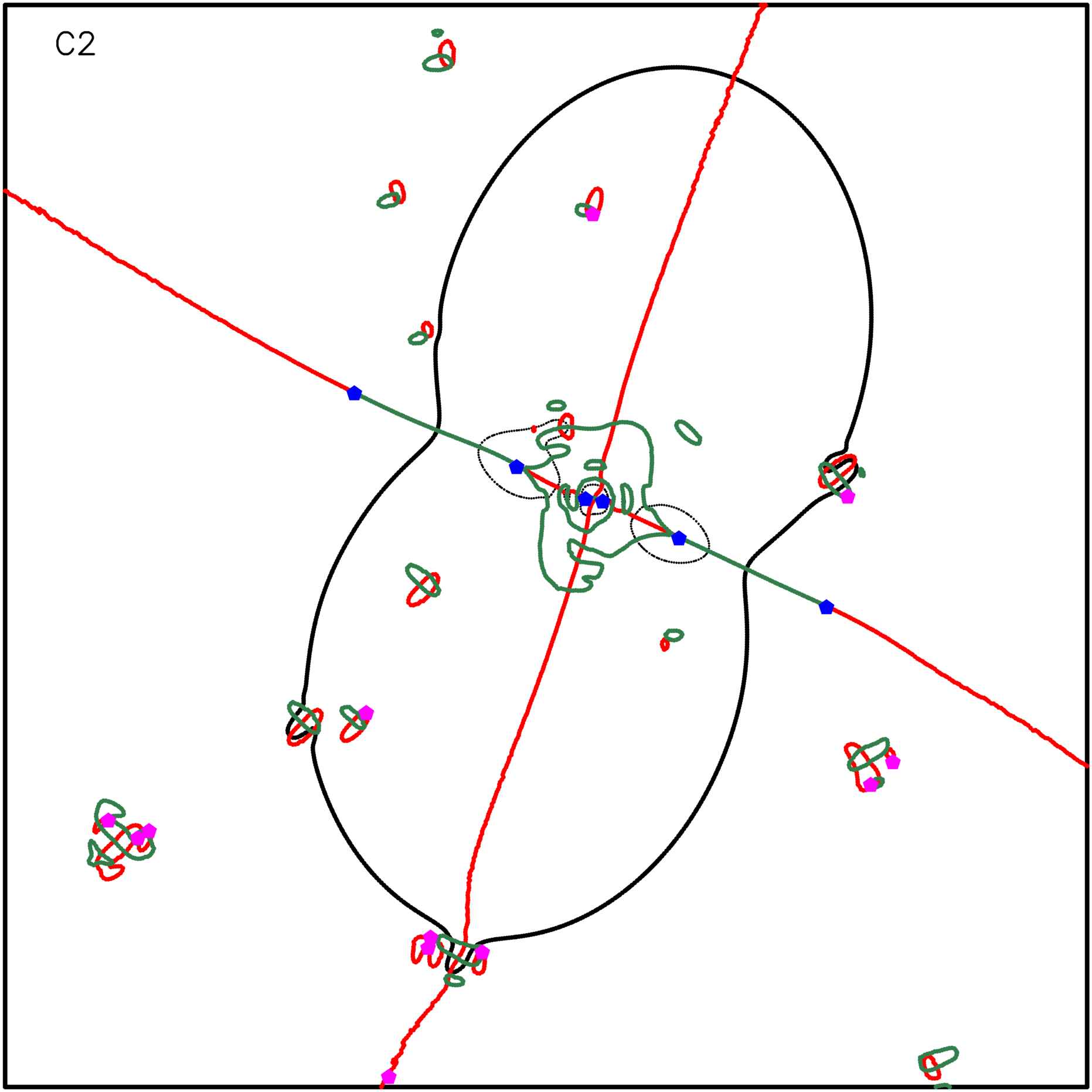}};
\node[anchor=south west,inner sep=0] (image) at (11.0,0) {\includegraphics[width=\textwidth,height=5.2cm,width=5.2cm]{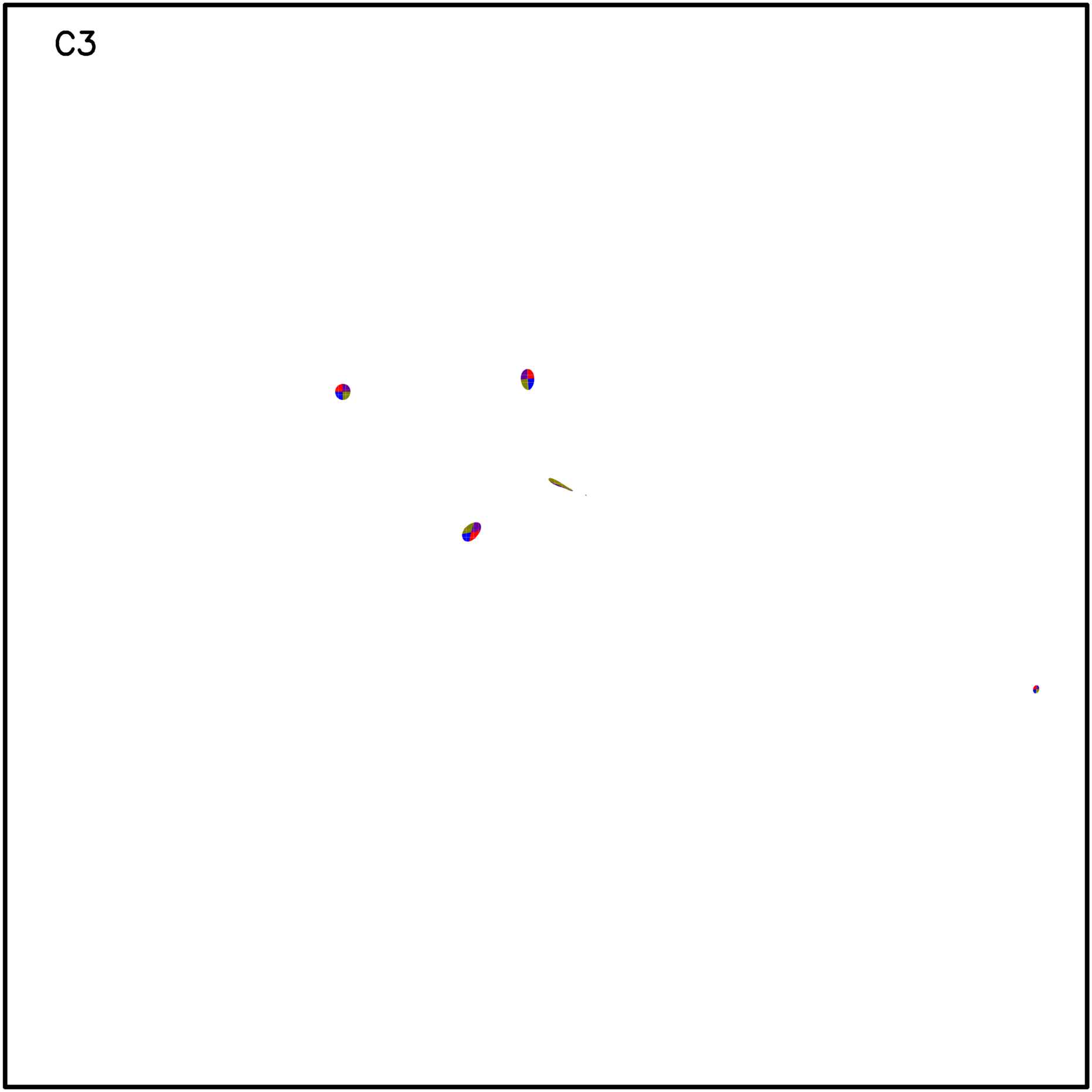}};
\end{tikzpicture}
\end{figure*}
\begin{figure*}
\centering
\begin{tikzpicture}
\node[anchor=south west,inner sep=0] (image) at (0,0) {\includegraphics[width=\textwidth,height=5.2cm,width=5.2cm]{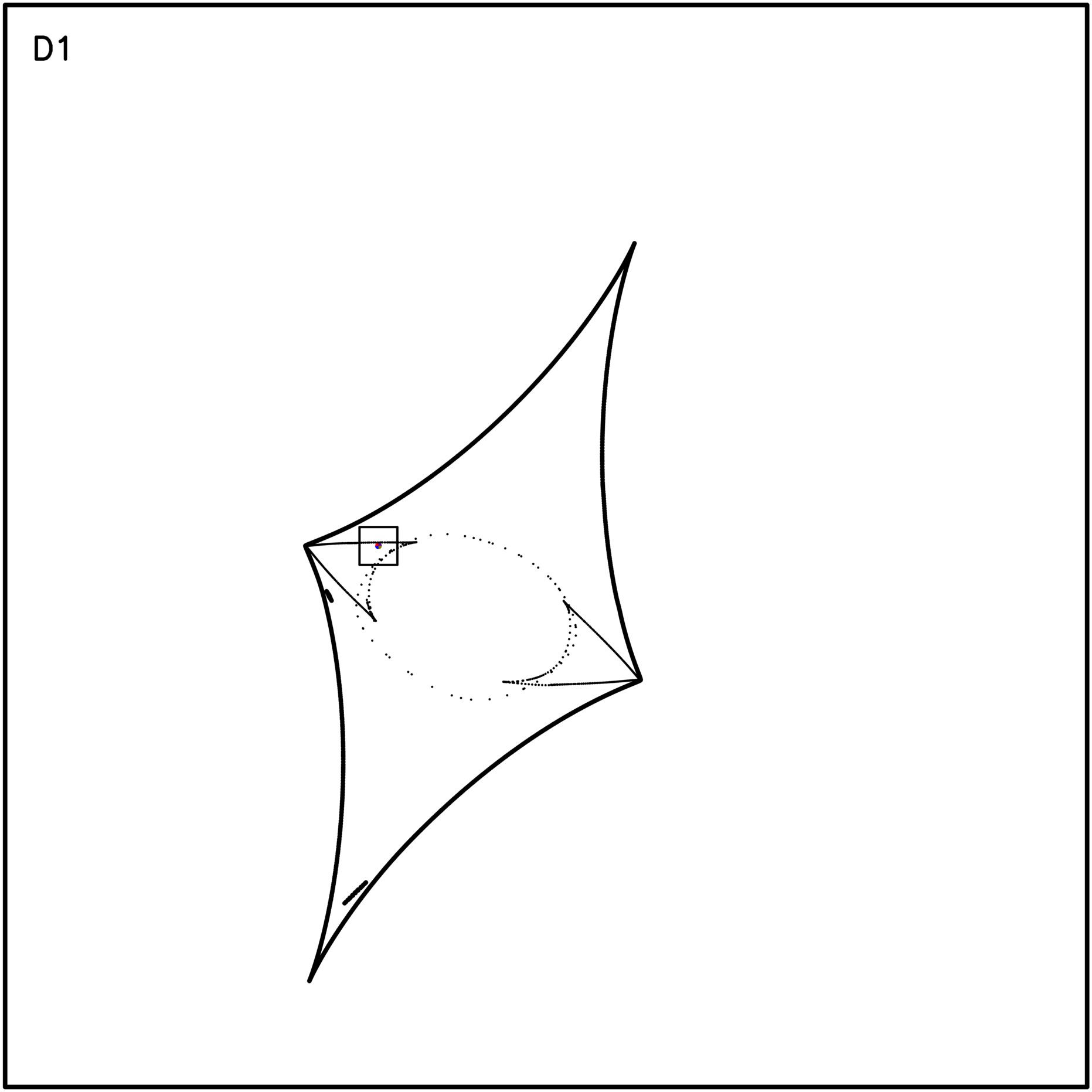}};
\begin{scope}[x={(image.south east)},y={(image.north west)}]
\node[anchor=south west,inner sep=0] (image) at (0.7,0.12) {\includegraphics[width=\textwidth,height=0.8cm,width=0.8cm]{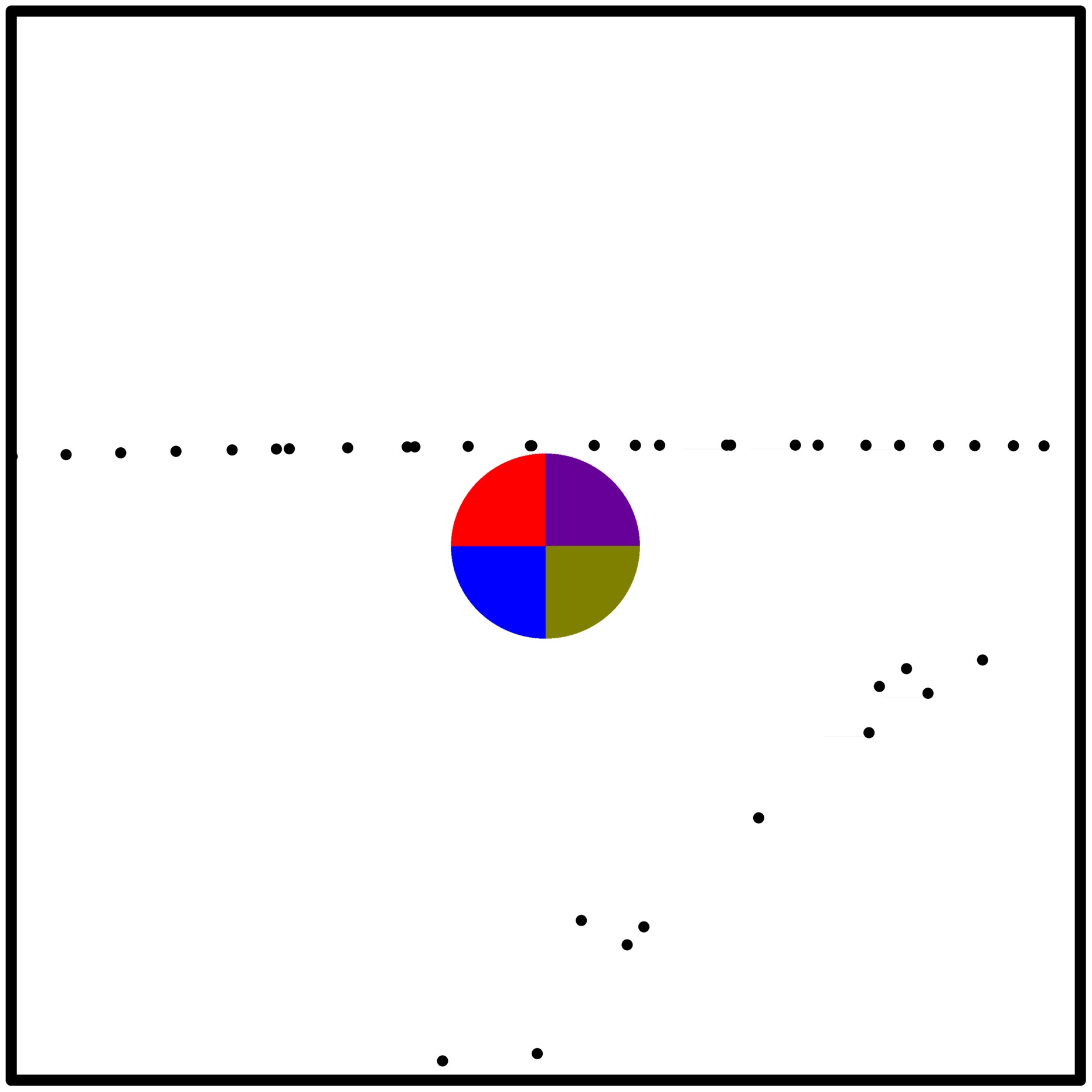}};
\end{scope}
\node[anchor=south west,inner sep=0] (image) at (5.5,0) {\includegraphics[width=\textwidth,height=5.2cm,width=5.2cm]{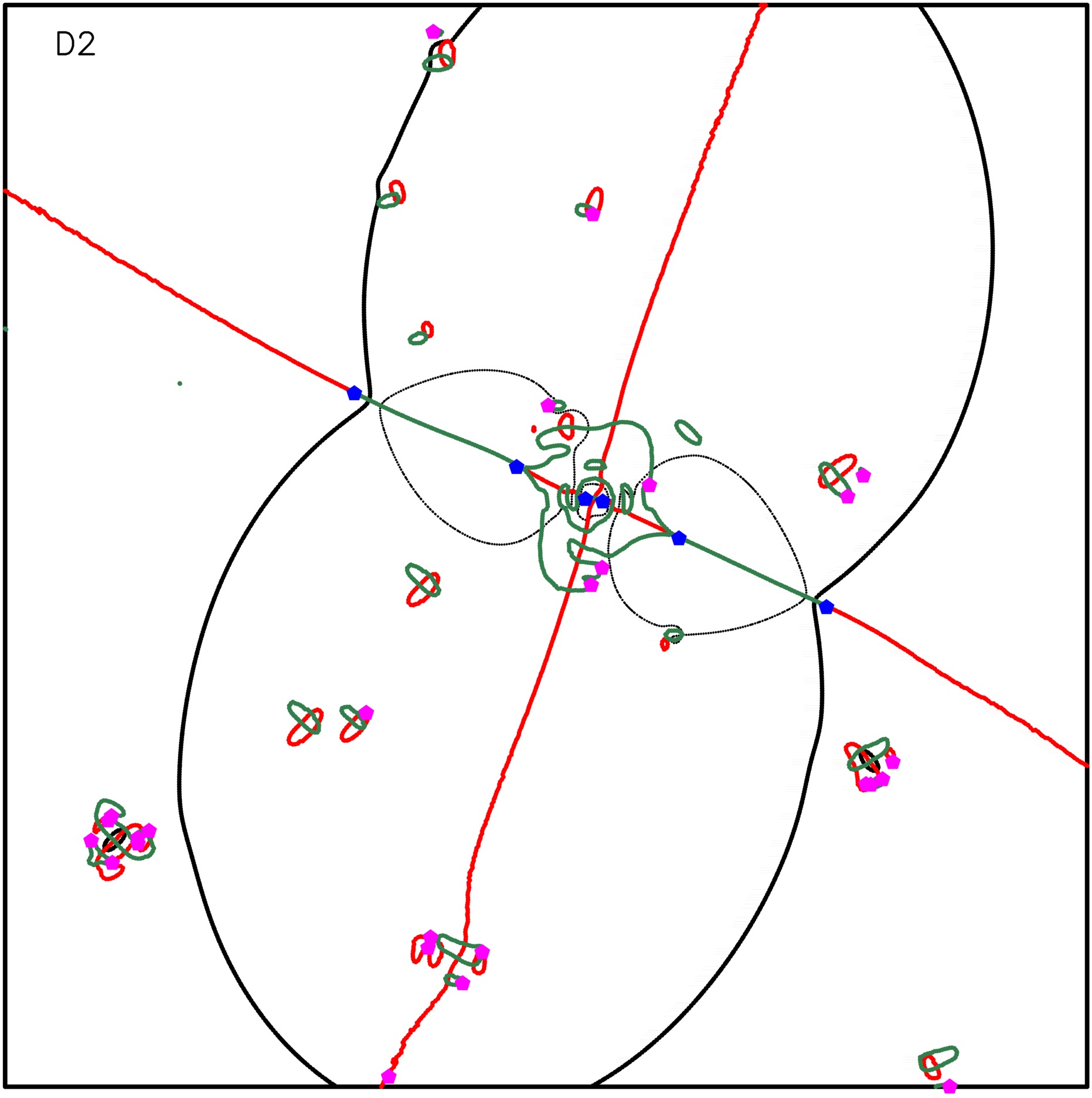}};
\node[anchor=south west,inner sep=0] (image) at (11.0,0) {\includegraphics[width=\textwidth,height=5.2cm,width=5.2cm]{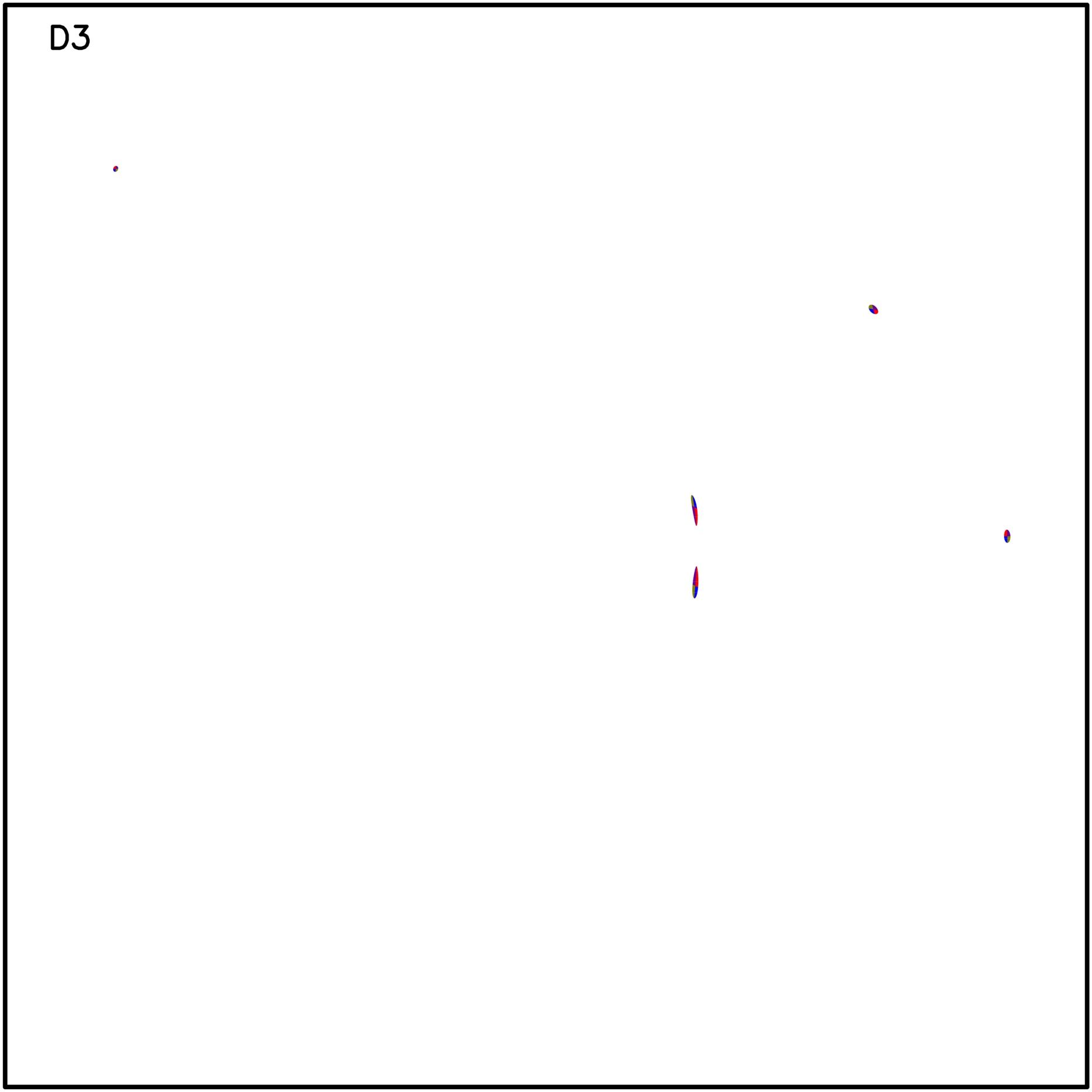}};
\end{tikzpicture}
\caption{Image formation and singularity map for Abell 697:  The
    left panel shows caustics, the 
    middle column shows critical curves in the image plane and the
    right column shows the lensed images.  The top panel represent
    image formation near a swallowtail singularity at source redshift
    $z_s=0.67$. The two middle panels represent image formation for a
    source at $z_s=0.82$ and the bottom panel represent image
    formation for a source at redshift $z_s=2.0$.}
\label{fig:abell697}
\end{figure*}

\subsection{One-Component Elliptical Lens}
\label{ssec:One component}

We first consider a one component elliptical lens.
This is a good model for an isolated lens that is dynamically relaxed,
e.g., an isolated galaxy or a cluster of galaxies. 
The elliptical isothermal lens with a finite core has a potential of
the form:
\begin{equation}
\psi(x_{1},x_{2}) = \psi_{0}\sqrt{r_{0}^{2} +
  \left(1-\epsilon\right)\left(x_{1}-x_{01}\right)^{2} +
  \left(1+\epsilon\right)\left(x_{2}-x_{02}\right)^{2}}, 
\label{eq:(one component)}
\end{equation}
where $r_{0}$ is the core radius of the lens, $\epsilon$ is the
ellipticity, $\left(x_{01},x_{02}\right)$ are the coordinates of the
centre of the lens with respect to the optical axis and $\psi_{0}$
describes the strength of the lens. 

The characteristics of this lens are described in detail in
\citet{BlandfordNarayan1986} (See Figure~10).  
The singularity map in lens plane for this lens model is shown in
figure~(\ref{fig:purse_figure}) (middle column (A2,B2,C2)).  
It has two $A_{3}$-lines (represented in red and dark green) for two
eigenvalues of the deformation tensor, running along the major and the
minor axis of the lens potential.  
This lens model also has two hyperbolic umbilics along the minor axis
and their position depends on the lens parameters, primarily on the
core radius $r_0$.  
Because of the elliptical symmetry in the lens model, both umbilics
lie at the same distance from the centre of the lens. 
As a result, both umbilics become critical at the same source redshift.
If we change the core radius, this distance from the centre and
the redshift at which these hyperbolic umbilics becomes critical, also
change. 

\subsection{Two-Component Elliptical Lens}
\label{ssec:Two component}

Most realistic lenses have several components, though one of the
components may dominate over others.
In this section we consider two component lenses.
We consider one primary (dominant) and one secondary component. 
Presence of the secondary component significantly affects lensing due
to primary lens.
In order to include the effects of the secondary lens, one has to
modify the lensing potential in equation~(\ref{eq:(magnification)}).
If the secondary lens also an elliptical lens (for simplicity), then
the lens potential becomes: 
\begin{equation}
\psi(x_{1},x_{2}) = \psi_{p} + \psi_{s},
\label{eq:(two component)}
\end{equation}
with $\psi_p$ and $\psi_s$ are one component elliptical lenses
centered at different points in the image plane with different core
radii and ellipticities and the major axes of the two components can
be at an angle.  
($r_{01}\, \left(r_{02}\right)$ are the corresponding core radii for
the two potentials,
$\epsilon_{1}\, \left(\epsilon_{2}\right)$ is the ellipticity and
$\left(x_{11},x_{12}\right) \,
\left(\left(x_{21},x_{22}\right)\right)$ are the coordinates of the
centre of the primary (secondary) lens with respect to the optical
axis.)

Different sets of lens parameters give different kinds of singularity
map and image formation.  
For example, figure~(\ref{fig:swallowtail_figure}) and
figure~(\ref{fig:pyramid_figure}) represent singularity map for
two-component lens models with two different set of lens parameters.  
Figure~(\ref{fig:map_figure}) shows some other possible singularity
maps for two-component elliptical lens with a fixed primary lens and
different (randomly picked) position and orientation of the secondary
lens.  
As before, red and dark green lines represent the $A_{3}$-lines with
swallowtail and umbilic points denoted by violet and blue points,
respectively.  
One can see the dependency of unstable singularities on lens
parameters: as we change the secondary lens the position and critical
redshift for the unstable singularities also changes.  
From figure~(\ref{fig:map_figure}), we also gain some knowledge about
the sensitivity of the unstable singularities to the lens parameters. 
All panels in figure~(\ref{fig:map_figure}) have hyperbolic
umbilics and swallowtail (except first panel), whereas only three
panels show elliptic umbilics. 
We infer that elliptic umbilics are more sensitive to the lens
parameters than the swallowtail and the hyperbolic umbilic. 

\subsection{Stability}
\label{ssec:Stability}

In general, finding an isolated gravitational lens with one or two
components is highly unlikely. 
Real gravitational lenses reside in an environment made of several 
structures. 
These external local structure perturb the lensing potential, by
introducing (constant) external convergence ($\kappa_{ext}$) and shear
($\gamma_{ext}$).
As a result, the perturbed lensing potential is given by,
\begin{equation}
\psi(x_{1},x_{2}) = \psi_{p} +
\frac{\kappa_{ext}}{2}\left(x_{1}^2+x_{2}^2\right) +
\frac{\gamma_{1}'}{2}\left(x_{1}^2-x_{2}^2\right)+\gamma_{2}'x_{1}x_{2},   
\label{eq:(perturb)}
\end{equation}
where $\psi_{p}$ is the potential of primary lens, given by
equation~(\ref{eq:(one component)}) or (\ref{eq:(two component)}) in
case of elliptical lenses or given by some other profile and
$\left(\gamma_{1}',\gamma_{2}'\right)$ denotes the component of
external shear ($\gamma_{ext}$).  

The effect of the external convergence ($\kappa_{ext}$) is equivalent
to the addition of a constant mass sheet in the lens model, which
simply changes the total strength of the primary lens.  
As a result, the critical redshift for unstable singularities changes,
but neither the unstable singularities vanish nor the location
of $A_{3}$-lines in lens plane changes due to the presence of external
convergence.  
On the other hand, the presence of external shear ($\gamma_{ext}$)
shifts the location of $A_{3}$-lines significantly and as a result it
changes the singularity map for a given lens model. 
The presence of external shear can also introduce or remove point
singularities.  
The effect of external shear with a fixed value of external
convergence in case of a one-component elliptical lens model,
(equation~(\ref{eq:(one component)})) is shown in
figure~(\ref{fig:stability1}).  
One can see that, for non-zero external shear, two extra
hyperbolic umbilics occur in the lens plane along the major or minor
axis depending on the values of shear components.  
As we increase the amount of external shear, this extra pair of
umbilics move towards already existing umbilics and merge with them.  
This implies that introducing a finite amount of external shear can
also remove the already existing point singularities from the
singularity map and it is possible (in highly symmetric case) to have
a singularity map without any point singularities. 
The amount of external shear
$\gamma_{ext}=\left(\sqrt{\gamma_{1}'^2+\gamma_{2}'^2}\right)$, under
which point singularities shift but remain in the lens plane depends
on the type of the singularity. 
In case of hyperbolic umbilic, it is of the order of $10^{-3}$.
Similarly, the amount of external shear for which a swallowtail
(elliptic umbilic) shifts but survives in the lens plane is of the
order of $10^{-4}\left(10^{-5}\right)$.  
But for some particular directions of external shear, the swallowtail
and elliptic umbilics show extra stability, i.e., the magnitude of
external shear under which these singularities remain in the lens
plane attain a higher value than the other cases.
This reinforces the impression from the qualitative study in the
last subsection that an elliptic umbilic is less stable as compared to
the hyperbolic umbilic and swallowtail. 

\subsection{Abell 697}
\label{ssec:Abell697}

After testing our approach with simple model lenses, we apply the
algorithm to a real lens to illustrate the utility and efficacy of our
approach.  
We work with the cluster lens Abell 697 $\left(z\, =\, 0.282\right)$.  
We use the data for the lens from RELICS~\citep{Cibirka_2018,
  Coe_2019}. 
The reason for choosing the Abell 697 for the preliminary analysis is 
the relative simplicity of the critical lines in the lens plane.
The study of more complicated lenses is under consideration, and the
results for a large set of clusters will be presented in a forthcoming
paper along with a statistical analysis of occurrence of point
singularities.  
The cosmological parameters used in the calculation of different
angular diameter distances are: $H_{0}= 70\, km s^{-1} Mpc^{-1}$,
$\Omega_{\Lambda}=0.7$, $\Omega_{m} = 0.3$.  
Figure~\ref{fig:abell697} shows the singularity map along with the
critical lines and caustics in image and source plane for Abell 697.   
Here we only considered the central region of Abell 697 with size $440
\times 440 $pixels (1 pixel =$0.06"$) ~\citep{Cibirka_2018}.  
We can see that the dominant component here is like an elliptical lens
and there is a lot of small scale structure contributed by other
components in the lens.
The role of other components is to increase the length of
$A_{3}$-lines and also to introduce point singularities.

The top panel in figure~\ref{fig:abell697} shows the image formation
near a swallowtail singularity for a source at redshift $z_s=0.67$.  
The second and third panel shows the image formation for a source at
redshift $z_s=0.82$ for two different source positions.  
The bottom panel shows the image formation for a source at redshift
$z_s=2.0$.
Here the source position is chosen in such a way so
  that it can reproduce the image formation for system 1
  in~\citep{Cibirka_2018}.  
One can see that we were able to reproduce the four images for system
1 along with the fifth image, which was not observed due to the
contamination from BCG, as mentioned in~\citep{Cibirka_2018}. Since we
considered a circular source, the shape of the images can be different
from~\citep{Cibirka_2018}.  
As one can see from the bottom panel, one pair of hyperbolic umbilic
is still outside the critical curves.  
This means that the critical redshift for this pair is higher than the
$1.1$.
Locations of these singularities are optimal sites for searching for
faint sources at high redshifts. 

\section{Conclusions}
\label{sec:Conclusions}

We have analyzed stable and unstable singularities that can occur in
strong gravitational lensing.
In order to locate these singularities, we have implemented algorithms
which take lens potential as an input.
We have applied our algorithm in the case of simple lens models as well as
a real lens. 
Singularity map, which comprises all these singularities provides a
compact representation of the given lens model in the lens plane.  
The presence of these unstable singularities in the singularity map
can be used to constrain the lens model if we can find a lensed source
in the vicinity.
Magnification is very large in the vicinity of these singularities and
each of these singularities has a characteristic image formation that
can be used to identify the singularities. Multiple
  images in these characteristic image formations lie in a very
  compact region (of the order of a few arcsec around the singular
  point) of the lens plane.  
Further, the regions with $A_{3}$-lines and point singularities are
obvious targets for deep surveys that use gravitational lenses to
search for very faint sources at high redshifts. 

The singularities can be identified using the characteristic image
forms.
In case of $A_4$ points or swallowtail, we get four images in a
straight line: the images form an arc with a radius of curvature much
larger than the distance from the cluster centre.
Abell 370 has an image system of this type.  
The hyperbolic umbilic (purse) has an image formation of a ring or a
cross centered away from the centre of the lens.
Further, in this case the radius of curvature of the ring is much
smaller than the characteristic radius of the lens system.
Such an image system has been seen in Abell 1703 \citep{Orban 2009}. 
The elliptic umbillic (pyramid) has images radiating out from a
centre, these do not show any tangential distortion.
The centre of the image system need not coincide with the centre of
the lens system.
To the best of our knowledge such an image system has not been seen so
far. 

We have studied the dependency of unstable singularities on lens
parameters as well as on the external shear.
The magnitude of external shear under which these singularities remain
in the singularity map is different for different singularities.
This is of the order of, $10^{-3}$, $10^{-4}$, $10^{-5}$ in case of
hyperbolic umbilic, swallowtail and elliptic umbilic, respectively.
Thus the elliptic umbilic is most sensitive to perturbations in
lensing potential and hence is the most unstable.

The somewhat unstable nature of such singularities can be put to good
use in two ways: finding characteristic image formations can be used
to constrain lens models, and, with multiple constraints on the lens
model we can potentially invert the problem and constrain redshifts of
sources to better than what can be achieved with photometric
redshifts.
These aspects will be investigated and addressed in more detail in a
follow up study of known lens systems. 

We have applied our approach in case of simple lens models and one
real lens: Abell 697.
We are studying other lens systems using our approach and an atlas of
lens singularities and their statistical analysis will be presented in
a forthcoming paper. 
Such an atlas can be of use for refinement of lens models with further
observations and also for targeting specific regions in searches for
very faint sources at high redshifts.
Along with an atlas of lens models we also propose to construct an
atlas of variations around the characteristic image forms.
Such an atlas of image forms can be used for training machine learning
programs, e.g., see \citet{2019MNRAS.tmp.1298D}. However, for a
complete analysis of the lens system, one requires detailed modeling
of the gravitational lens.  

\section{Acknowledgements}

AKM would like to thank CSIR for financial support through research
fellowship  No.524007.
This research has made use of NASA's Astrophysics  Data  System
Bibliographic Services.
We acknowledge the HPC@IISERM, used for some of the computations
presented here.
Authors thank Professor D Narasimha for useful discussions and comments. 
JSB thanks Professor Varun Sahni and Professor Sergei Shandarin for
insightful discussions on singularities and caustics, and also for
useful comments on the manuscript.
Authors thank Richard Ellis, Prasenjit Saha and Liliya Williams for
detailed comments on the manuscript. 
Authors also thank Ms Soniya Sharma who worked on some aspects of this
problem for her MS thesis.


\bsp	
\label{lastpage}
\end{document}